**Estimates of Population Highly Annoyed from Transportation Noise in the United States: An Unfair Share of the Burden by Race and Ethnicity**

Ching-Hsuan Huang [a*], Edmund Seto [a]

[a] Department of Environmental and Occupational Health Sciences, School of Public Health, University of Washington

*Corresponding author.
E-mail address: hsuan328@uw.edu (C. -H. Huang)

**Abstract**
Transportation is one of the most pervasive sources of community noise. In this study, we used a spatially-resolved model of transportation-related noise with established transportation noise exposure-response functions to estimate the population highly annoyed (HA) due to aviation, road, and railway traffic sources in the United States. Additionally, we employed the use of the Fair Share Ratio to assess race/ethnicity disparities in traffic noise exposures. Our results estimate that in 2020, 7.8 million (2.4%) individuals were highly annoyed by aviation noise, while 5.2 million (1.6%) and 7.9 million (2.4%) people were highly annoyed by rail and roadway noise, respectively, across the US. The Fair Share Ratio revealed that Non-Hispanic Asian, Black, NHPI, and Other, and Hispanic populations were disproportionally highly annoyed by transportation noise nationwide. Notably, Hispanic populations experienced the greatest share of high annoyance from aviation noise (1.69 times their population share). Non-Hispanic Black populations experienced the greatest share of high annoyance from railway noise (1.48 times their population share). Non-Hispanic Asian populations experienced the greatest share of high annoyance from roadway noise (1.51 times their population share). Analyses at the state and Urban Area levels further highlighted varying disparities in transportation noise exposure and annoyance across different race ethnicity groups, but still suggested that Non-Hispanic White populations were less annoyed by all sources of transportation noise compared to non-White populations. Our findings indicate widespread presence of transportation noise annoyance across the US and emphasize the need for targeted source-specific noise mitigation strategies and policies to minimize the disproportionate impact of transportation noise in the US.

## 1. Introduction

### 1.1 Background on noise-induced annoyance

Environmental noise exposure has been associated with various adverse health effects, including elevated cardiovascular disease morbidity and mortality (Hahad et al., 2019; Münzel et al., 2021), obesity (Cramer et al., 2019), sleep disturbances (Muzet, 2007), compromised mental health, and heightened annoyance levels (Gong et al., 2022). Noise-induced annoyance, commonly identified as a stress reaction consisting of emotional, cognitive, psychological, and behavioral responses to long-term environmental noise exposure, has been adopted as one of the primary indicators to quantify community response to, and impact from noise (Ouis, 2001; Stallen, 1999). Given its acoustical (e.g., perceived noise loudness) and non-acoustical (e.g., subjective perception of noise) nature, robust assessment is essential in evaluating community noise annoyance. Moreover, population noise exposure and annoyance assessments enable comparison across different community settings and times in relationship to transportation and other built environment developments. To facilitate such comparisons, the term "highly annoyed" was introduced by Schultz in 1978, defining population highly annoyed as the people who respond to the upper 27% of an 11-point numeric annoyance scale or the upper 29% of a 7-point numeric annoyance scale (Schultz, 1978). Later, the International Commission on Biological Effects of Noise (ICBEN) proposed an alternate definition, identifying the population highly annoyed as those who respond to the upper 40% of a 5-step verbal response scale. This definition was formalized and published as a Technical Specification (TS) in 2003 (ISO/TS 15666: 2003) by the International Standards Organization (ISO). To enhance data harmonization, the updated ISO/TS 15666 in 2021 (ISO/TS 15666: 2021) introduced a standardized scoring and naming conventions for “highly annoyed” (Clark et al., 2021; International Organization for Standardization, 2021). Despite variations in noise metrics, definitions of highly annoyed were commonly adopted by scientific studies and governmental agencies as the critical health outcome of long-term noise exposure, including in the United States (US) by the Federal Interagency Committee on Noise (FICON) and internationally by the World Health Organization (WHO) (Federal Interagency Committee on Noise, 1992; World Health Organization, 2018).

### 1.2 Exposure-response functions (ERFs) for assessing community noise annoyance

The earliest development of exposure-response functions (ERFs) for traffic noise exposure and community annoyance in the US can be traced back to 1978, when Schultz introduced the concept of high annoyance (Schultz, 1978). The “Schultz curves” synthesized in this study were later updated by Fidell et al. (Fidell et al., 1991) and FICON (Fidell, 2003). Several European studies have also synthesized annoyance curves from socio-acoustic surveys based on different noise metrics, including day-night-level (Ldn/DNL) and day-night-evening level (Lden/DENL). In late 1990, Miedema and Vos (Miedema & Vos, 1998) augmented the data examined by Schultz and Fidell et al. (Fidell et al., 1991; Schultz, 1978) and derived curves for the relationship between Ldn/DNL and percentage of population highly annoyed (%HA) by traffic noise. Later, Miedema and Oudshoorn (Miedema & Oudshoorn, 2001) presented annoyance curves with Ldn/DNL and Lden/DENL used as the main noise metric. These curves were previously used in a US-based study assessing traffic-induced noise annoyance (Seto et al., 2007), and were used by different agencies to develop suitable noise limit. For example, the US Federal Aviation Administration (FAA) uses a 65 dB DNL noise threshold for aviation noise policy assessment (Federal Aviation Administration, 2022).

With the goal of protecting human health from the harmful effects of environmental noise exposure, the WHO has recently revised its Environmental Noise Guidelines for the European Region to provide recommendations on protecting individuals from exposure to environmental noise from the sources of transportation, including aircraft, railway, road, wind turbines, and leisure noise. The guidelines address five key health outcomes associated with environmental noise impacts, encompassing cardiovascular diseases, sleep disturbance, annoyance, cognitive and hearing impairment, and reading skills and oral comprehension in children (World Health Organization, 2018). These recommendations are based on a meta-analysis conducted by Guski et al., which models ERFs for aircraft, road, railway, and wind turbine noise levels and %HA, drawing from 57 field research studies conducted between 2000 and 2014 in Europe and Asia (Guski et al., 2017). Although these ERFs faced controversy due to their exclusive focus on acoustic factors (Fidell et al., 2022; Gjestland, 2018; UK Civil Aviation Authority, 2021), they have been applied in different studies to estimate the impact of traffic noise, mainly in European countries (Tobollik et al., 2019; Veber et al., 2022).

**1.3 Environmental noise mapping**

Mapping environmental noise levels and assessing population exposures to noise was mandated by the European Union (EU) Environmental Noise Directive (END) for the member states as part of action plans to implement potential noise mitigation. A 55 dB Lden was defined by the END for reporting excess exposure (Murphy & King, 2010). Several European countries, such as Italy, have established land-use based noise limits in alignment with the END (Biondi, 2000). In the US, however, no such requirement was made possibly due to the lack of city, state, and federal mandates and funding often leads to the absence of such nationwide population noise exposure and health impact assessments (Bronzaft, 2017; Hammer et al., 2014). Although the US Environmental Protection Agency (EPA) conducted the first nationwide noise pollution study in 1981, previous research in the US has primarily focused on mapping noise exposure and evaluating its health effects within specific cities and regions (Kim et al., 2012; Lee et al., 2014; Seto et al., 2007).

Recent efforts to map noise levels and exposures across the continental US have involved the development of spatial models based on noise monitoring data, particularly within national parks, as mandated by the US National Park Service (NPS) (Collins et al., 2020; Mennitt & Fristrup, 2016; Mennitt et al., 2014; Sherrill, 2012; US EPA, 1981). The advancement in wireless sensor networks (Alías & Alsina-Pagès, 2019; Liu et al., 2020; López et al., 2020), and their integration with artificial intelligence and machine learning approaches (Ascari et al., 2023;

Asdrubali & D'Alessandro, 2018; Fredianelli et al., 2022; Nourani et al., 2020; Shah et al., 2020; Steinbach & Altinsoy, 2019) have shown promising applications in various European studies. This integration holds the potential to enhance noise prediction and facilitate real-time noise detection, enabling comprehensive data collection and analysis (Bolognese et al., 2023). Moreover, novel noise assessment methodologies, such as Statistical Pass-By (Ascari et al., 2022; Moreno et al., 2023), provide a more detailed understanding of noise sources and their impact. In the context of the current development of electric vehicles (EV), the European Common NOise aSSessment methOdS (CNOSSOS-EU) has been proposed to harmonize assessment of EV noise emission (Licitra et al., 2023; Pallas et al., 2016). Together, these methods could pave the way for nationwide noise mapping, emerging as invaluable tools in environmental noise management, particularly in the US context.

The US Bureau of Transportation Statistics (BTS) developed the National Transportation Noise Map (NTNM), which modeled the traffic noise level across the continental US, Alaska, and Hawaii (US Bureau of Transportation Statistics, 2022). With the purpose of facilitating stakeholders in planning and prioritizing transportation investment, it provides noise level estimates of different transportation sources, including aviation, rail, and road traffic. A recent study by Seto and Huang describes the development of the National Transportation Noise Exposure (NTNE) map, which overlayed the most recent BTS NTNM (2020) with population estimates from the US census data at the census block group levels with improved spatial accuracy (Seto & Huang, 2023). Compared to previous studies that have utilized the BTS NTNM data, the NTNE map provides computation of block group level population exposures, which accounts for uneven spatial distributions of populations and noise levels within census tracts. This enables estimating population noise exposures and applying noise ERFs in health impact assessments with improved spatial granularity. While transportation noise is pervasive, some population groups, including children and the elderly, are found to be more vulnerable to its adverse health effects (Carrier et al., 2016; Collins et al., 2020; Stansfeld & Clark, 2015; van Kamp & Davies, 2013). There is a growing concern that communities of color and individuals with low socioeconomic status may experience disproportionate exposure to transportation noise. These disparities in exposure could be influenced by factors such as land-use management practices and socioeconomic conditions (Morello-Frosch, 2002). Therefore, assessing population exposures to transportation noise and estimates of population annoyance at finer spatial resolutions is crucial for understanding the impact of noise on community health.

### 1.4 Study aims

The main objective of this study is to utilize the BTS NTNM and NTNE map data, along with established ERFs, to estimate the population highly annoyed by transportation noise from different sources, including aviation, road, and railway traffic, and to explore and racial/ethnic disparities in transportation noise exposures in the US. We provide summaries of population numbers and proportions exposed to and highly annoyed by various noise levels at state, national, and Urban Area levels. In the Discussion, we describe potential limitations associated with the use of BTS NTNM and NTNE map data, as well as the existing ERFs, for estimating the population highly annoyed by transportation noise, and opportunities for further study of noise health impacts for US communities.

## 2. Materials and Methods

### 2.1 Data preparation

We obtained the NTNE map dataset developed from a previous study by Seto and Huang (Seto & Huang, 2023). Briefly, the dataset was prepared by overlaying the most recent (2020) spatial raster files of noise levels at 30 m spatial resolution related to aviation, rail, and roadway traffic from the BTS NTNM with the spatial population estimates from the 5-year (2016 - 2020) American Community Survey (ACS) at the census block-group level. The exposed population of each census block group was overlapped with LAeq noise categories of 45 - 50, 50 - 60, 60 - 70, 70 - 80, 80 - 90, and 90+ dB (greater than or equal applied to the lower bound of each category, and less than applied to the upper bound), and aggregated to the census tract level. The BTS NTNM files were organized by state and incorporated the modeled noise levels attributed to aviation, rail, and roadway traffic. Detailed descriptions of the modeling procedures can be found in the BTS documentation (US Bureau of Transporation Statistics, 2020). Noise level files for separate transportation sources are also available from the BTS. In the current analysis, we retrieved the BTS NTNM and the NTNE map data separately for each transportation source, including aviation, rail, and roadway.

### 2.2 Estimations of the number of populations highly annoyed by traffic noise and exploration of race/ethnicity disparities

We first calculated the %HA of each noise exposure category based on the WHO ERFs derived by Guski et al. shown in **Table S1** for each transportation noise source (Guski et al., 2017; World Health Organization, 2018). The method of the calculation was described in Supplementary Materials **Table S2**. We multiplied the population numbers acquired from the NTNE map data by the corresponding %HA at the census tract level to estimate the number of people that are highly annoyed under each exposure level category and for each transportation source. To estimate the %HA of each exposure level category by state, we aggregated the number of people that are highly annoyed and divided it by the total population number at the state level. We also aggregated the data by the top 5% most populous Urban Areas (i.e., the territory with 50,000 or more people residing in defied by the US Census Bureau) (US Census Bureau, 2021). The Census' Urban Area classification is meant to represent densely developed areas, and encompass residential, commercial, and other non-residential urban land uses. Urban Areas include the settled urban core that meet a minimum population density requirement along with adjacent densely settled areas that are linked to the urban core. The Census' Urban Areas correspond to the urban cores of Metropolitan Statistical Areas.

The race/ethnicity groups considered in the analysis included Hispanic individuals and Non-Hispanic groups (Asian, Black, White, American Indian and Alaskan Native (AIAN), Native Hawaiian and Pacific Islander (NHPI), and individuals with other or more than two races/ethnicities). To quantify the racial/ethnicity disparity for traffic noise exposure, the following ratio (denoted as "Fair Share Ratio" hereafter) was calculated at the state and the Urban Area levels:

$$\text{Fair share } ratio_i = \frac{\#HA_i / total\ \#HA}{\#tot_pop_i / \#tot_pop}$$

where $\#HA_i$ is the population number highly annoyed by source-specific traffic noise (aviation, rail, or roadway) of race/ethnicity group $i$; $total\ \#HA$ is the total population number highly annoyed of all race/ethnicity groups; $\#tot_pop_i$ is the total population number of race/ethnicity group $i$; and $\#tot_pop$ is the total population number of all race/ethnicity groups. This Ratio metric is conceptually similar to the index of dissimilarity (ID), which also compares a group's share of population health and the group's population share (as a difference, rather than a ratio), but unlike the ID retains the inequity information for each race/ethnic group *I*, rather than collapsing it into a single metric like the ID (Wagstaff et al., 1991).

To illustrate the Ratio approach, consider **Figure 1**, which shows three different hypothetical scenarios for how the burden of high annoyance to noise might be distributed among race/ethnicity groups in a population. For each of these scenarios, the inner sections of the nested pie chart illustrate the percentage share of each race/ethnicity group in the population. In all three scenarios, the hypothetical population is made up of 40% White, 20% Black, 8% Asian, 2% Other, and 30% Hispanic individuals. The outer ring of the nested pie chart illustrates the percentages of affected and non-affected (i.e., highly annoyed vs not) for each group. The percentage affected in each race/ethnicity group out of the total number of affected individuals is also shown as the percent affected.

In Scenario A (**Figure 1(A)**), within each of the race/ethnicity groups, there are equal numbers of affected (i.e., highly annoyed) and unaffected (i.e., not highly annoyed) individuals. The White group has the greatest percentage of affected individuals, but it matches the group's population share (40% of the underlying population and 40% out of the affected population). This scenario illustrates a situation in which there is a fair share of health burden experienced by each group in the population.

Next, in Scenario B (**Figure 1(B)**), within each of the race/ethnicity groups, there is a larger proportion of affected versus unaffected individuals. But the percentage affected in each group is the same as each group's population share. This scenario illustrates a situation in which there is a large amount of health burden experienced within the population, but each race/ethnicity group has a fair share of the burden.

Finally, in Scenario C (**Figure 1(C)**), within each of the race/ethnicity groups, there are varying proportions of affected versus unaffected individuals. White affected individuals only account for 36.4% of the total health burden in the population despite White individuals accounting for 40% population share (36.4%/40% < 1). Conversely, Asian and Hispanic affected individuals account for 10.9% and 32.7% of the total health burden, which is more than their population share of 8% and 30%, respectively (10.9%/8% > 1 and 32.7%/30% > 1). This scenario illustrates a situation in which some race/ethnicity groups (Asian and Hispanic) are disproportionately impacted. It also illustrates how the Fair Share Ratio (i.e., the ratio of the percentage affected in each group to each group's population share) serves as an indicator of health inequality.

Data analysis was conducted using Rstudio version 2022.12.09 with packages sf (version 1.0-9), stars (version 0.6-0), tidycensus (version 1.3.2), dplyr (version 1.1.0)", and ggplot2 (version 3.4.1).

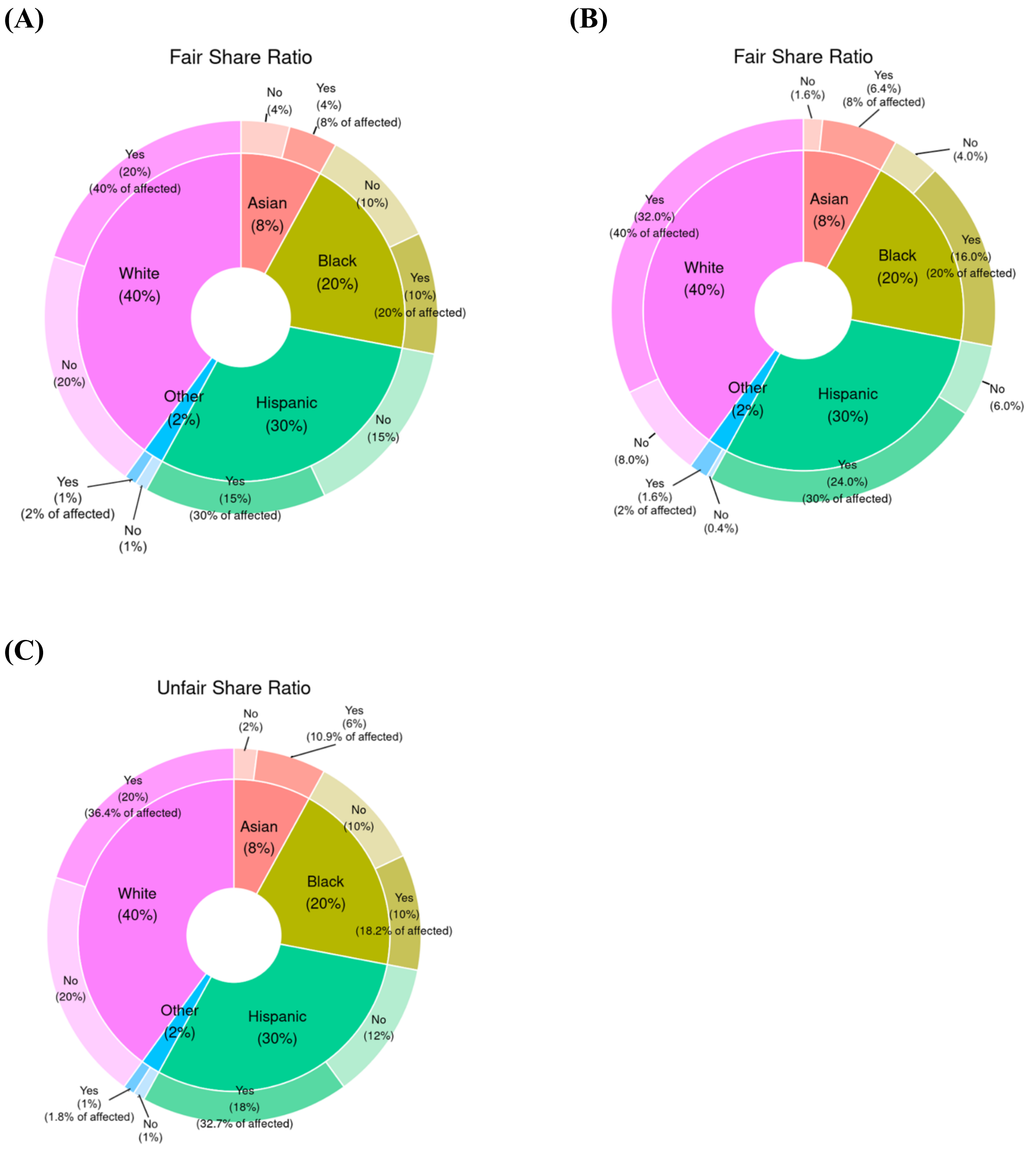

**Figure 1. Three hypothetical scenarios illustrating fair (scenarios A and B) and unfair sharing (scenario C) of health burden among race/ethnicity groups in a population.**

## 3. Results

### 3.1 Population highly annoyed by states and by race/ethnic groups

Population numbers and percentages of the population highly annoyed by aviation, rail, and roadway noise are presented by states and the District of Columbia, and for the nation as a whole in **Table 1**. In 2020, an estimated 7.8 million (2.40% of the total population), 5.2 million (1.60% of the total population), and 7.9 million (2.43 % of the total population) individuals were highly annoyed by aviation, rail, and roadway noise, respectively, in the US. The census tract level map of the %HA exposed to aviation, rail, and roadway noise for the US is shown in **Figure 2** (maps of Alaska and Hawaii are provided in Supplemental Materials as **Figures S1** and **S2**). Overall, Nevada had the highest %HA exposed to aviation noise, whereas Illinois and Washington DC had the highest %HA exposed to rail and roadway noise, respectively (**Table 3** and **Figure S3)**.

While the population numbers exposed to traffic noise were observed to be related to the total population of each state (e.g., populous states such as California and Texas tend to have larger numbers of people exposed to transportation noise), the %HA exposed to traffic noise was not directly related to the total population number of each state. For example, despite California having a population over three times larger than Illinois, the percentage of highly annoyed individuals exposed to aviation and rail noise in California was not higher than that in Illinois. This suggests the importance of spatial relationships between local transportation infrastructure and the population distributions in shaping the effects of transportation noise on community annoyance. Furthermore, the spatial patterns of populations exposed vary between different noise sources, with populations exposed to noise being concentrated where there are major aviation traffic sites, while those exposed to rail and roadway noise can be seen to follow the networks of inter and intrastate rail and roadway lines, respectively (**Figure 2 (B) and (C)**). Compared to aviation and rail noise exposure, the spatial pattern of the population highly annoyed by roadway noise exposure was found to be more dispersed, with the population highly annoyed by roadway traffic noise being greater than 1 % in almost all states (**Figures 2 (C) and S3**).

Population numbers and percentages of the population highly annoyed by aviation, rail, and roadway noise estimated using noise metric LAeq are presented at the state level and for the nation in **Table S3.** An estimated 4.8 million (1.49% of the total population), 2.9 million (0.89 % of the total population), and 5.6 million (1.73 % of the total population) people were highly annoyed by aviation, rail, and roadway noise, respectively.

**Table 1. Population numbers and proportions exposed and highly annoyed by traffic noise [a] by state.**

| State | Aviation | | | Rail | | | Roadway | | | Total population |
|---|---|---|---|---|---|---|---|---|---|---|
| | # exposed | # HA | %HA [b] | # exposed | # HA | %HA [b] | # exposed | # HA | %HA [b] | |
| Alabama | 160,984 | 49,045 | 1.00 | 318,962 | 47,400 | 0.97 | 412,832 | 76,142 | 1.56 | 4,893,186 |
| Alaska | 89,198 | 27,508 | 3.73 | 1,267 | 207 | 0.03 | 50,768 | 9,304 | 1.26 | 736,990 |
| Arizona | 809,654 | 244,220 | 3.40 | 182,656 | 26,433 | 0.37 | 1,107,919 | 211,571 | 2.95 | 7,174,064 |
| Arkansas | 73,322 | 22,206 | 0.74 | 301,011 | 43,780 | 1.45 | 211,297 | 36,522 | 1.21 | 3,011,873 |
| California | 4,629,028 | 1,404,679 | 3.57 | 4,353,460 | 640,191 | 1.63 | 8,549,337 | 1,701,228 | 4.32 | 39,346,023 |
| Colorado | 468,329 | 135,764 | 2.39 | 247,293 | 36,644 | 0.64 | 684,332 | 121,291 | 2.13 | 5,684,926 |
| Connecticut | 60,885 | 17,927 | 0.50 | 286,843 | 44,835 | 1.26 | 433,272 | 76,334 | 2.14 | 3,570,549 |
| Delaware | 63,940 | 18,865 | 1.95 | 65,367 | 9,854 | 1.02 | 97,045 | 18,051 | 1.87 | 967,679 |
| District of Columbia | 54,693 | 16,197 | 2.31 | 132,407 | 18,743 | 2.67 | 223,474 | 38,012 | 5.42 | 701,974 |
| Florida | 2,799,135 | 851,762 | 4.01 | 849,541 | 124,203 | 0.59 | 2,443,448 | 464,037 | 2.19 | 21,216,924 |
| Georgia | 530,715 | 160,057 | 1.52 | 627,419 | 95,551 | 0.91 | 823,884 | 149,958 | 1.43 | 10,516,579 |
| Hawaii [c] | 76,582 | 22,715 | 1.60 | NA | NA | NA | 201,337 | 40,059 | 2.82 | 1,420,074 |
| Idaho | 91,338 | 26,037 | 1.48 | 72,205 | 10,271 | 0.59 | 167,126 | 27,644 | 1.58 | 1,754,367 |
| Illinois | 2,215,975 | 655,961 | 5.16 | 4,049,438 | 619,756 | 4.87 | 1,833,532 | 330,485 | 2.60 | 12,716,164 |
| Indiana | 200,023 | 56,533 | 0.84 | 1,037,049 | 156,063 | 2.33 | 720,522 | 127,723 | 1.91 | 6,696,893 |
| Iowa | 75,323 | 21,684 | 0.69 | 381,843 | 55,380 | 1.76 | 294,781 | 53,147 | 1.69 | 3,150,011 |
| Kansas | 102,596 | 29,235 | 1.00 | 519,396 | 75,890 | 2.61 | 245,271 | 41,791 | 1.43 | 2,912,619 |
| Kentucky | 237,594 | 71,935 | 1.61 | 410,675 | 60,433 | 1.35 | 451,347 | 86,389 | 1.94 | 4,461,952 |
| Louisiana | 138,269 | 41,857 | 0.90 | 399,383 | 56,838 | 1.22 | 369,128 | 66,645 | 1.43 | 4,664,616 |
| Maine | 29,362 | 8,780 | 0.65 | 25,368 | 3,993 | 0.30 | 87,187 | 15,610 | 1.16 | 1,340,825 |
| Maryland | 344,848 | 101,223 | 1.68 | 408,102 | 61,024 | 1.01 | 757,101 | 136,859 | 2.27 | 6,037,624 |
| Massachusetts | 688,721 | 198,170 | 2.88 | 998,287 | 148,550 | 2.16 | 1,327,633 | 241,444 | 3.51 | 6,873,003 |
| Michigan | 283,559 | 80,872 | 0.81 | 273,011 | 40,493 | 0.41 | 1,211,684 | 219,416 | 2.20 | 9,973,907 |
| Minnesota | 379,459 | 111,726 | 2.00 | 632,152 | 92,349 | 1.65 | 532,477 | 93,753 | 1.67 | 5,600,166 |
| Mississippi | 194,765 | 58,324 | 1.96 | 192,856 | 28,238 | 0.95 | 176,841 | 30,670 | 1.03 | 2,981,835 |
| Missouri | 218,824 | 63,303 | 1.03 | 610,416 | 90,298 | 1.47 | 582,758 | 105,419 | 1.72 | 6,124,160 |
| Montana | 95,702 | 27,043 | 2.55 | 82,033 | 11,702 | 1.10 | 78,377 | 13,420 | 1.26 | 1,061,705 |
| Nebraska | 30,270 | 8,834 | 0.46 | 409,692 | 62,831 | 3.27 | 188,655 | 32,409 | 1.68 | 1,923,826 |
| Nevada | 556,796 | 170,749 | 5.63 | 36,633 | 5,465 | 0.18 | 326,131 | 62,837 | 2.07 | 3,030,281 |
| New Hampshire | 46,975 | 13,266 | 0.98 | 19,090 | 2,974 | 0.22 | 104,676 | 19,159 | 1.41 | 1,355,244 |
| New Jersey | 782,007 | 224,019 | 2.52 | 1,908,904 | 271,157 | 3.05 | 1,346,432 | 237,475 | 2.67 | 8,885,418 |
| New Mexico | 70,282 | 19,329 | 0.92 | 78,851 | 11,216 | 0.53 | 208,676 | 38,048 | 1.81 | 2,097,021 |
| New York | 2,850,300 | 806,750 | 4.13 | 4,379,198 | 653,141 | 3.35 | 4,250,947 | 731,900 | 3.75 | 19,514,849 |
| North Carolina | 497,338 | 156,830 | 1.51 | 196,572 | 29,640 | 0.29 | 829,020 | 160,131 | 1.54 | 10,386,227 |
| North Dakota | 43,388 | 12,015 | 1.58 | 94,754 | 12,955 | 1.70 | 64,458 | 10,775 | 1.42 | 760,394 |
| Ohio | 401,220 | 114,337 | 0.98 | 1,628,311 | 239,405 | 2.05 | 1,133,347 | 202,249 | 1.73 | 11,675,275 |
| Oklahoma | 221,688 | 65,386 | 1.66 | 373,410 | 54,604 | 1.38 | 303,905 | 53,170 | 1.35 | 3,949,342 |
| Oregon | 286,825 | 82,546 | 1.98 | 387,998 | 56,836 | 1.36 | 652,899 | 115,891 | 2.77 | 4,176,346 |
| Pennsylvania | 286,720 | 82,482 | 0.64 | 2,625,312 | 392,473 | 3.07 | 1,510,122 | 257,159 | 2.01 | 12,794,885 |
| Rhode Island | 33,989 | 10,196 | 0.96 | 59,478 | 9,148 | 0.86 | 166,361 | 28,687 | 2.71 | 1,057,798 |
| South Carolina | 310,166 | 91,825 | 1.80 | 188,399 | 29,571 | 0.58 | 368,272 | 65,606 | 1.29 | 5,091,517 |
| South Dakota | 66,894 | 20,014 | 2.28 | 33,352 | 4,567 | 0.52 | 72,487 | 12,852 | 1.46 | 879,336 |
| Tennessee | 401,831 | 121,175 | 1.79 | 225,059 | 32,765 | 0.48 | 628,464 | 114,146 | 1.69 | 6,772,268 |
| Texas | 2,472,543 | 742,927 | 2.59 | 3,041,805 | 455,480 | 1.59 | 3,589,199 | 667,926 | 2.33 | 28,635,442 |
| Utah | 248,177 | 69,474 | 2.20 | 259,584 | 39,269 | 1.25 | 442,667 | 82,329 | 2.61 | 3,151,239 |
| Vermont | 14,385 | 4,295 | 0.69 | 7,789 | 1,235 | 0.20 | 35,058 | 5,943 | 0.95 | 624,340 |
| Virginia | 506,869 | 148,813 | 1.75 | 710,621 | 106,907 | 1.26 | 932,114 | 173,307 | 2.04 | 8,509,358 |
| Washington | 877,683 | 270,096 | 3.60 | 467,682 | 67,329 | 0.90 | 998,193 | 176,478 | 2.35 | 7,512,465 |
| West Virginia | 27,602 | 8,222 | 0.45 | 69,908 | 10,588 | 0.59 | 115,906 | 20,911 | 1.16 | 1,807,426 |
| Wisconsin | 235,497 | 70,283 | 1.21 | 385,836 | 57,880 | 1.00 | 669,636 | 118,153 | 2.03 | 5,806,975 |
| Wyoming | 47,613 | 14,556 | 2.50 | 52,504 | 7,219 | 1.24 | 45,574 | 7,641 | 1.31 | 581,348 |
| US nationwide [d] | 26,429,879 | 7,852,047 | 2.40 | 35,099,183 | 5,213,774 | 1.60 | 43,057,905 | 7,928,106 | 2.43 | 326,569,308 |

[a] Greater than or equal to 54 dB Lden.

[b] Percentage out of the total population of the state.

[c] No rail data available.

[d] Includes the continental US, the District of Columbia, and the states of Alaska and Hawaii.

(A)

(B)

©

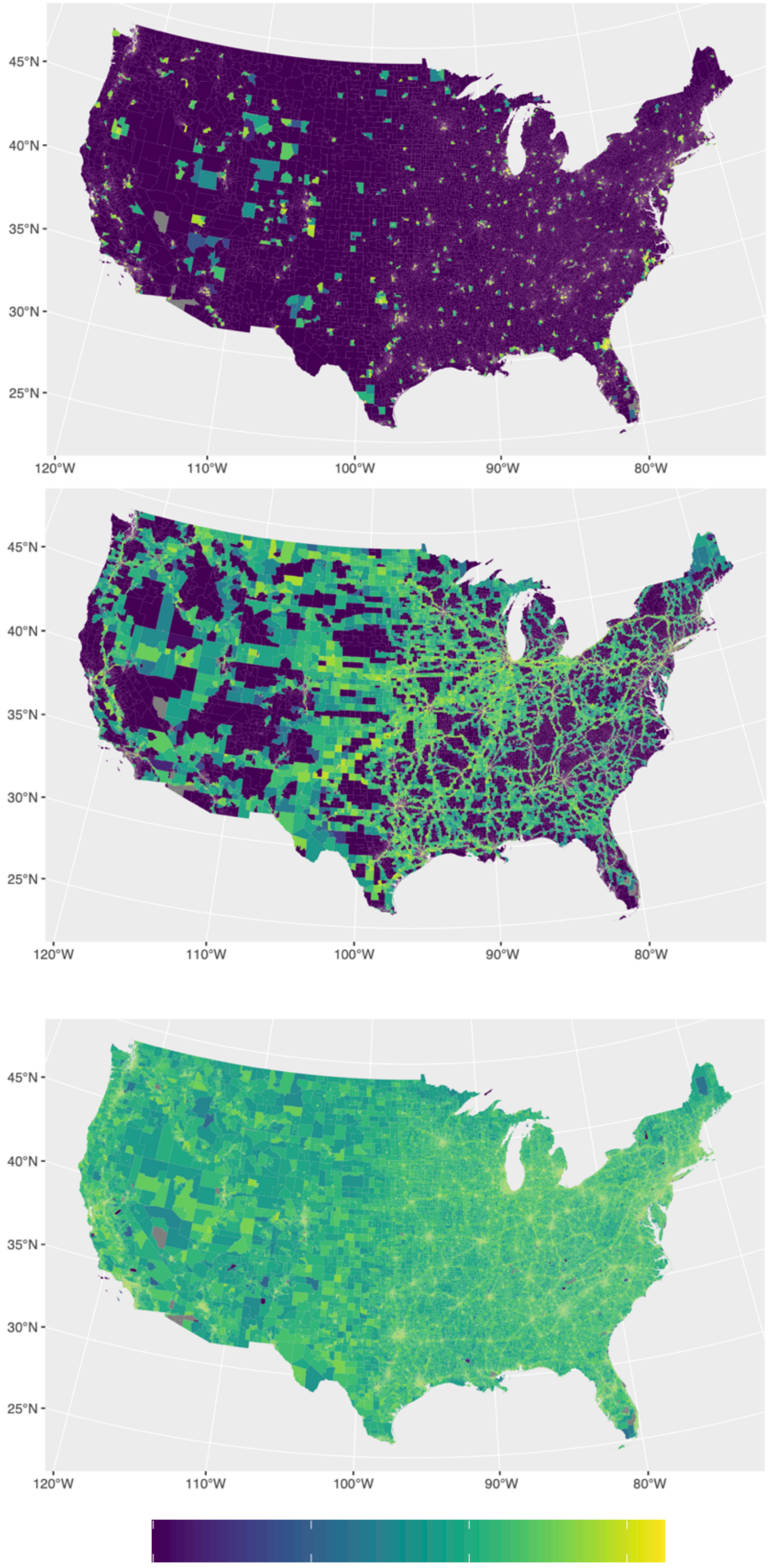

**Figure 2. The census-tract level maps of the percent population highly annoyed (%HA) by (a) aviation, (b) rail, and (c) roadway traffic noise in the continental US.**

The race/ethnicity-specific population numbers and proportions highly annoyed by transportation noise by states in the continental US and the states of Alaska and Hawaii are presented in **Tables S4 – S6.** Based on these numbers, the race/ethnicity-specific Fair Share Ratios were calculated and presented **in Figures 3, 4,** and **5**. Across the entire US, the Ratios for Non-Hispanic Asian, Non-Hispanic Black, Hispanic, Non-Hispanic NHPI, and Non-Hispanic Other were larger than one for aviation and roadway noise, indicating populations of these race/ethnicity groups were disproportionally affected by aviation and roadway noise (**Figures 3** and **5**). At the state level, Hispanic and Non-Hispanic Black populations also appeared to be disproportionally highly annoyed by all types of traffic noise, with more than 80% of the 51 states having the Fair Share Ratios for aviation, rail, and roadway noise larger than 1. On the other hand, the Non-Hispanic Asian population seemed to be disproportionally highly annoyed by aviation and roadway noise, with more than 78% of the 51 states having the Ratios for aviation and roadway noise that were larger than 1 (**Figures 3 and 5**). In some states, some race/ethnicity groups seem to be extremely disproportionately affected by traffic noise. For example, the Non-Hispanic Black population was found to be especially disproportionally annoyed by aviation noise in the states of Hawaii (HI), Kentucky (KT), Maine (ME), Missouri (MO), New Hampshire (MH), South Dakota (SD), Tennessee (TN), Vermont (VT), Washington (WA), and West Virginia (WV) (Ratios > 2) (**Figure 3**). The Non-Hispanic NHPI and other populations also seemed to be disproportionally highly annoyed by all types of traffic noise (**Figures 3, 4, and 5**).

Conversely, the Fair Share Ratios of the Non-Hispanic White population for aviation, rail, and roadway noise were generally lower than one for most states, suggesting they are less likely to be disproportionally highly annoyed by all sources of traffic noise.

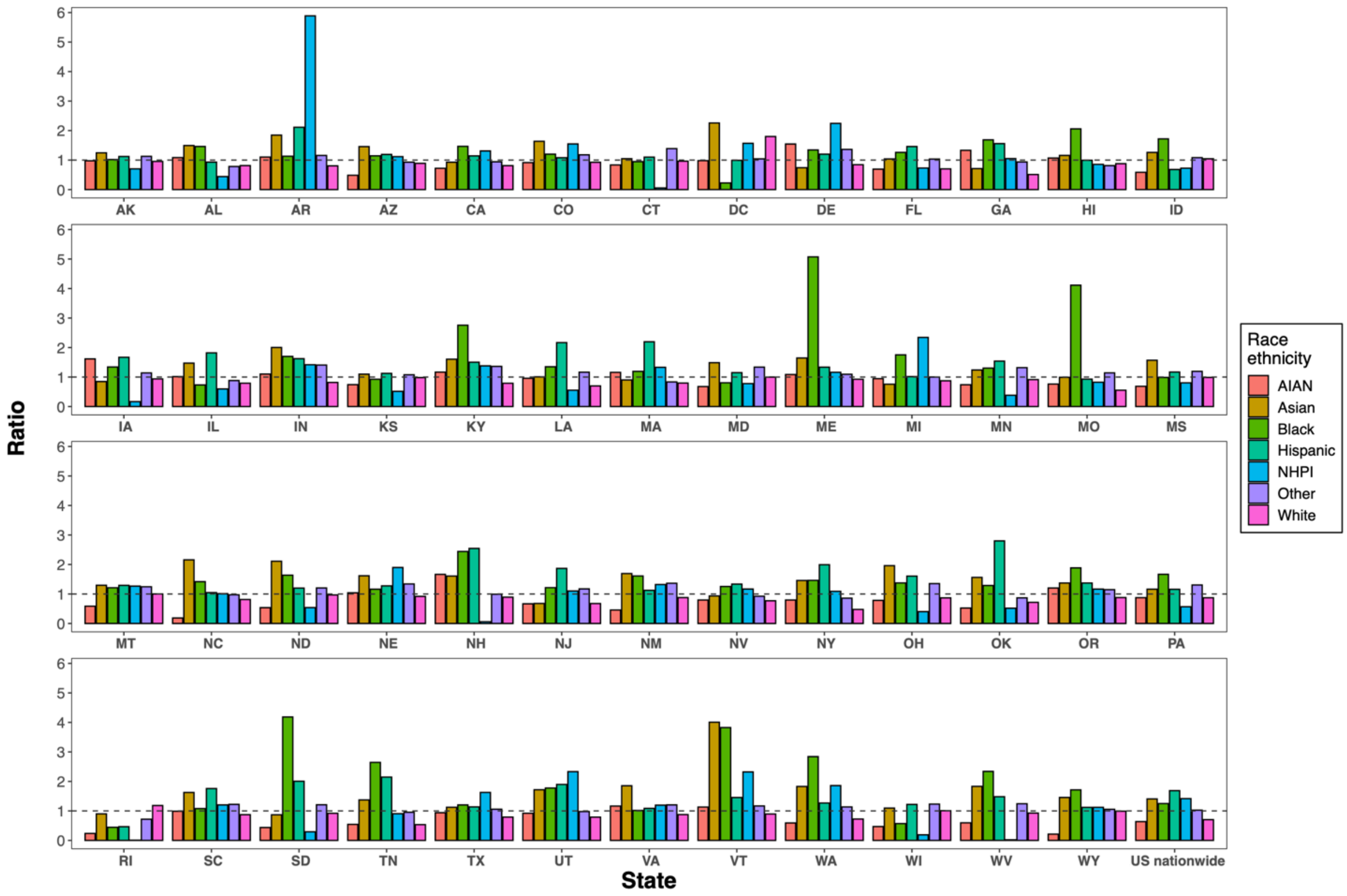

**Figure 3. Race/ethnicity-specific Fair Share Ratio of aviation noise by state.** The "US nationwide" category includes the data for the continental US and the states of Alaska and Hawaii. The dashed lines represent a ratio of 1.

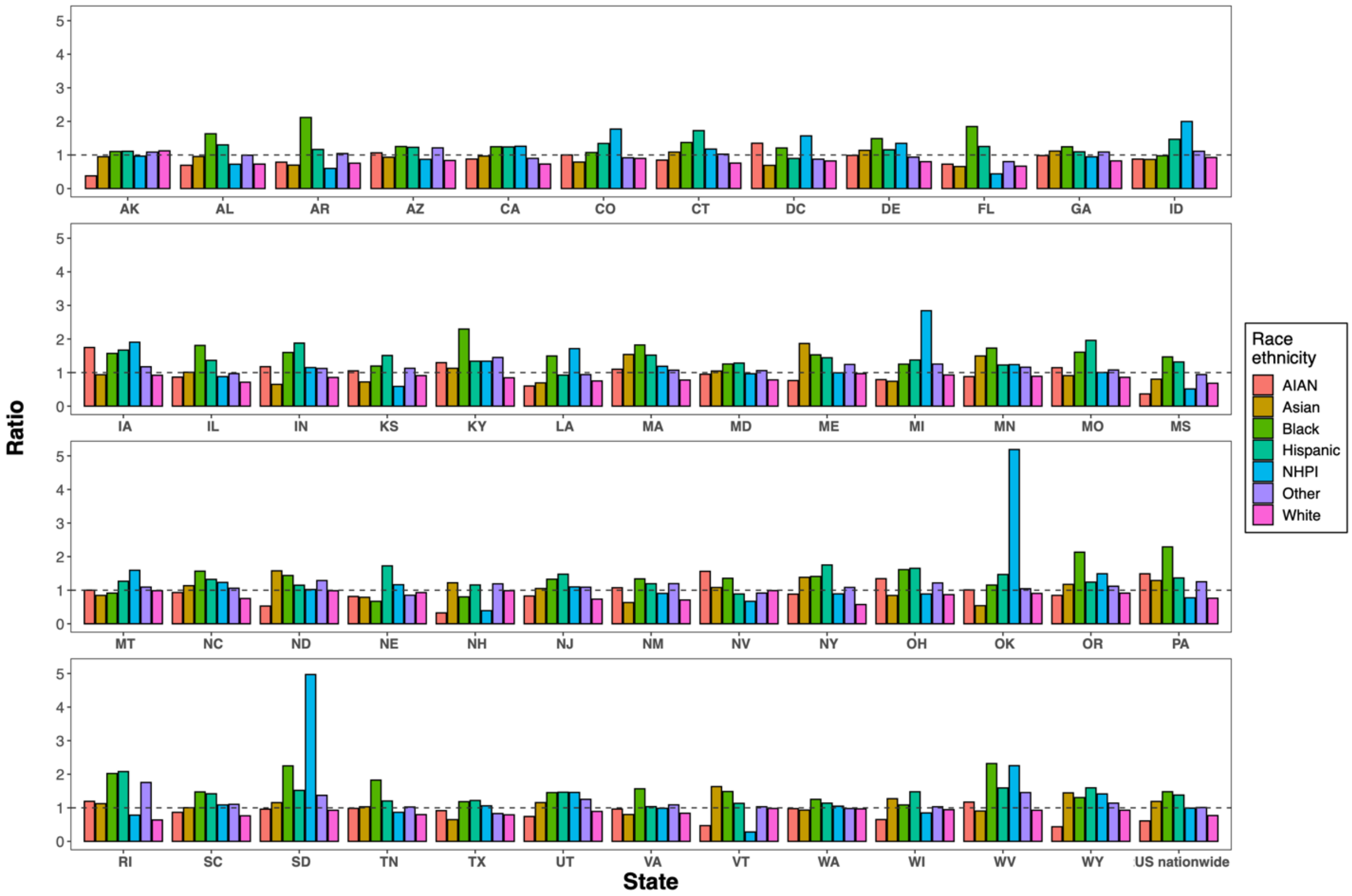

**Figure 4. Race/ethnicity-specific Fair Share Ratio of rail noise by state.** The "US nationwide" category includes the data for the continental US and the states of Alaska and Hawaii. The dashed lines represent a ratio of 1.

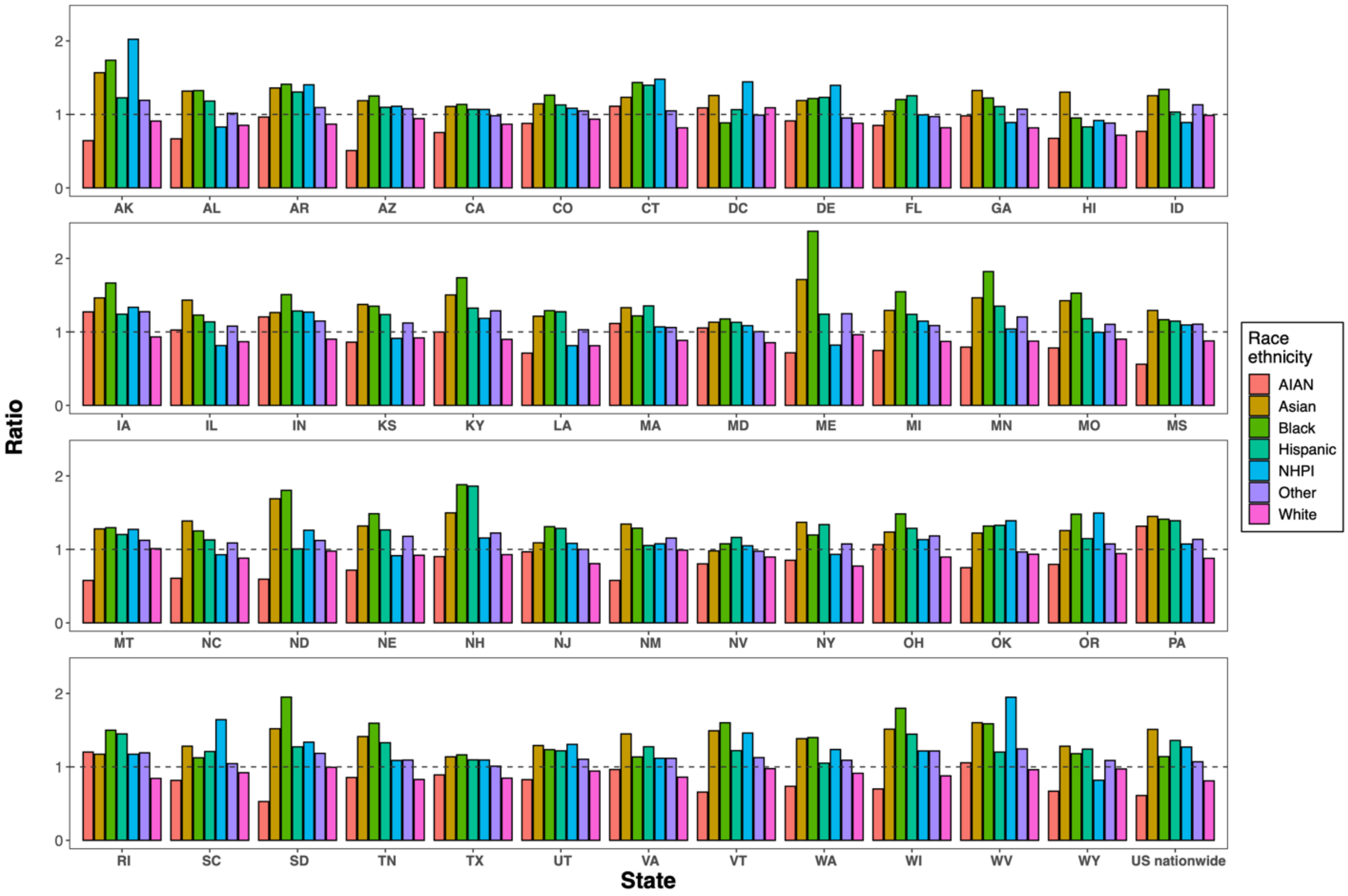

**Figure 5. Race/ethnicity-specific Fair Share Ratio of roadway noise by state.** The "US nationwide" category includes the data for the continental US and the states of Alaska and Hawaii. The dashed lines represent a ratio of 1.

### 3.2 Population highly annoyed by Urban Areas and by race/ethnic groups

Race/ethnicity-specific population numbers and percentages of the population highly annoyed by aviation, rail, and roadway noise for the 24 regions that account for the top 5% most populous Urban Areas in the US are presented in **Tables S7 – S9**. In 2020, approximately 121.2 million (37.1%) individuals lived in the top 24 Urban Areas nationwide. Among these 24 regions, it was estimated that 4.8 million, 2.8 million, and 4.2 million individuals were highly annoyed by aviation, rail, and roadway noise, respectively. These numbers account for 1.50%, 0.86%, and 1.31% of the total population nationwide. The spatial patterns of high %HA for different transportation noise sources were observed to overlap with the US Census Bureau's Urban Area classification (**Figures S1, S2, and S4**).

Based on the numbers shown in **Tables S7 – S9**, the race/ethnicity-specific Fair Share Ratios were calculated and are presented in **Figures 6 – 8** by the 24 regions that account for the top 5% most populous Urban Areas. In terms of aviation noise, the Non-Hispanic AIAN, Non-Hispanic Black, Non-Hispanic NHPI, and Hispanic population were found to be disproportionally highly annoyed, with more than 50% of the 24 regions having a Fair Share Ratio > 1. The Non-Hispanic Black population was most disproportionally highly annoyed in the St. Louis, Missouri Urban

Area, with a ratio indicating that they were 2.67 times more affected by noise annoyance than their population share. The Non-Hispanic NHPI population was most disproportionally highly annoyed in the San Francisco – Oakland, California Urban Area (Ratio = 2.33), followed by the Riverside – San Bernadino, California Urban Area (Ratio = 2.10). The Hispanic population was estimated to be most disproportionally highly annoyed in Boston, Massachusetts – New Hampshire – Rhode Island Urban Area (Ratio = 2.46). Non-Hispanic Asian and other populations in some of the 24 regions were also found to be disproportionally highly annoyed by the aviation noise. For example, the Non-Hispanic Asian population was found to be most disproportionally highly annoyed by aviation noise in the Riverside – San Bernardino, California Urban Area (Ratio = 2.58), whereas the Non-Hispanic Other population was estimated to be most disproportionally highly annoyed in the Riverside – San Bernadino, California Urban Area (Ratio = 1.42). The Non-Hispanic White population appeared to be less disproportionally highly annoyed by the aviation noise, with only those populations in the Baltimore, Maryland, Denver – Aurora, Colorado, Phoenix – Mesa, Arizona, San Antonio, Texas, and Washington, DC – Virginia – Maryland having a Ratio > 1 (**Figure 6**).

For rail traffic noise, over 50% of the 24 regions had more than 50% of the Non-Hispanic AIAN, Non-Hispanic Black, and Hispanic populations disproportionally highly annoyed. The Minneapolis – St. Paul, Minnesota – Wisconsin and the St. Louis, Missouri Urban Area had the most Non-Hispanic AIAN population (Ratio = 1.69) and Non-Hispanic Black population disproportionally highly annoyed (Ratio = 1.76) by rail traffic noise, respectively. The Hispanic population was found to be most highly annoyed in the Boston, Massachusetts – New Hampshire – Rhode Island (Ratio = 1.50), followed by San Francisco – Oakland, California (Ratio = 1.49). Though the Non-Hispanic Asian and Non-Hispanic Other populations appeared to be less impacted by rail traffic noise, in some of these 24 regions, they were still estimated to be disproportionally highly annoyed by rail, such as in the Minneapolis – St. Paul, Minnesota – Wisconsin (Ratio for Non-Hispanic Asian population = 1.36) and Detroit, Michigan Urban Area (Ratio for Non-Hispanic Other population = 1.32). Similar to annoyance to aviation noise, Non-Hispanic White populations were less disproportionally highly annoyed by the rail traffic noise, with only the Non-Hispanic White populations in the Detroit, Michigan and San Diego, California Urban Areas having Fair Share Ratios > 1(**Figure 7**).

In all 24 regions, the non-Hispanic Black population was found to be disproportionally highly annoyed by the road traffic noise (Fair Share Ratio > 1). The Hispanic population was also estimated to be disproportionally highly annoyed by road traffic noise in all 24 regions, except for the Chicago, Illinois Urban Area (Fair Share Ratio = 0.99). Overall, none of these 24 regions had Non-Hispanic White population being disproportionally highly annoyed by the road traffic noise (**Figure 8**).

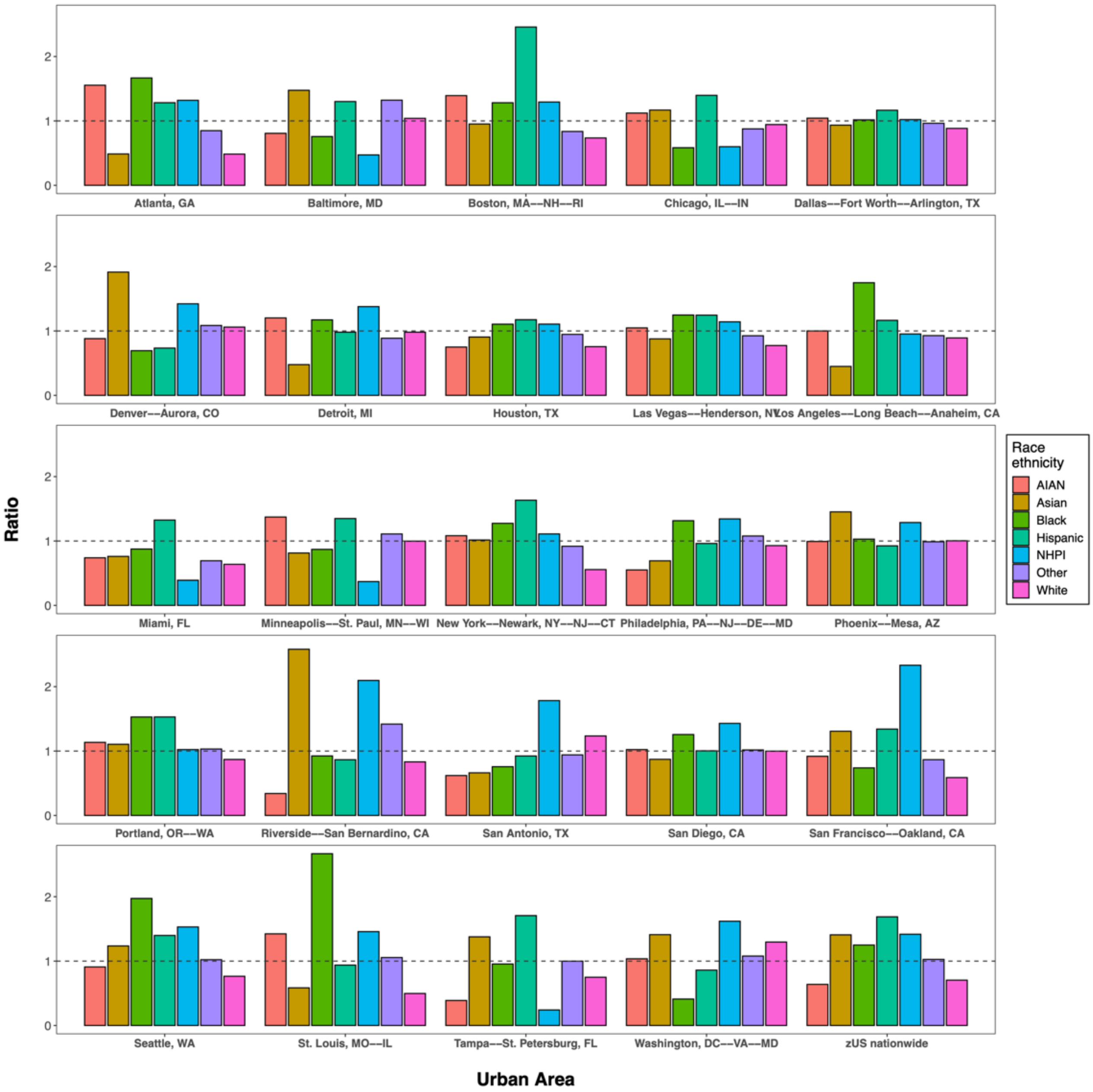

**Figure 6**. Race/ethnicity-specific Fair Share Ratio of aviation noise by the 24 regions that account for the top 5% most populous Urban Areas in the US. The "US nationwide" category includes the data for the continental US and the states of Alaska and Hawaii. The dashed lines represent a ratio of 1.

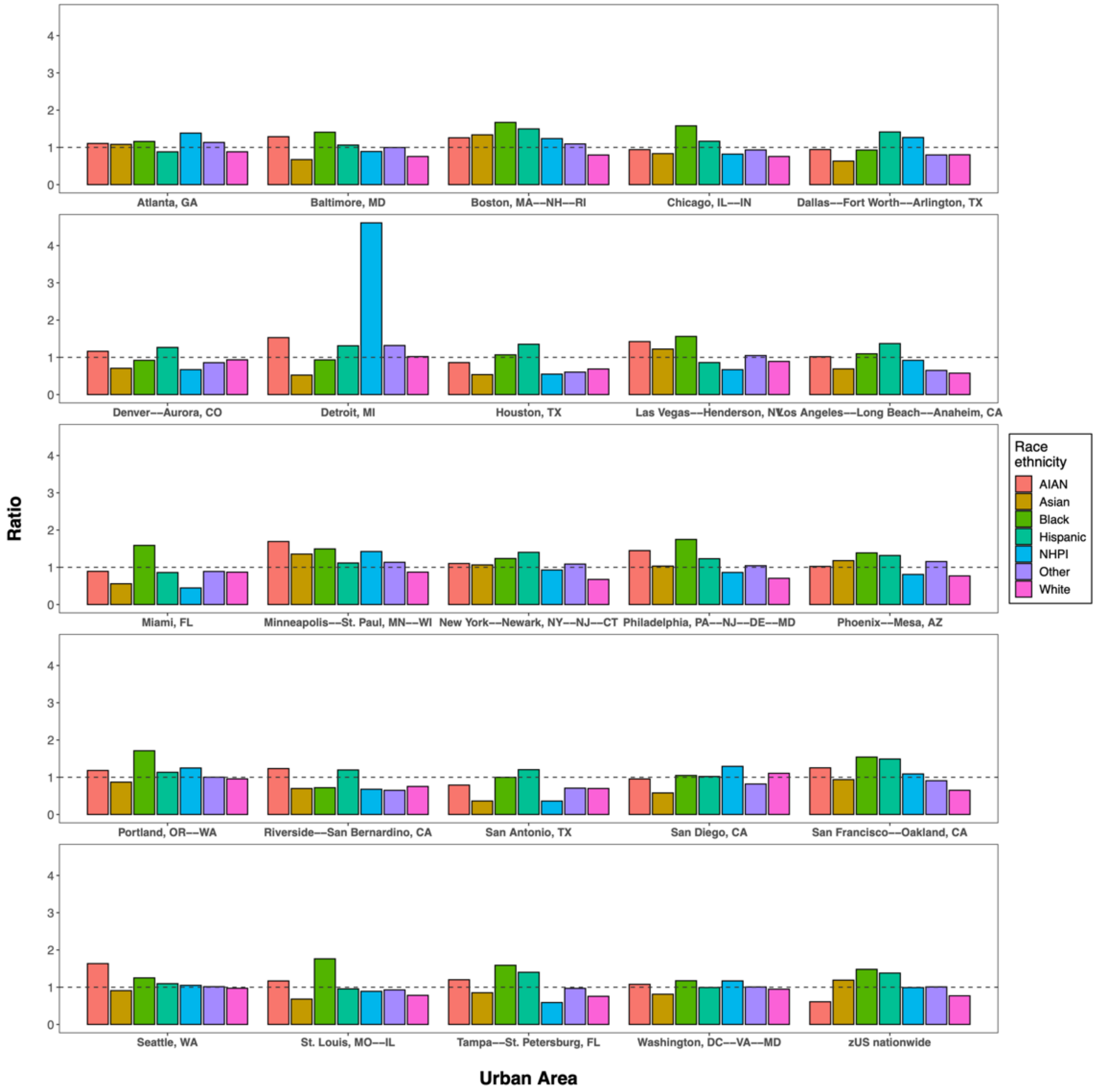

**Figure 7. Race/ethnicity-specific Fair Share Ratio of rail noise by the 24 regions that account for the top 5% most populous Urban Areas in the US.** The "US nationwide" category includes the data for the continental US and the states of Alaska and Hawaii. The dashed lines represent a ratio of 1.

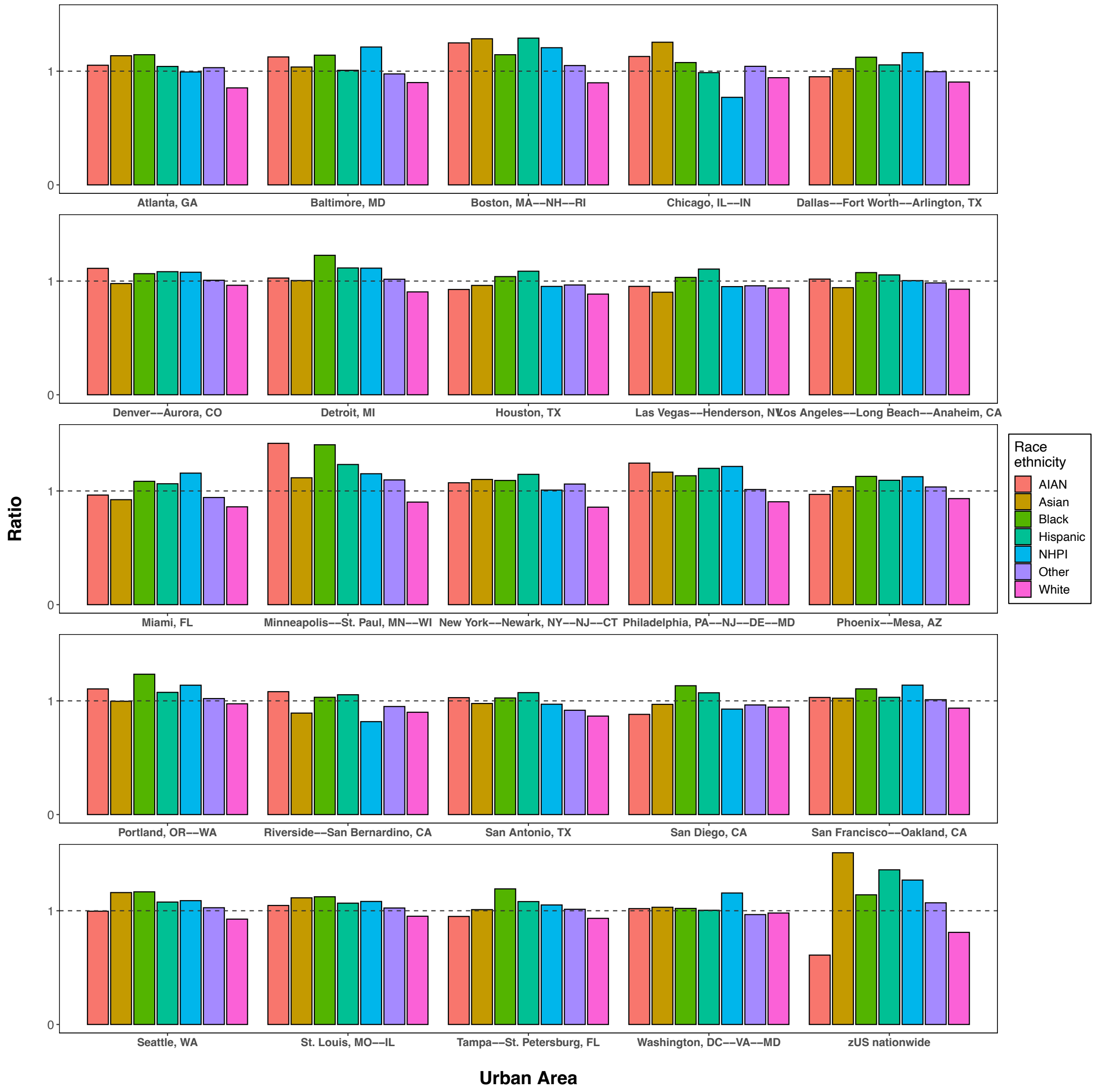

**Figure 8. Race/ethnicity-specific Fair Share Ratio of road noise by the 24 regions that account for the top 5% most populous Urban Areas in the US.** The "US nationwide" category includes the data for the continental US and the states of Alaska and Hawaii. The dashed lines represent a ratio of 1.

## 4. Discussion

In the study, we present the utilization of the BTS NTNM and NTNE map data, and propose the methodology of using the Fair Share Ratio to assess the race/ethnicity disparities in traffic noise exposure, with annoyance as the primary outcome of interest. The improved spatial resolution the NTNE map provides enables us to compute block-group level population exposure, which potentially accounts for uneven spatial distributions of populations within each census tract and variations in noise levels. Our findings reveal that in 2020, 7.8 million individuals were highly annoyed by aviation noise, while 5.2 million and 7.9 million people were highly annoyed by rail and roadway noise in the continental US and the states of Alaska and Hawaii. These numbers correspond to 2.40%, 1.60%, and 2.43 % of the total population in the US, respectively, emphasizing the significance of transportation noise as a potential public health concern.

The race/ethnicity-specific Fair Share Ratio indicated that populations of Non-Hispanic Asian, Non-Hispanic Black, Hispanic, Non-Hispanic NHPI, and Non-Hispanic Other race ethnicity groups were disproportionally highly annoyed by transportation noise across the nation. Notably, across the US, Hispanic populations experience the greatest share of high annoyance from aviation noise – 1.69 times their population share. Non-Hispanic Black populations experience the greatest share of high annoyance from railway noise – 1.48 times their population share. Non-Hispanic Asian populations experience the greatest share of high annoyance from roadway noise – 1.51 times their population share. The analysis at the state and Urban Area levels further highlighted the disparities in transportation noise exposure and annoyance across different ethnic groups, which both suggested that the Non-Hispanic White populations were less annoyed by all sources of transportation noise compared to non-White populations.

This study makes use of modeled noise data and existing ERFs for high annoyance, and therefore is based on assumptions. First, the ERFs used to estimate the percentage of the population highly annoyed (%HA) were originally developed by Guski et al. based on the European standard noise metric (Lden), whereas the BTS NTNM and NTNE map data employed the noise metric LAeq. To reconcile this discrepancy, we converted LAeq to Lden by assuming a constant LAeq level throughout the entire day. Because of the added penalties during evening and night hours, the resulting equivalent Lden values are consistently higher than LAeq, which will probably lead to a high estimate of %HA. As a sensitivity analysis, we have also provided the results using LAeq as the input for the %HA ERFs (i.e., without added 5 and 10 dB evening and night penalties) (**Table S3**). The true value of the population highly annoyed by different traffic noise should fall within the range between the estimates obtained using LAeq and their equivalent Lden values. Second, the ERFs for %HA are applicable up to 70 dB Lden for aviation noise and 80 dB Lden for rail and roadway noise. However, because noise exposures can be above the applicable ranges of the ERFs, and we limited the exposure values not to exceed the applicable range, our estimates of %HA may be underestimated for high exposure areas. Third, this analysis focuses on assessing the population highly annoyed by specific traffic noise sources individually (i.e., aviation, rail, and roadway noise). It is possible that individuals may experience overlapping exposures to different traffic sources (e.g., being highly annoyed by both aviation and roadway noise simultaneously). However, we are unaware of methods to estimate the cumulative annoyance across multiple transportation sources.

Several studies have attempted to synthesize the traffic noise ERFs/curves for annoyance over the past decades. Notably, these ERFs were based on varying noise metrics (Fidell, 2003; Guski et al., 2017; Miedema & Oudshoorn, 2001; Miedema & Vos, 1998; Schultz, 1978). **Figure S3** presents the traffic noise ERFs/curves for traffic noise annoyance based on the Ldn/DNL noise metric from Schultz, FICON, Miedema and Oudshoorn (M+O), and Miedema and Vos (M+V). The figure shows that the M+O and M+V curves exhibit similar fits for aircraft, rail, and roadway noise annoyance. Aircraft/aviation noise induces the highest %HA when compared to rail and roadway noise at the same exposure level. The Schultz and FICON curves fitted for the aircraft/aviation %HA also exhibit a similar trend with the M+O and M+V aircraft/aviation curves, despite wider ranges of Ldn/DNL were used. In **Figure S4**, we present the ERFs/curves for traffic noise annoyance based on the Lden/DENL noise metric, including the one synthesized by WHO and Miedema and Oudshoorn (M+O). Interestingly, the WHO curves exhibit higher %HA for the rail noise compared to road noise, which contrasts with the findings of M+O and M+V. This discrepancy might be attributed to the utilization of different social acoustics surveys during the process of curve synthesis. For this analysis, we selected the ERFs/curves developed by WHO, which were included as part of the Environmental Noise Guidelines for the European Region. These ERFs, published in 2017, are the most current and incorporate data from environmental noise annoyance surveys that employed different definitions of high annoyance. The meta-analysis process in this study corrected these discrepancies, ensuring the robustness of the WHO ERFs/curves for assessing transportation noise annoyance.

Among the three transportation sources assessed in this study, the impact of aviation noise on community annoyance was found to be the most significant, followed by roadway noise. This is reflected in the ERFs used, where higher annoyance is associated with aircraft noise than road traffic or railway noise at the same exposure level. It should be noted that airports explicitly for military use were not included in the synthesis of the BTS NTNM data. Noise associated with military aviation activity has long been understudied due to regulations in regions surrounding the military area and airbase, and difficulties in characterizing noise from intermittent flight activities (Waitz et al., 2005). For example, a study at Whidbey Island, Washington State where a naval air station is located has shown that military aircraft activity can produce noise with level up to 118 dBA in-air (Kuehne et al., 2020). Such noise level has been known to cause sleep disruption and other negative health effects on humans as well as wildlife (Pepper et al., 2003).

Given the large numbers of people estimated to be exposed to noise levels that induce high annoyance, and the disproportionate impacts experienced by some groups, future studies assessing the trend of transportation noise exposure are warranted for gaining a more comprehensive understanding of the long-term health impacts of community noise exposure, and inequality in noise-related burden as an environmental justice issue. While our study provides insights into traffic noise annoyance based on the most recent available data, it is important to consider the unique circumstances during the pandemic that may have influenced noise exposure and annoyance levels, specifically for our study that utilized 2020 data. The COVID-19 pandemic was officially declared a national emergency in the US on March 13, 2020, followed by shelter-in-place orders that suspended social and commercial activities in each state (The White House, 2022). Studies on modeling and monitoring traffic noise have reported decreased levels during the pandemic compared to pre-pandemic periods worldwide (Amoatey et al., 2022; Basu et al., 2021; Greco et al., 2022). According to the BTS, the monthly nationwide traffic was reduced by 253%, 282%, and 81% for aviation, passenger rail, and roadway traffic, respectively, in 2020 compared to 2019. In terms of environmental noise complaints, several studies have noted a decrease in reported complaints related to traffic noise (Amoatey et al., 2022; Dümen & Şaher, 2020; Michaud et al., 2022). However, it is crucial to note that the pandemic-induced changes in noise levels may not represent a permanent shift. As restrictions ease and normal activities resume, traffic-related noise levels may gradually return to pre-pandemic levels or even exceed them due to pent-up demand and potential changes in travel patterns. Therefore, future studies should assess the temporal trends of transportation noise beyond the immediate impact of the pandemic to capture any potential fluctuations and long-term changes in noise exposure.

Despite the aforementioned limitations, our study demonstrates several strengths. First, by leveraging the availability of BTS NTNM data and the NTNE map, which offer source-specific traffic noise exposure information, our study comprehensively evaluates the effects of different transportation noise sources, including aviation, passenger rail, and roadway, on population annoyance at the US census tract level. Prior studies rarely distinguished the impact of different transportation noise sources on annoyance, often focusing on assessing the impact of individual traffic sources. The source-specific transportation noise impacts our study provides can guide the development of noise mitigation strategies and policies (Ongel & Sezgin, 2016), particularly in communities that are also cumulatively and disproportionately impacted by other environmental exposures and hazards. Additionally, our study's quantitative health effects estimation, combined with qualitative assessments from previous studies, could aid health impact assessments on judging the impact of transportation policies and programs (Bhatia & Seto, 2011). Second, our study addresses health inequalities by race/ethnicity groups, which have been less explored in previous research that mainly focused on socioeconomic class or status. We propose and employ the Fair Share Ratio to assess race/ethnicity disparities in transportation annoyance, which shares a similar concept with a commonly used measure of inequality and segregation – the Index of Dissimilarity (ID) – that also accounts for population share in its computation (Duncan & Duncan, 1955; Wagstaff et al., 1991). However, studies have highlighted a major drawback of the ID in assessing segregation/inequality as it is not decomposable (i.e., a summary measure of inequality/segregation across all population groups assessed, rather than specific to a group) (Reardon & Firebaugh, 2002; Reardon & O'Sullivan, 2004). In contrast, the Fair Share Ratio allows us to compare inequality between two groups and spatial units, enabling the assessment of race/ethnicity-specific disparities.

## 5. Conclusions

In this study, we leverage the most recent NTNE map data and introduce an approach using the Fair Share Ratio to evaluate the race/ethnicity disparities in traffic noise exposure. Our study outcomes highlighted the widespread presence of transportation noise annoyance across the US, examining its prevalence at the national, state, and Urban Area levels. These findings contribute to the existing body of knowledge on transportation noise and its impact on public health, emphasizing the need for targeted source-specific noise mitigation strategies and policies to minimize the disproportionate impact of traffic noise in the US.

**Funding sources**

This work was supported by the University of Washington EDGE Center of the National Institutes of Health [award number P30ES007033].

## Supplementary Materials

**Estimates of Population Highly Annoyed from Transportation Noise in the United States: An Unfair Share of the Burden by Race and Ethnicity**

Ching-Hsuan Huang [a*], Edmund Seto [a]

[a] Department of Environmental and Occupational Health Sciences, School of Public Health, University of Washington

*Corresponding author.
E-mail address: hsuan328@uw.edu (C. -H. Huang)

(A)

(B)

(C)

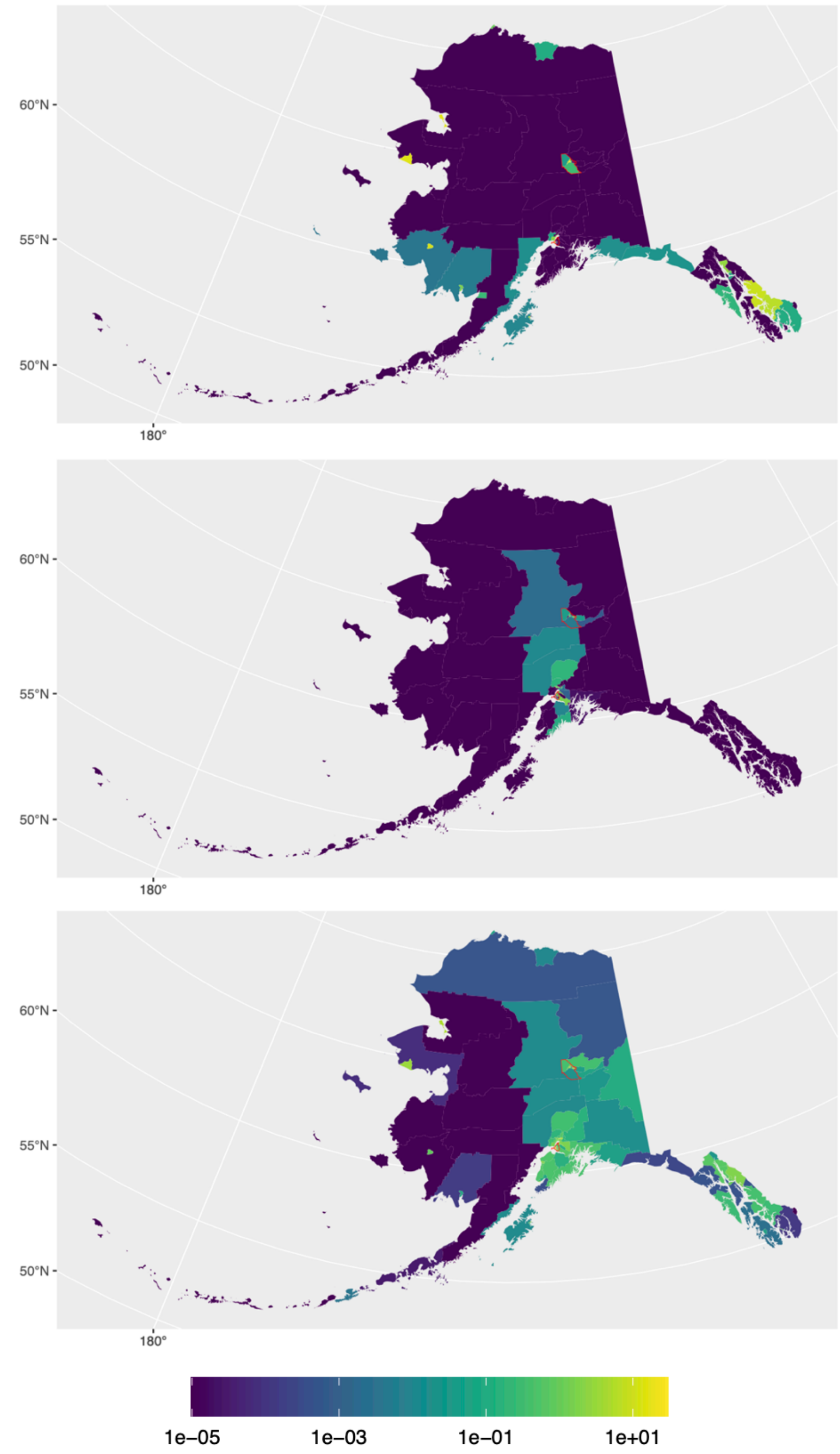

**Figure S1. The census-tract level maps of the percent population highly annoyed (%HA) by (a) aviation (b) rail (c) roadway transportation noise in the state of Alaska, overlayed with urban areas defined by the US Census Bureau.**

**(A)**

**(B)**

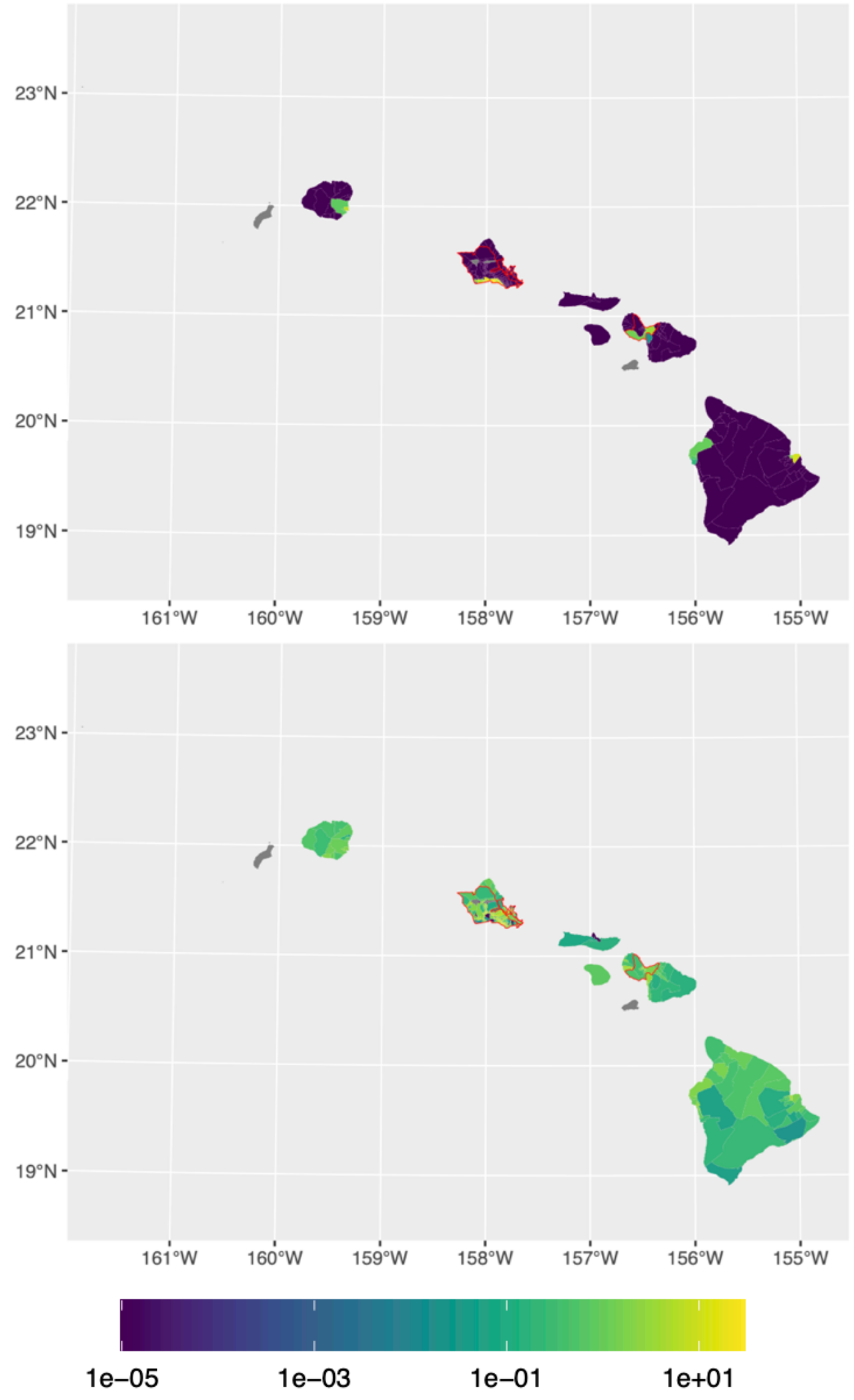

**Figure S2. The census-tract level maps of the percent population highly annoyed (%HA) by (a) aviation and (b) roadway transportation noise in the state of Hawaii, overlayed with urban areas defined by the US Census Bureau. No rail data is available from the BTS NTNM.**

(A)

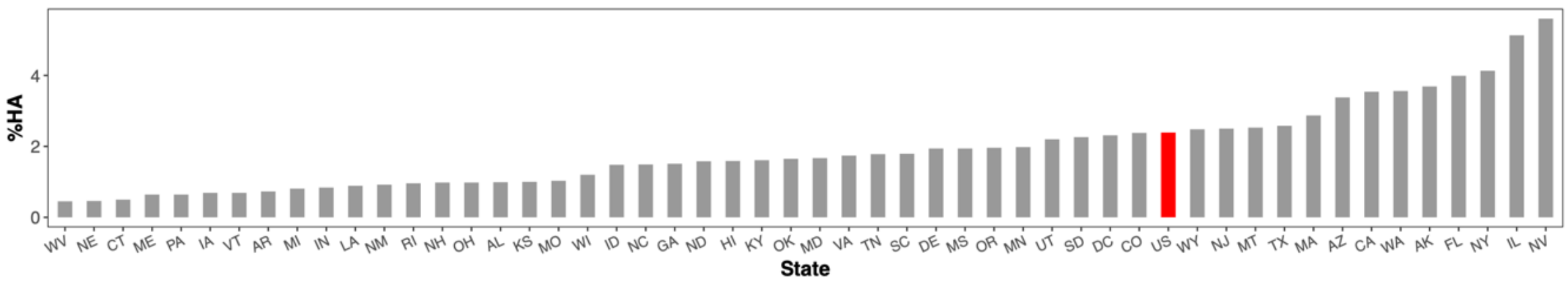

(B)

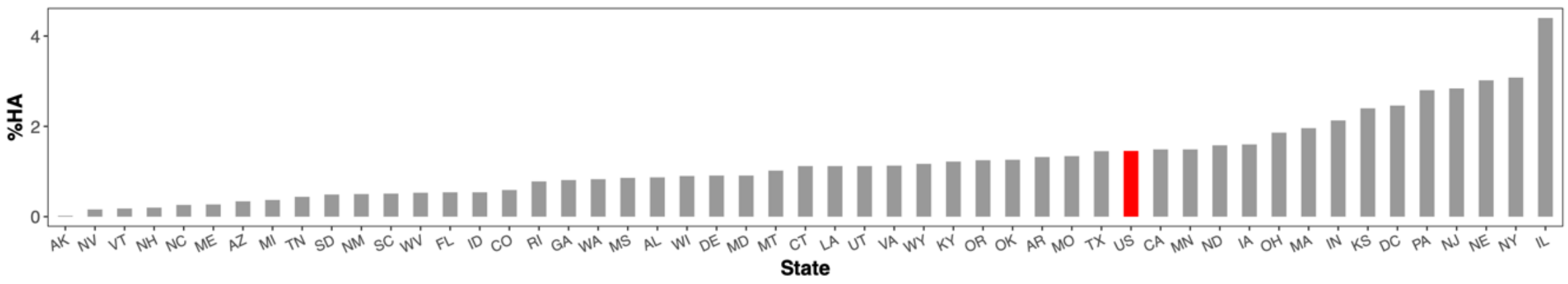

(C)

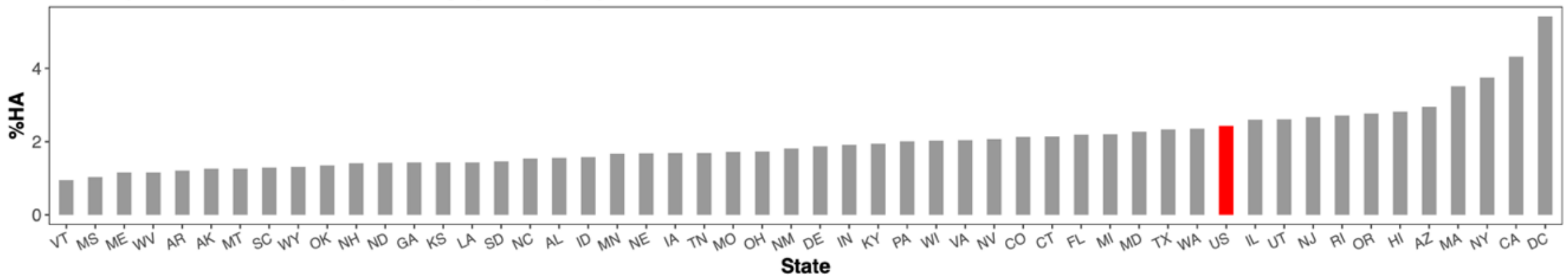

**Figure S3. The percent population highly annoyed (%HA) by (a) aviation, (b) rail, and (c) roadway traffic noise by state. US nationwide %HA shown in red.**

(A)

(B)

(C)

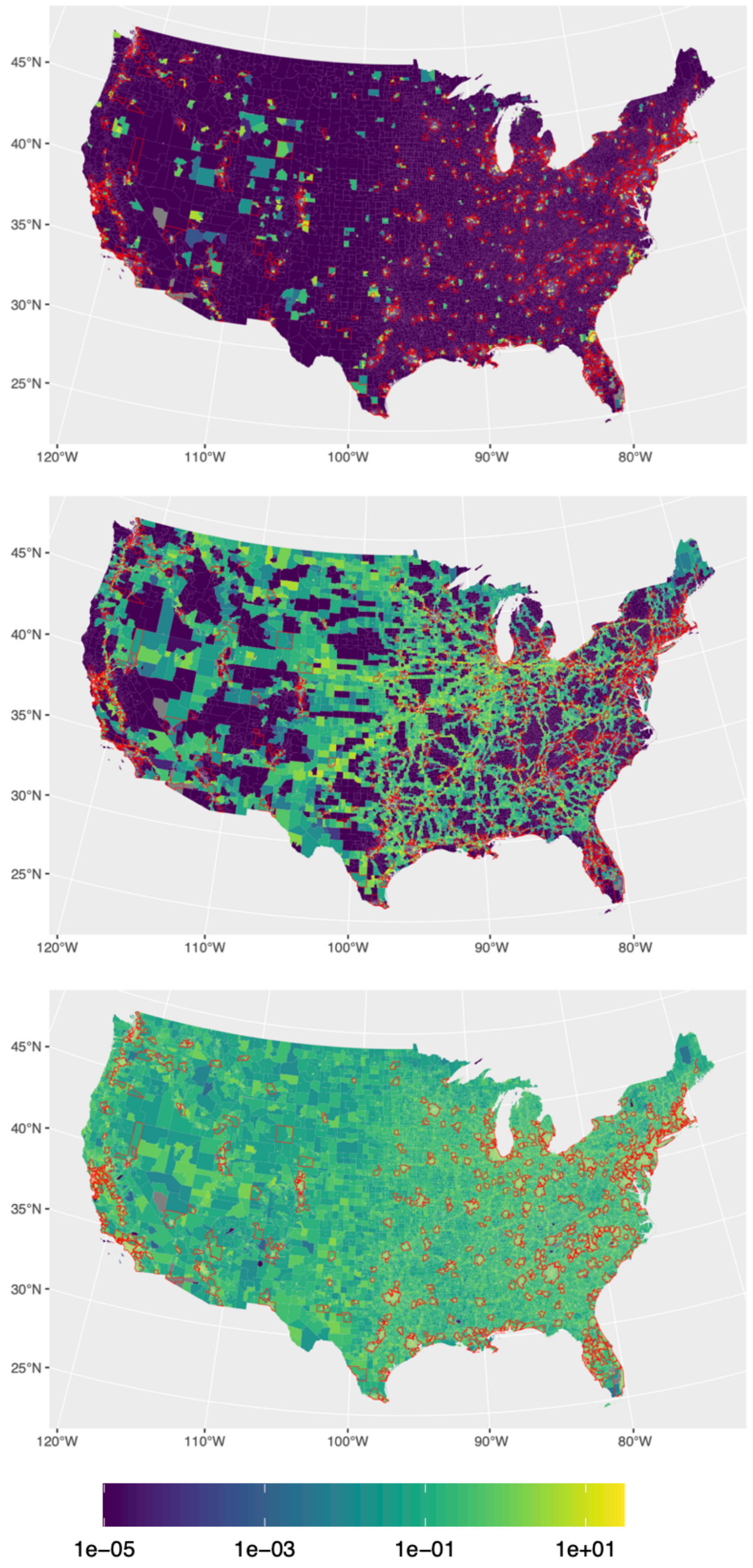

**Figure S4. The census-tract level maps of the percent population highly annoyed (%HA) by (a) aviation (b) rail (c) roadway traffic noise, overlayed with the 24 regions that account for the top 5% most populous Urban Areas in the continental US (with red outlines).**

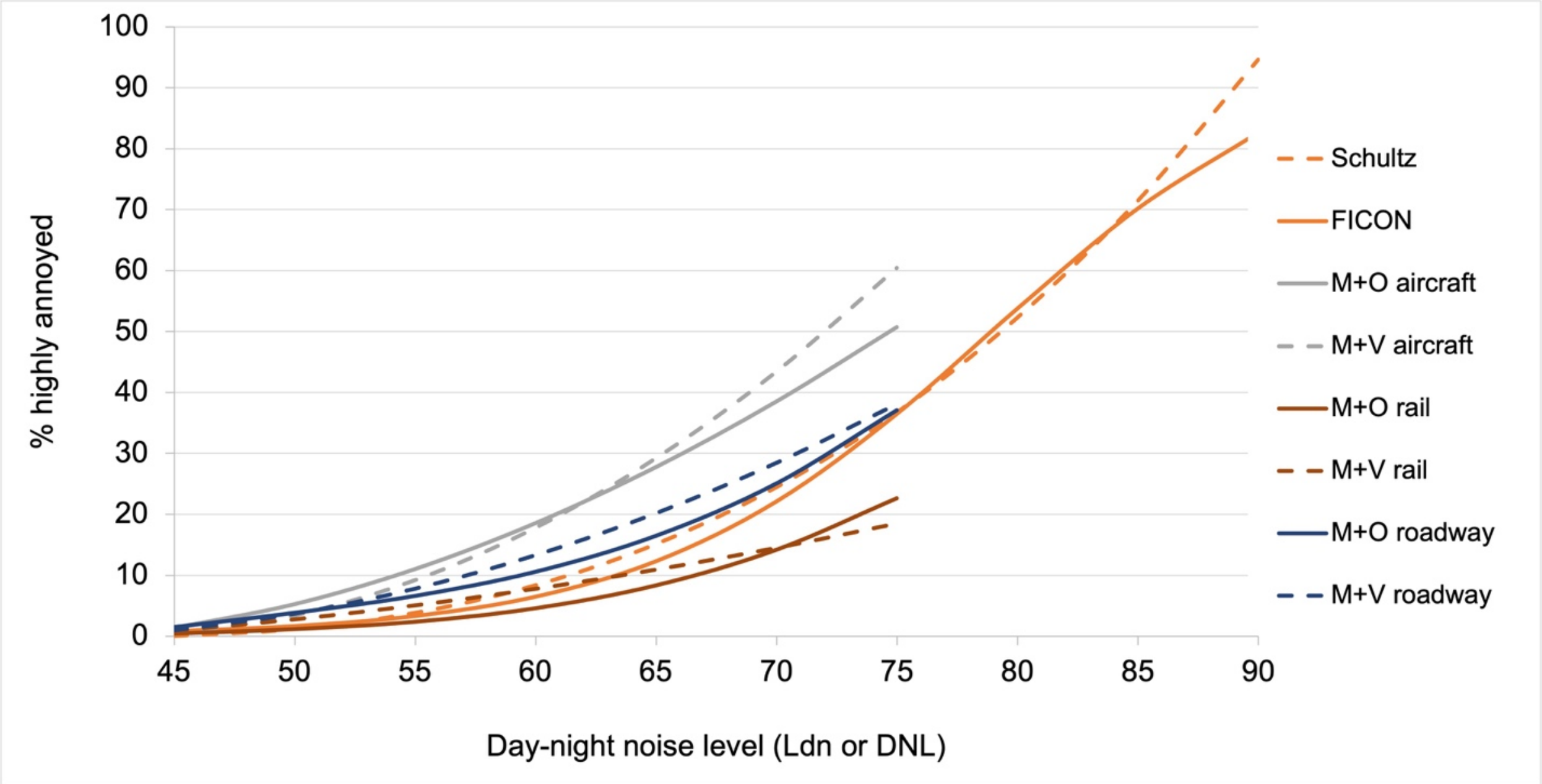

**Figure S5. Comparison of the Ldn-based high annoyance to noise exposure response curves by Schultz, FICON, Miedema and Oudshroon, and Miedema and Vos.** Definition of abbreviations: M+O: Miedema and Oudshoorn; M+V: Miedema and Vos.

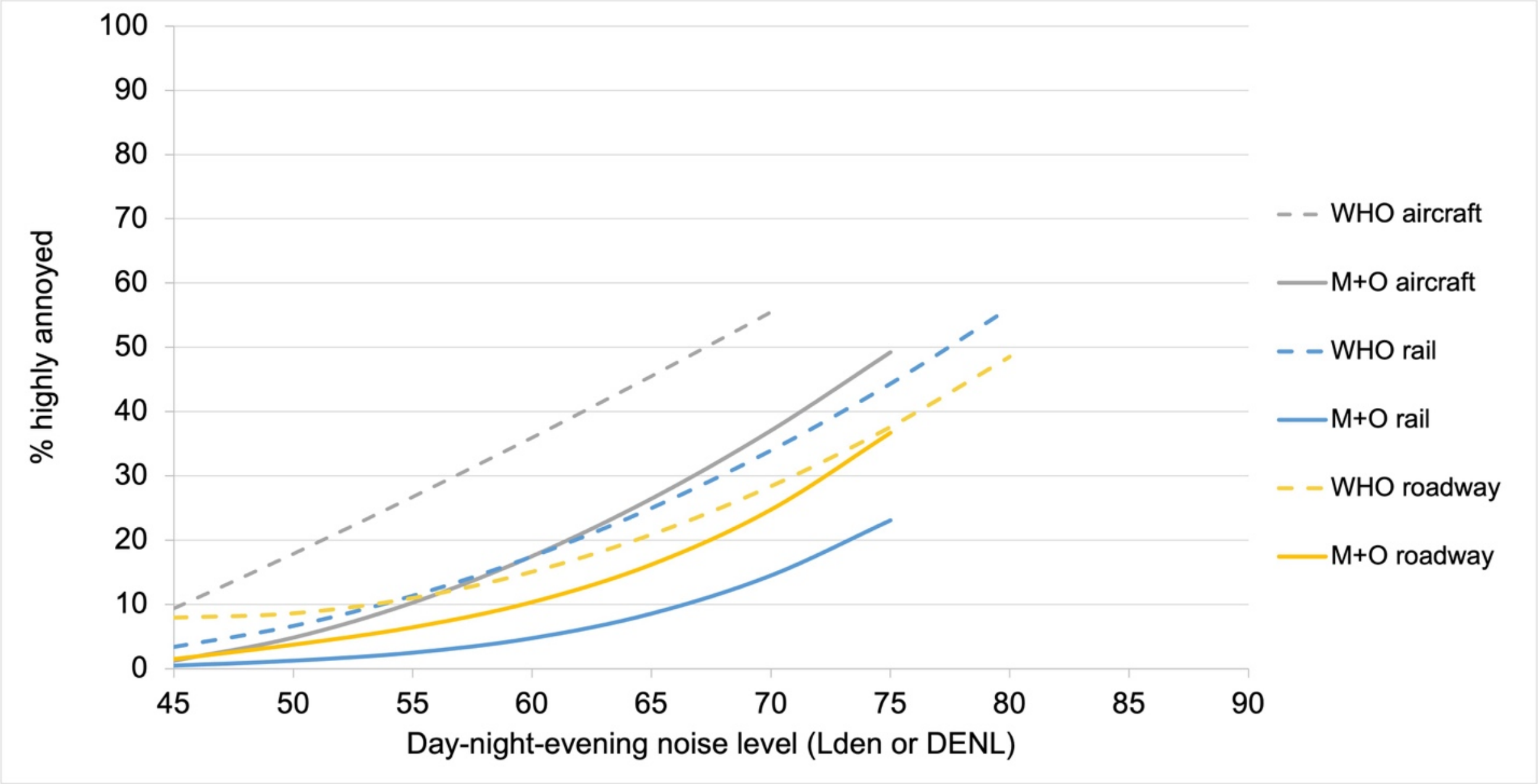

**Figure S6. Comparison of the Lden-based high annoyance to noise exposure response curves by Schultz, FICON, Miedema and Oudshroon.** Definition of abbreviations: M+O: Miedema and Oudshoorn.

(a)

(b)

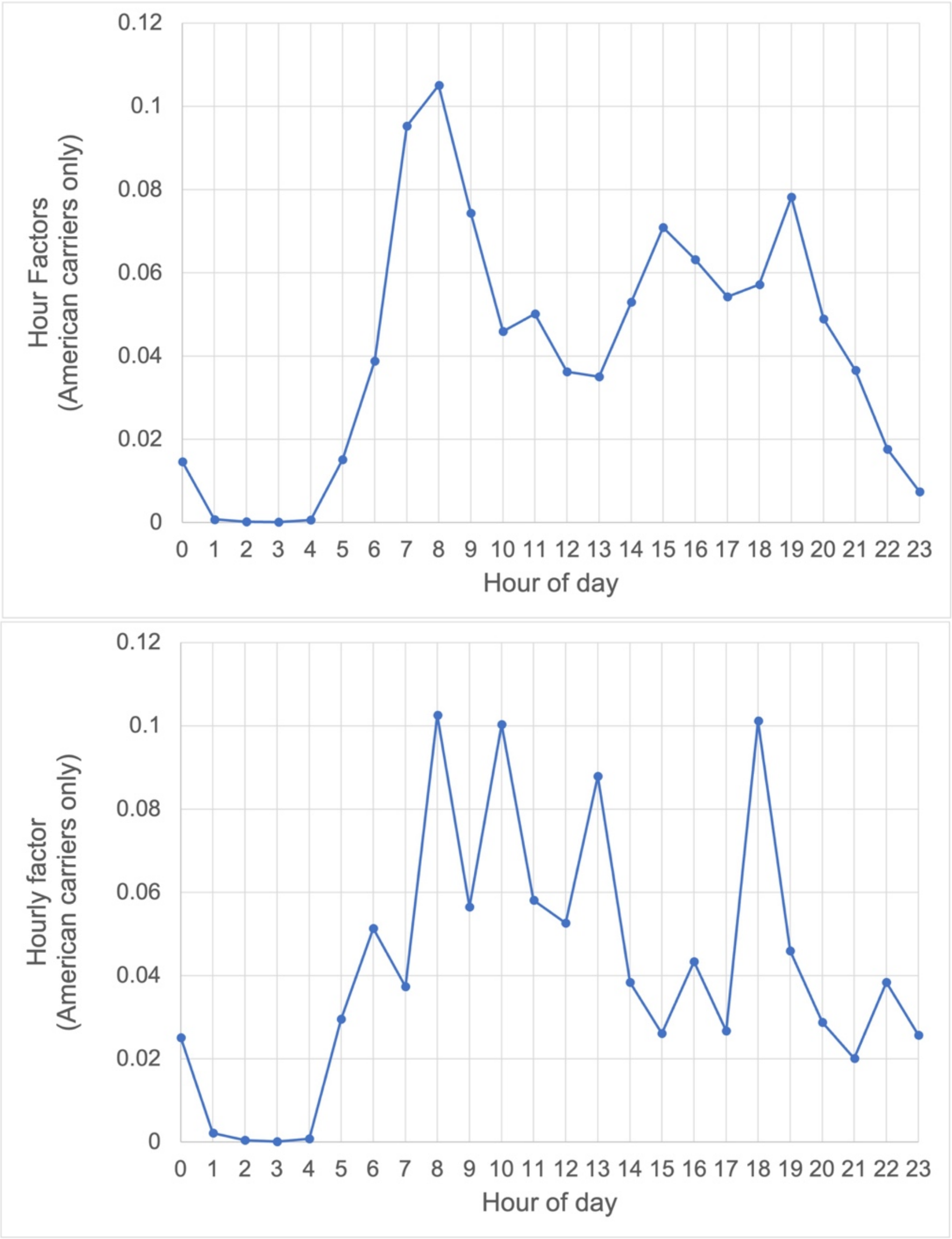

**Figure S7. Patterns of aircraft traffic (American carriers only) at (a) John F. Kennedy International Airport and (b) San Francisco International Airport.**

(a)

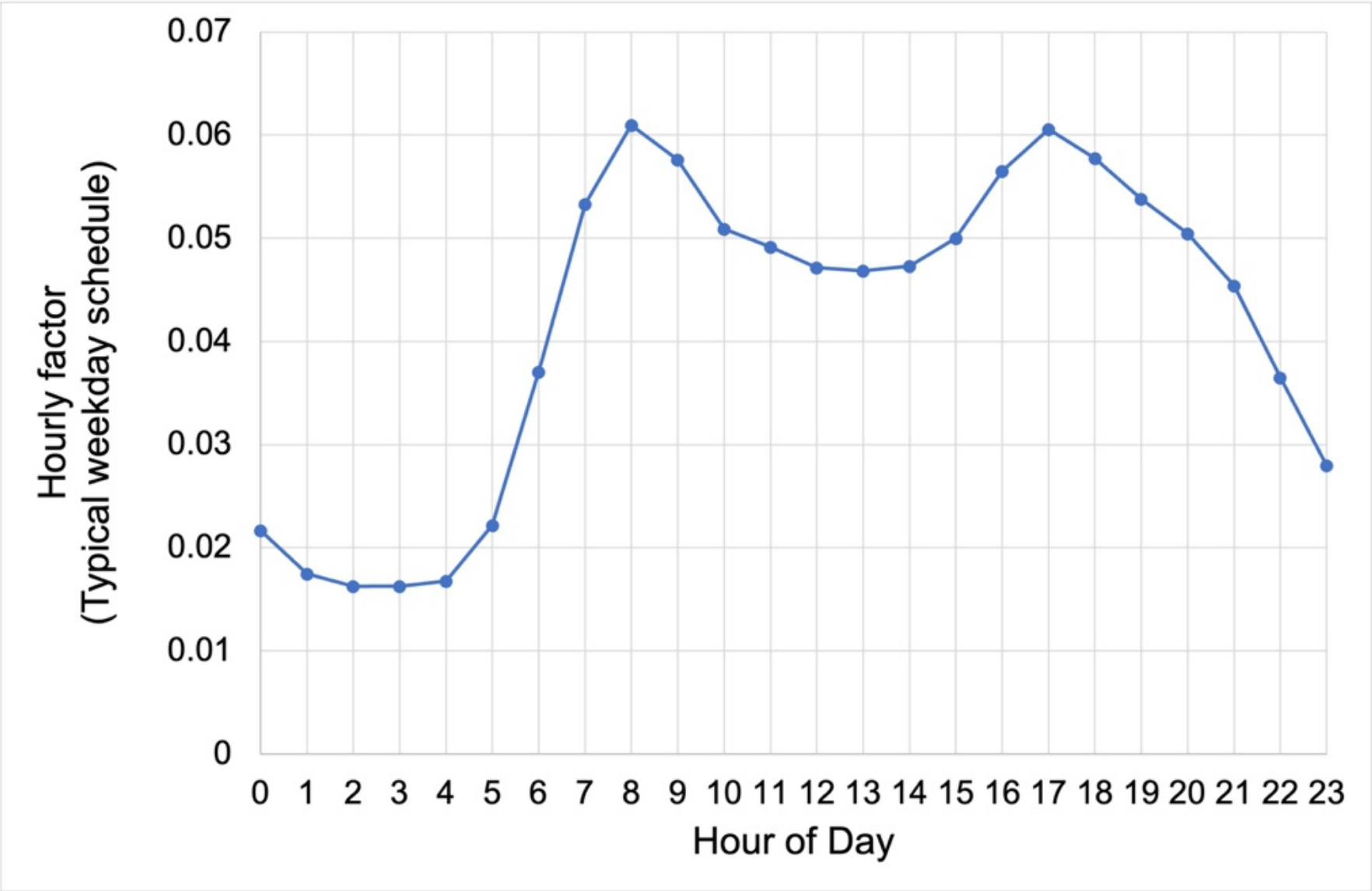

(b)

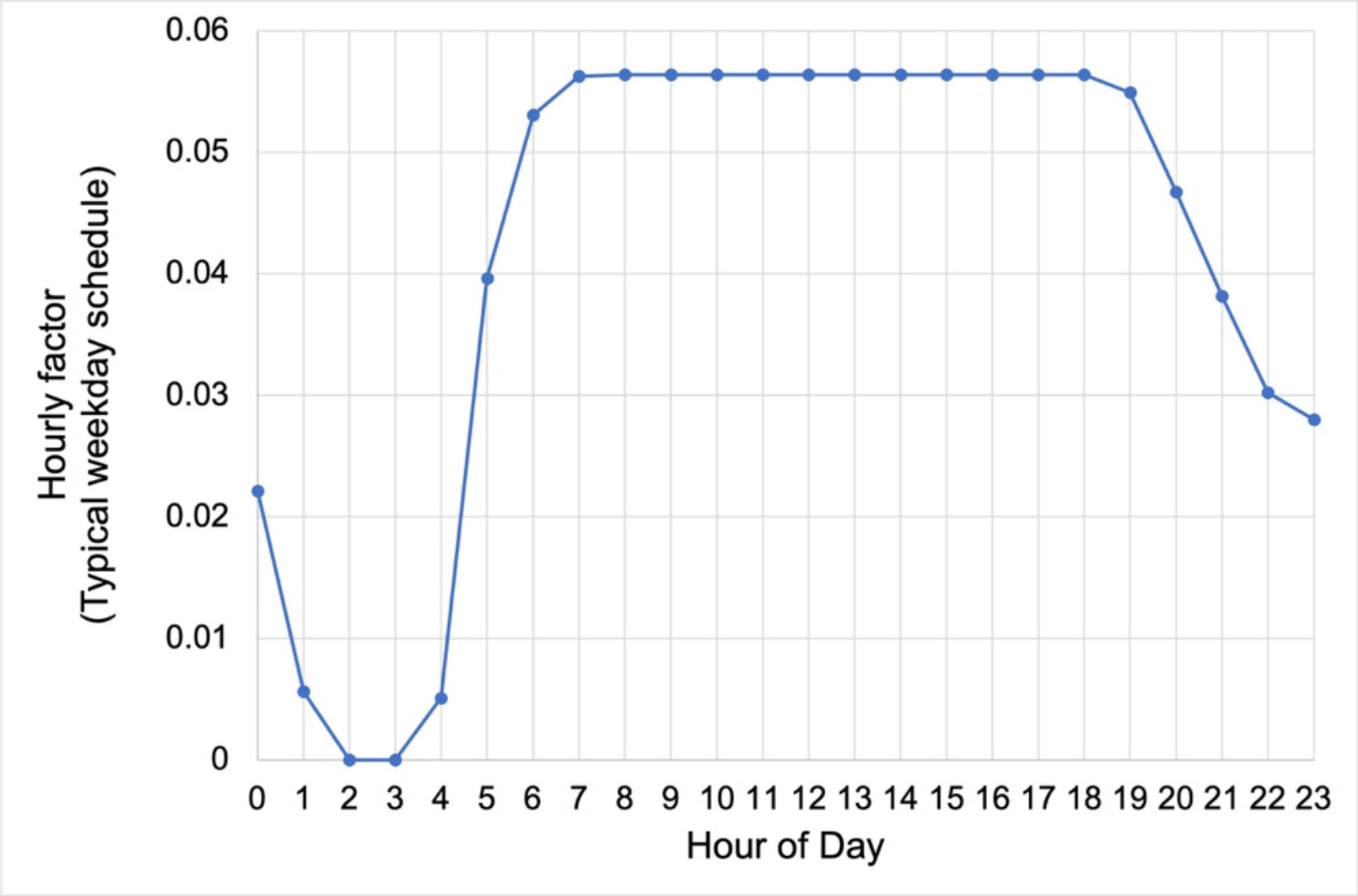

**Figure S8. Patterns of rail traffic of (a) New York Metropolitan Transportation Authority (MTA) network and (b) California Bay Area Rapid Transit (BART) network, based on typical weekday schedule.**

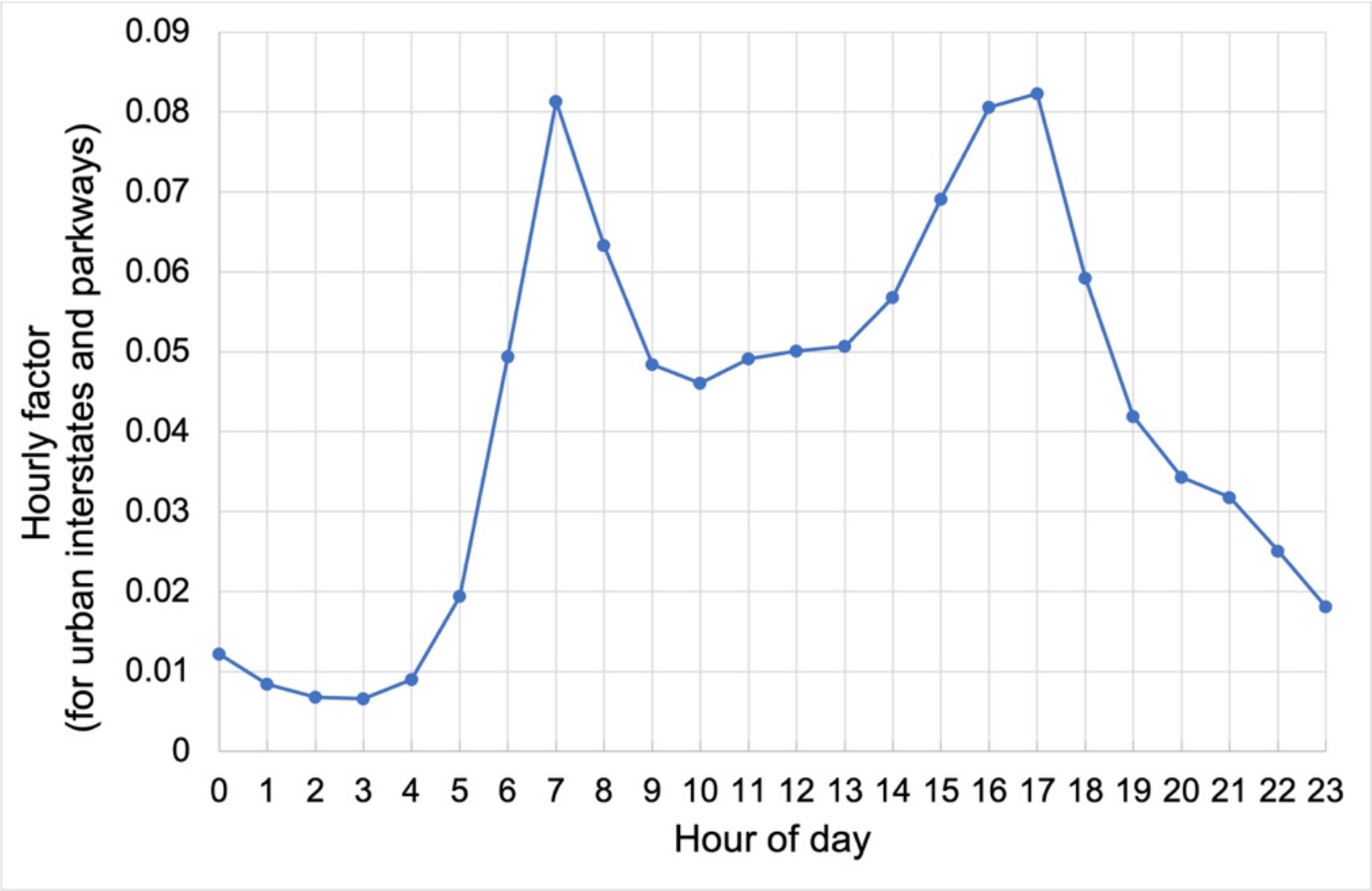

**Figure S9. Patterns of roadway traffic of the US nationwide urban interstates and parkways.**

**Table S1. Exposure-response functions (ERFs) for the percentage of the population highly annoyed for aviation, rail, and road traffic noise (Guski et al., 2017; World Health Organization, 2018).**

| Noise source | ERF of %HA | Applicable range of noise level |
|---|---|---|
| Aircraft/aviation | $-50.9693 + 1.0168 \times L_{den} + 0.0072 \times {L_{den}}^2$ | 40-70 Lden |
| Rail | $38.1596 - 2.05538 \times L_{den} + 0.0285 \times {L_{den}}^2$ | 40-80 Lden |
| Road traffic | $78.9270 - 3.1162 \times L_{den} + 0.0342 \times {L_{den}}^2$ | 40-80 Lden |

The ERFs provided in **Table 1** are based on the standard day-evening-night sound level metric (Lden) and are applicable to specific ranges of Lden. The estimation of %HA by aviation noise applies to a Lden range of 40-70 dB, whereas the estimations of %HA by rail and roadway noise apply to a Lden range of 40-80 dB. To allow estimation, we first converted the LAeq categories from the NTNE map to their equivalent Lden values. Using the midpoint of each NTNE LAeq category and the assumption of constant LAeq level over a whole day, a penalty of 5 dB for evening noise (i.e., 7 pm to 11 pm) and a penalty of 10 dBA for nighttime noise (i.e., 11 pm to 7 am) was applied. The conversion was based on the following formula (International Organization for Standardization, 2016):

$$\mathrm{Lden} = 10\log_{10}\left(\frac{1}{24}\left(12 \cdot 10^{\frac{L_{day}}{10}} + 4 \cdot 10^{\frac{L_{evening}+5}{10}} + 8 \cdot 10^{\frac{L_{night}+10}{10}}\right)\right)$$

where $L_{day}$ , $L_{evening}$, and $L_{night}$ is the A-weighted averaged sound level over the day (7 am to 7 pm), evening (7 pm to 11 pm), and night (11 pm to 7 am) period, respectively. The same definition of Lden was also adopted for the 2002 European standard for Lden for EU noise assessments.

The %HA of each LAeq category was calculated using its equivalent Lden for different traffic sources according to the ERFs in **Table S1**. In cases where the equivalent Lden values fell outside the specified ranges for which the ERFs were developed, the %HA corresponding to the highest Lden level was utilized. For example, when estimating the %HA attributable to aviation noise above the 60 - 70 LAeq category, the ERF corresponding to a Lden value of 61.4 dB was applied (**Table S2**). More details on the conversion between LAeq and Lden are discussed in the footnote of **Table S2**.

For additional sensitivity analyses, we also used the LAeq midpoint noise levels from the NTNE map, as is, without applying 5 or 10 dB penalties for evening and nighttime hours, in the ERFs to estimate %HA. Similarly, in cases where the equivalent LAeq values fell outside the specified ranges for which the ERFs were developed (**Table 1**), the %HA corresponding to the highest LAeq level was utilized. The results of this sensitivity analysis are presented in Supplemental Information.

**Table S2. Conversions of the equivalent Lden from the NTNE LAeq categories and the Lden input for estimating the percentage of population highly annoyed (%HA).**

| NTNE noise LAeq category (dB) | Midpoint LAeq (dB) | Equivalent Lden (dB) [a] | Lden input for %HA calculation | | |
|---|---|---|---|---|---|
| | | | Aviation noise | Rail noise | Roadway noise |
| 45-50 | 47.5 | 53.9 | 53.9 | 53.9 | 53.9 |
| 50-60 | 55 | 61.4 | 61.4 | 61.4 | 61.4 |
| 60-70 | 65 | 71.4 | 70 | 71.4 | 71.4 |
| 70-80 | 75 | 81.4 | 70 | 80 | 80 |
| 80-90 | 85 | 91.4 | 70 | 80 | 80 |
| 90+ | 90+ | 90+ | 70 | 80 | 80 |

[a] Based on the ISO 1996-1:20161. A penalty of 5 dBA and 10 dBA was added to evening time (7 pm – 11 pm) and nighttime (11 pm – 7 am), respectively.

Our conversion of the equivalent Lden from the LAeq assumes constant LAeq levels over a whole day, as mentioned in the footnote of Table S1. Brink et al. derived an equation for the difference between Lnight and Lden (Brink et al., 2018). Similarly, we derived the difference between Lden and LAeq:

$$L_{Aeq} = 10 \cdot log_{10}\left(\frac{1}{24} \cdot \sum_{H=0}^{23} 10^{\frac{LAeqH}{10}}\right)$$

$$L_{den} = 10 \cdot log_{10}\left(\frac{1}{24}\left(\sum_{H=7}^{18} 10^{\frac{LAeqH}{10}} + \sum_{H=19}^{22} 10^{\frac{LAeqH+5}{10}} + \sum_{H=23}^{6} 10^{\frac{LAeqH+10}{10}}\right)\right)$$

$$L_{den} - L_{Aeq} = 10 \cdot log_{10}\left(\frac{\frac{1}{24}\left(\sum_{H=7}^{18} F_H + 3.16\sum_{H=19}^{22} F_H + 10\sum_{H=23}^{6} F_H\right)}{\frac{1}{24} \cdot \sum_{H=0}^{23} F_H}\right)$$

where $L_{den}$ is the day-evening-night sound level (dB) over the day; $L_{Aeq}$ is the A-weighted averaged sound level over the day; $H$ is the hour of day, and $F_H$ is the hourly distribution factor for traffic.

As an illustrative example, we explored the relationship between LAeq and Lden for aviation, rail, and roadway noise retrieved from different data sources.

For aviation noise, we utilized the hourly factors for aircraft traffic (American carriers) at John F. Kennedy International Airport (JFK) and San Francisco International Airport (SFO) in 2021, derived from the flight data provided by US Bureau of Transportation Statistics (United States Bureau of Transporation Statistics, 2023). These factors were integrated into the conversion equation above. Visualizations of the hourly factors illustrating the patterns in aircraft traffic at these airports are shown in Figure S7a and S7b. The calculation shows the estimated difference between Lden and LAeq for aircraft traffic at JFK and SFO are +3.2 dB and +4.0 dB, respectively.

Using the same approach, we applied the hourly factors for rail traffic of the New York Metropolitan Transportation Authority (MTA) network and California Bay Area Rapid Transit (BART) network sourced from the General Transit Feed Specification (GTFS) platform for LAeq to Lden conversion. Visualizations of the hourly factors illustrating the patterns in rail traffic of these two networks are shown in Figure S8a and S8b. The calculation shows the estimated difference between Lden and LAeq for rail traffic at MTA and BART are +4.8 dB and +4.4 dB, respectively.

For roadway noise, we applied the hourly factors for US national urban interstates and parkways roadway traffic retrieved from Allen et al. (Allen et al., 1998) to conversion equation. A visualization of the hourly factors illustrating the patterns in roadway traffic is shown in Figure S9. The calculation shows the estimated difference between Lden and LAeq for roadway traffic is +3.9 dB.

With these examples utilizing transportation data from the US, applying the conversion equation derived by Brink et al. consistently yielded higher Lden values for aviation, rail, and roadway noise. These outcomes align with our conversion from LAeq to Lden, where we observed an average difference between Lden and LAeq of approximately +6 dB.

**Table S3. Population numbers and proportions exposed to and highly annoyed by traffic noise [a] by state and nationwide based on the LAeq noise metric.**

| State | Aviation | | | Rail | | | Roadway | | | Total population |
|---|---|---|---|---|---|---|---|---|---|---|
| | # exposed | #HA | % HA [b] | # exposed | #HA | % HA [b] | # exposed | #HA | % HA [b] | |
| Alabama | 160,984 | 30,875 | 0.63% | 318,962 | 26,664 | 0.54% | 412,832 | 53,858 | 1.10% | 4,893,186 |
| Alaska | 89,198 | 17,415 | 2.36% | 1,267 | 121 | 0.02% | 50,768 | 6,529 | 0.89% | 736,990 |
| Arizona | 809,654 | 152,214 | 2.12% | 182,656 | 14,674 | 0.20% | 1,107,919 | 149,275 | 2.08% | 7,174,064 |
| Arkansas | 73,322 | 13,932 | 0.46% | 301,011 | 24,400 | 0.81% | 211,297 | 25,987 | 0.86% | 3,011,873 |
| California | 4,629,028 | 879,138 | 2.23% | 4,353,460 | 357,839 | 0.91% | 8,549,337 | 1,241,139 | 3.15% | 39,346,023 |
| Colorado | 468,329 | 82,798 | 1.46% | 247,293 | 20,542 | 0.36% | 684,332 | 85,753 | 1.51% | 5,684,926 |
| Connecticut | 60,885 | 11,070 | 0.31% | 286,843 | 25,596 | 0.72% | 433,272 | 54,191 | 1.52% | 3,570,549 |
| Delaware | 63,940 | 11,621 | 1.20% | 65,367 | 5,566 | 0.58% | 97,045 | 12,740 | 1.32% | 967,679 |
| District of Columbia | 54,693 | 9,950 | 1.42% | 132,407 | 10,323 | 1.47% | 223,474 | 26,499 | 3.77% | 701,974 |
| Florida | 2,799,135 | 533,222 | 2.51% | 849,541 | 69,251 | 0.33% | 2,443,448 | 330,157 | 1.56% | 21,216,924 |
| Georgia | 530,715 | 99,910 | 0.95% | 627,419 | 54,219 | 0.52% | 823,884 | 106,402 | 1.01% | 10,516,579 |
| Hawaii [c] | 76,582 | 14,055 | 0.99% | NA | NA | NA | 201,337 | 28,335 | 2.00% | 1,420,074 |
| Idaho | 91,338 | 15,730 | 0.90% | 72,205 | 5,663 | 0.32% | 167,126 | 19,467 | 1.11% | 1,754,367 |
| Illinois | 2,215,975 | 404,261 | 3.18% | 4,049,438 | 350,930 | 2.76% | 1,833,532 | 231,997 | 1.82% | 12,716,164 |
| Indiana | 200,023 | 33,983 | 0.51% | 1,037,049 | 87,698 | 1.31% | 720,522 | 90,027 | 1.34% | 6,696,893 |
| Iowa | 75,323 | 13,136 | 0.42% | 381,843 | 30,824 | 0.98% | 294,781 | 37,647 | 1.20% | 3,150,011 |
| Kansas | 102,596 | 17,645 | 0.61% | 519,396 | 42,192 | 1.45% | 245,271 | 29,670 | 1.02% | 2,912,619 |
| Kentucky | 237,594 | 44,770 | 1.00% | 410,675 | 33,888 | 0.76% | 451,347 | 61,007 | 1.37% | 4,461,952 |
| Louisiana | 138,269 | 26,156 | 0.56% | 399,383 | 31,398 | 0.67% | 369,128 | 46,852 | 1.00% | 4,664,616 |
| Maine | 29,362 | 5,499 | 0.41% | 25,368 | 2,284 | 0.17% | 87,187 | 10,913 | 0.81% | 1,340,825 |
| Maryland | 344,848 | 62,029 | 1.03% | 408,102 | 34,321 | 0.57% | 757,101 | 97,059 | 1.61% | 6,037,624 |
| Massachusetts | 688,721 | 120,119 | 1.75% | 998,287 | 83,399 | 1.21% | 1,327,633 | 170,099 | 2.47% | 6,873,003 |
| Michigan | 283,559 | 48,790 | 0.49% | 273,011 | 22,704 | 0.23% | 1,211,684 | 154,811 | 1.55% | 9,973,907 |
| Minnesota | 379,459 | 68,654 | 1.23% | 632,152 | 51,602 | 0.92% | 532,477 | 66,596 | 1.19% | 5,600,166 |
| Mississippi | 194,765 | 36,260 | 1.22% | 192,856 | 15,785 | 0.53% | 176,841 | 21,677 | 0.73% | 2,981,835 |
| Missouri | 218,824 | 38,528 | 0.63% | 610,416 | 50,615 | 0.83% | 582,758 | 75,011 | 1.22% | 6,124,160 |
| Montana | 95,702 | 16,270 | 1.53% | 82,033 | 6,456 | 0.61% | 78,377 | 9,425 | 0.89% | 1,061,705 |
| Nebraska | 30,270 | 5,419 | 0.28% | 409,692 | 35,396 | 1.84% | 188,655 | 22,937 | 1.19% | 1,923,826 |
| Nevada | 556,796 | 107,353 | 3.54% | 36,633 | 3,070 | 0.10% | 326,131 | 45,177 | 1.49% | 3,030,281 |
| New Hampshire | 46,975 | 7,954 | 0.59% | 19,090 | 1,698 | 0.13% | 104,676 | 13,461 | 0.99% | 1,355,244 |
| New Jersey | 782,007 | 135,656 | 1.53% | 1,908,904 | 149,317 | 1.68% | 1,346,432 | 168,151 | 1.89% | 8,885,418 |
| New Mexico | 70,282 | 11,430 | 0.55% | 78,851 | 6,177 | 0.29% | 208,676 | 26,810 | 1.28% | 2,097,021 |
| New York | 2,850,300 | 483,744 | 2.48% | 4,379,198 | 365,836 | 1.87% | 4,250,947 | 520,291 | 2.67% | 19,514,849 |
| North Carolina | 497,338 | 100,444 | 0.97% | 196,572 | 16,720 | 0.16% | 829,020 | 112,826 | 1.09% | 10,386,227 |
| North Dakota | 43,388 | 7,119 | 0.94% | 94,754 | 7,050 | 0.93% | 64,458 | 7,606 | 1.00% | 760,394 |
| Ohio | 401,220 | 68,978 | 0.59% | 1,628,311 | 133,903 | 1.15% | 1,133,347 | 143,331 | 1.23% | 11,675,275 |
| Oklahoma | 221,688 | 40,212 | 1.02% | 373,410 | 30,494 | 0.77% | 303,905 | 37,841 | 0.96% | 3,949,342 |
| Oregon | 286,825 | 50,288 | 1.20% | 387,998 | 31,664 | 0.76% | 652,899 | 81,375 | 1.95% | 4,176,346 |
| Pennsylvania | 286,720 | 50,046 | 0.39% | 2,625,312 | 220,340 | 1.72% | 1,510,122 | 181,151 | 1.42% | 12,794,885 |
| Rhode Island | 33,989 | 6,320 | 0.60% | 59,478 | 5,192 | 0.49% | 166,361 | 20,206 | 1.91% | 1,057,798 |
| South Carolina | 310,166 | 56,791 | 1.12% | 188,399 | 16,964 | 0.33% | 368,272 | 46,292 | 0.91% | 5,091,517 |
| South Dakota | 66,894 | 12,417 | 1.41% | 33,352 | 2,480 | 0.28% | 72,487 | 9,075 | 1.03% | 879,336 |
| Tennessee | 401,831 | 75,478 | 1.11% | 225,059 | 18,249 | 0.27% | 628,464 | 80,758 | 1.19% | 6,772,268 |
| Texas | 2,472,543 | 462,220 | 1.61% | 3,041,805 | 255,626 | 0.89% | 3,589,199 | 476,222 | 1.66% | 28,635,442 |
| Utah | 248,177 | 41,421 | 1.31% | 259,584 | 22,197 | 0.70% | 442,667 | 58,525 | 1.86% | 3,151,239 |
| Vermont | 14,385 | 2,651 | 0.42% | 7,789 | 710 | 0.11% | 35,058 | 4,174 | 0.67% | 624,340 |
| Virginia | 506,869 | 91,391 | 1.07% | 710,621 | 60,402 | 0.71% | 932,114 | 122,576 | 1.44% | 8,509,358 |
| Washington | 877,683 | 170,441 | 2.27% | 467,682 | 37,290 | 0.50% | 998,193 | 124,916 | 1.66% | 7,512,465 |
| West Virginia | 27,602 | 5,112 | 0.28% | 69,908 | 5,986 | 0.33% | 115,906 | 14,852 | 0.82% | 1,807,426 |
| Wisconsin | 235,497 | 43,644 | 0.75% | 385,836 | 32,648 | 0.56% | 669,636 | 83,557 | 1.44% | 5,806,975 |
| Wyoming | 47,613 | 9,160 | 1.58% | 52,504 | 3,921 | 0.67% | 45,574 | 5,389 | 0.93% | 581,348 |
| US nationwide [d] | 26,429,881 | 4,853,719 | 1.49% | 35,099,182 | 2,922,282 | 0.89% | 43,057,909 | 5,650,623 | 1.73% | 326,569,308 |

[a] Greater than or equal to 45 dB LAeq

[b] Percentage out of the total population of the state.

[c] No rail data available.

[d] Includes the continental US and the states of Alaska and Hawaii.

**Table S4. Population numbers and proportions highly annoyed by different transportation noise sources [a] by state and nationwide, stratified by race ethnicity.**

| State | Total | | | Asian | | | | Black | | | |
|---|---|---|---|---|---|---|---|---|---|---|---|
| | Pop | #HA | % | Pop | % | #HA | %HA | Pop | % | #HA | %HA |
| **Aviation** | | | | | | | | | | | |
| Alabama | 4,893,186 | 49,045 | 1.00% | 67,317 | 1.38% | 1,008 | 2.05% | 1,292,950 | 26.42% | 18,932 | 38.60% |
| Alaska | 736,990 | 27,508 | 3.73% | 45,855 | 6.22% | 2,133 | 7.75% | 22,720 | 3.08% | 864 | 3.14% |
| Arizona | 7,174,064 | 244,220 | 3.40% | 233,048 | 3.25% | 11,555 | 4.73% | 305,973 | 4.26% | 11,926 | 4.88% |
| Arkansas | 3,011,873 | 22,206 | 0.74% | 45,389 | 1.51% | 618 | 2.78% | 455,485 | 15.12% | 3,798 | 17.10% |
| California | 39,346,023 | 1,404,679 | 3.57% | 5,743,983 | 14.60% | 190,099 | 13.53% | 2,142,371 | 5.44% | 112,093 | 7.98% |
| Colorado | 5,684,926 | 135,764 | 2.39% | 178,265 | 3.14% | 6,969 | 5.13% | 224,190 | 3.94% | 6,430 | 4.74% |
| Connecticut | 3,570,549 | 17,927 | 0.50% | 161,787 | 4.53% | 850 | 4.74% | 352,036 | 9.86% | 1,675 | 9.34% |
| Delaware | 967,679 | 18,865 | 1.95% | 38,249 | 3.95% | 553 | 2.93% | 208,126 | 21.51% | 5,450 | 28.89% |
| District of Columbia | 701,974 | 16,197 | 2.31% | 28,347 | 4.04% | 1,477 | 9.12% | 312,661 | 44.54% | 1,625 | 10.03% |
| Florida | 21,216,924 | 851,762 | 4.01% | 579,476 | 2.73% | 24,158 | 2.84% | 3,231,108 | 15.23% | 163,506 | 19.20% |
| Georgia | 10,516,579 | 160,057 | 1.52% | 430,473 | 4.09% | 4,663 | 2.91% | 3,275,581 | 31.15% | 84,050 | 52.51% |
| Hawaii | 1,420,074 | 22,715 | 1.60% | 522,125 | 36.77% | 9,677 | 42.60% | 25,130 | 1.77% | 827 | 3.64% |
| Idaho | 1,754,367 | 26,037 | 1.48% | 23,603 | 1.35% | 441 | 1.69% | 10,558 | 0.60% | 269 | 1.03% |
| Illinois | 12,716,164 | 655,961 | 5.16% | 702,457 | 5.52% | 53,351 | 8.13% | 1,766,586 | 13.89% | 66,778 | 10.18% |
| Indiana | 6,696,893 | 56,533 | 0.84% | 157,076 | 2.35% | 2,656 | 4.70% | 622,317 | 9.29% | 8,925 | 15.79% |
| Iowa | 3,150,011 | 21,684 | 0.69% | 78,873 | 2.50% | 460 | 2.12% | 114,312 | 3.63% | 1,055 | 4.86% |
| Kansas | 2,912,619 | 29,235 | 1.00% | 85,734 | 2.94% | 944 | 3.23% | 159,797 | 5.49% | 1,483 | 5.07% |
| Kentucky | 4,461,952 | 71,935 | 1.61% | 67,524 | 1.51% | 1,746 | 2.43% | 356,048 | 7.98% | 15,822 | 21.99% |
| Louisiana | 4,664,616 | 41,857 | 0.90% | 79,976 | 1.71% | 722 | 1.73% | 1,489,071 | 31.92% | 18,008 | 43.02% |
| Maine | 1,340,825 | 8,780 | 0.65% | 15,157 | 1.13% | 163 | 1.86% | 17,761 | 1.32% | 590 | 6.72% |
| Maryland | 6,037,624 | 101,223 | 1.68% | 382,027 | 6.33% | 9,519 | 9.40% | 1,773,702 | 29.38% | 23,959 | 23.67% |
| Massachusetts | 6,873,003 | 198,170 | 2.88% | 462,831 | 6.73% | 12,024 | 6.07% | 466,288 | 6.78% | 16,027 | 8.09% |
| Michigan | 9,973,907 | 80,872 | 0.81% | 314,736 | 3.16% | 1,930 | 2.39% | 1,342,592 | 13.46% | 19,073 | 23.58% |
| Minnesota | 5,600,166 | 111,726 | 2.00% | 273,100 | 4.88% | 6,749 | 6.04% | 354,540 | 6.33% | 9,220 | 8.25% |
| Mississippi | 2,981,835 | 58,324 | 1.96% | 29,561 | 0.99% | 907 | 1.56% | 1,118,689 | 37.52% | 21,588 | 37.01% |
| Missouri | 6,124,160 | 63,303 | 1.03% | 122,506 | 2.00% | 1,254 | 1.98% | 692,510 | 11.31% | 29,441 | 46.51% |

| | | | | | | | | | | |
|---|---|---|---|---|---|---|---|---|---|---|
| Montana | 1,061,705 | 27,043 | 2.55% | 8,527 | 0.80% | 281 | 1.04% | 4,931 | 0.46% | 152 | 0.56% |
| Nebraska | 1,923,826 | 8,834 | 0.46% | 47,539 | 2.47% | 353 | 4.00% | 89,744 | 4.66% | 478 | 5.42% |
| Nevada | 3,030,281 | 170,749 | 5.63% | 246,904 | 8.15% | 12,991 | 7.61% | 271,744 | 8.97% | 19,214 | 11.25% |
| New Hampshire | 1,355,244 | 13,266 | 0.98% | 36,398 | 2.69% | 573 | 4.32% | 18,343 | 1.35% | 438 | 3.30% |
| New Jersey | 8,885,418 | 224,019 | 2.52% | 851,568 | 9.58% | 14,565 | 6.50% | 1,121,134 | 12.62% | 34,314 | 15.32% |
| New Mexico | 2,097,021 | 19,329 | 0.92% | 31,328 | 1.49% | 489 | 2.53% | 37,975 | 1.81% | 564 | 2.92% |
| New York | 19,514,849 | 806,750 | 4.13% | 1,657,284 | 8.49% | 99,920 | 12.39% | 2,737,471 | 14.03% | 165,483 | 20.51% |
| North Carolina | 10,386,227 | 156,830 | 1.51% | 306,140 | 2.95% | 9,979 | 6.36% | 2,182,623 | 21.01% | 46,722 | 29.79% |
| North Dakota | 760,394 | 12,015 | 1.58% | 11,831 | 1.56% | 395 | 3.28% | 23,459 | 3.09% | 607 | 5.05% |
| Ohio | 11,675,275 | 114,337 | 0.98% | 266,601 | 2.28% | 5,118 | 4.48% | 1,422,412 | 12.18% | 19,155 | 16.75% |
| Oklahoma | 3,949,342 | 65,386 | 1.66% | 85,858 | 2.17% | 2,221 | 3.40% | 282,134 | 7.14% | 6,018 | 9.20% |
| Oregon | 4,176,346 | 82,546 | 1.98% | 185,221 | 4.44% | 5,026 | 6.09% | 75,418 | 1.81% | 2,812 | 3.41% |
| Pennsylvania | 12,794,885 | 82,482 | 0.64% | 445,725 | 3.48% | 3,338 | 4.05% | 1,352,329 | 10.57% | 14,525 | 17.61% |
| Rhode Island | 1,057,798 | 10,196 | 0.96% | 35,856 | 3.39% | 311 | 3.05% | 58,680 | 5.55% | 252 | 2.47% |
| South Carolina | 5,091,517 | 91,825 | 1.80% | 82,800 | 1.63% | 2,429 | 2.65% | 1,337,126 | 26.26% | 26,038 | 28.36% |
| South Dakota | 879,336 | 20,014 | 2.28% | 12,275 | 1.40% | 243 | 1.21% | 18,436 | 2.10% | 1,756 | 8.78% |
| Tennessee | 6,772,268 | 121,175 | 1.79% | 121,808 | 1.80% | 2,994 | 2.47% | 1,118,600 | 16.52% | 52,964 | 43.71% |
| Texas | 28,635,442 | 742,927 | 2.59% | 1,396,953 | 4.88% | 40,761 | 5.49% | 3,367,449 | 11.76% | 105,188 | 14.16% |
| Utah | 3,151,239 | 69,474 | 2.20% | 72,061 | 2.29% | 2,729 | 3.93% | 34,982 | 1.11% | 1,371 | 1.97% |
| Vermont | 624,340 | 4,295 | 0.69% | 10,042 | 1.61% | 277 | 6.45% | 7,496 | 1.20% | 197 | 4.60% |
| Virginia | 8,509,358 | 148,813 | 1.75% | 564,706 | 6.64% | 18,303 | 12.30% | 1,590,974 | 18.70% | 28,288 | 19.01% |
| Washington | 7,512,465 | 270,096 | 3.60% | 656,578 | 8.74% | 43,214 | 16.00% | 279,720 | 3.72% | 28,572 | 10.58% |
| West Virginia | 1,807,426 | 8,222 | 0.45% | 13,948 | 0.77% | 116 | 1.42% | 63,133 | 3.49% | 672 | 8.18% |
| Wisconsin | 5,806,975 | 70,283 | 1.21% | 162,010 | 2.79% | 2,152 | 3.06% | 360,526 | 6.21% | 2,488 | 3.54% |
| Wyoming | 581,348 | 14,556 | 2.50% | 4,746 | 0.82% | 173 | 1.19% | 4,712 | 0.81% | 202 | 1.39% |
| **US nationwide [b]** | **326,569,308** | **7,852,047** | **2.40%** | **18,184,182** | **5.57%** | **615,577** | **7.84%** | **39,994,653** | **12.25%** | **1,201,884** | **15.31%** |
| **Rail** | | | | | | | | | | | |
| Alabama | 4,893,186 | 47,400 | 0.97% | 67,317 | 1.38% | 622 | 1.31% | 1,292,950 | 26.42% | 20,437 | 43.12% |
| Alaska | 736,990 | 207 | 0.03% | 45,855 | 6.22% | 12 | 5.89% | 22,720 | 3.08% | 7 | 3.40% |
| Arizona | 7,174,064 | 26,433 | 0.37% | 233,048 | 3.25% | 801 | 3.03% | 305,973 | 4.26% | 1,411 | 5.34% |
| Arkansas | 3,011,873 | 43,780 | 1.45% | 45,389 | 1.51% | 460 | 1.05% | 455,485 | 15.12% | 14,020 | 32.02% |
| California | 39,346,023 | 640,191 | 1.63% | 5,743,983 | 14.60% | 90,472 | 14.13% | 2,142,371 | 5.44% | 43,427 | 6.78% |
| Colorado | 5,684,926 | 36,644 | 0.64% | 178,265 | 3.14% | 905 | 2.47% | 224,190 | 3.94% | 1,551 | 4.23% |
| Connecticut | 3,570,549 | 44,835 | 1.26% | 161,787 | 4.53% | 2,206 | 4.92% | 352,036 | 9.86% | 6,061 | 13.52% |
| Delaware | 967,679 | 9,854 | 1.02% | 38,249 | 3.95% | 443 | 4.50% | 208,126 | 21.51% | 3,153 | 32.00% |
| District of Columbia | 701,974 | 18,743 | 2.67% | 28,347 | 4.04% | 523 | 2.79% | 312,661 | 44.54% | 10,088 | 53.82% |
| Florida | 21,216,924 | 124,203 | 0.59% | 579,476 | 2.73% | 2,224 | 1.79% | 3,231,108 | 15.23% | 34,936 | 28.13% |
| Georgia | 10,516,579 | 95,551 | 0.91% | 430,473 | 4.09% | 4,366 | 4.57% | 3,275,581 | 31.15% | 37,009 | 38.73% |
| Hawaii [c] | NA | NA | NA | NA | NA | NA | NA | NA | NA | NA | NA |
| Idaho | 1,754,367 | 10,271 | 0.59% | 23,603 | 1.35% | 120 | 1.17% | 10,558 | 0.60% | 60 | 0.59% |
| Illinois | 12,716,164 | 619,756 | 4.87% | 702,457 | 5.52% | 34,571 | 5.58% | 1,766,586 | 13.89% | 155,795 | 25.14% |
| Indiana | 6,696,893 | 156,063 | 2.33% | 157,076 | 2.35% | 2,377 | 1.52% | 622,317 | 9.29% | 23,192 | 14.86% |
| Iowa | 3,150,011 | 55,380 | 1.76% | 78,873 | 2.50% | 1,297 | 2.34% | 114,312 | 3.63% | 3,157 | 5.70% |
| Kansas | 2,912,619 | 75,890 | 2.61% | 85,734 | 2.94% | 1,611 | 2.12% | 159,797 | 5.49% | 4,977 | 6.56% |
| Kentucky | 4,461,952 | 60,433 | 1.35% | 67,524 | 1.51% | 1,031 | 1.71% | 356,048 | 7.98% | 11,075 | 18.33% |
| Louisiana | 4,664,616 | 56,838 | 1.22% | 79,976 | 1.71% | 676 | 1.19% | 1,489,071 | 31.92% | 27,123 | 47.72% |
| Maine | 1,340,825 | 3,993 | 0.30% | 15,157 | 1.13% | 84 | 2.11% | 17,761 | 1.32% | 81 | 2.03% |
| Maryland | 6,037,624 | 61,024 | 1.01% | 382,027 | 6.33% | 4,043 | 6.62% | 1,773,702 | 29.38% | 22,518 | 36.90% |
| Massachusetts | 6,873,003 | 148,550 | 2.16% | 462,831 | 6.73% | 15,392 | 10.36% | 466,288 | 6.78% | 18,369 | 12.37% |
| Michigan | 9,973,907 | 40,493 | 0.41% | 314,736 | 3.16% | 946 | 2.34% | 1,342,592 | 13.46% | 6,824 | 16.85% |
| Minnesota | 5,600,166 | 92,349 | 1.65% | 273,100 | 4.88% | 6,732 | 7.29% | 354,540 | 6.33% | 10,107 | 10.94% |
| Mississippi | 2,981,835 | 28,238 | 0.95% | 29,561 | 0.99% | 225 | 0.80% | 1,118,689 | 37.52% | 15,545 | 55.05% |
| Missouri | 6,124,160 | 90,298 | 1.47% | 122,506 | 2.00% | 1,640 | 1.82% | 692,510 | 11.31% | 16,412 | 18.18% |
| Montana | 1,061,705 | 11,702 | 1.10% | 8,527 | 0.80% | 79 | 0.68% | 4,931 | 0.46% | 50 | 0.42% |
| Nebraska | 1,923,826 | 62,831 | 3.27% | 47,539 | 2.47% | 1,224 | 1.95% | 89,744 | 4.66% | 1,961 | 3.12% |
| Nevada | 3,030,281 | 5,465 | 0.18% | 246,904 | 8.15% | 481 | 8.81% | 271,744 | 8.97% | 664 | 12.15% |
| New Hampshire | 1,355,244 | 2,974 | 0.22% | 36,398 | 2.69% | 98 | 3.28% | 18,343 | 1.35% | 32 | 1.08% |
| New Jersey | 8,885,418 | 271,157 | 3.05% | 851,568 | 9.58% | 27,287 | 10.06% | 1,121,134 | 12.62% | 45,420 | 16.75% |
| New Mexico | 2,097,021 | 11,216 | 0.53% | 31,328 | 1.49% | 106 | 0.94% | 37,975 | 1.81% | 272 | 2.42% |
| New York | 19,514,849 | 653,141 | 3.35% | 1,657,284 | 8.49% | 76,770 | 11.75% | 2,737,471 | 14.03% | 129,173 | 19.78% |
| North Carolina | 10,386,227 | 29,640 | 0.29% | 306,140 | 2.95% | 991 | 3.35% | 2,182,623 | 21.01% | 9,760 | 32.93% |
| North Dakota | 760,394 | 12,955 | 1.70% | 11,831 | 1.56% | 318 | 2.45% | 23,459 | 3.09% | 575 | 4.44% |
| Ohio | 11,675,275 | 239,405 | 2.05% | 266,601 | 2.28% | 4,614 | 1.93% | 1,422,412 | 12.18% | 47,048 | 19.65% |
| Oklahoma | 3,949,342 | 54,604 | 1.38% | 85,858 | 2.17% | 640 | 1.17% | 282,134 | 7.14% | 4,501 | 8.24% |

| | | | | | | | | | | |
|---|---|---|---|---|---|---|---|---|---|---|
| Oregon | 4,176,346 | 56,836 | 1.36% | 185,221 | 4.44% | 2,959 | 5.21% | 75,418 | 1.81% | 2,189 | 3.85% |
| Pennsylvania | 12,794,885 | 392,473 | 3.07% | 445,725 | 3.48% | 17,624 | 4.49% | 1,352,329 | 10.57% | 94,984 | 24.20% |
| Rhode Island | 1,057,798 | 9,148 | 0.86% | 35,856 | 3.39% | 349 | 3.81% | 58,680 | 5.55% | 1,026 | 11.22% |
| South Carolina | 5,091,517 | 29,571 | 0.58% | 82,800 | 1.63% | 483 | 1.63% | 1,337,126 | 26.26% | 11,436 | 38.67% |
| South Dakota | 879,336 | 4,567 | 0.52% | 12,275 | 1.40% | 74 | 1.61% | 18,436 | 2.10% | 215 | 4.72% |
| Tennessee | 6,772,268 | 32,765 | 0.48% | 121,808 | 1.80% | 607 | 1.85% | 1,118,600 | 16.52% | 9,874 | 30.14% |
| Texas | 28,635,442 | 455,480 | 1.59% | 1,396,953 | 4.88% | 14,370 | 3.15% | 3,367,449 | 11.76% | 63,403 | 13.92% |
| Utah | 3,151,239 | 39,269 | 1.25% | 72,061 | 2.29% | 1,038 | 2.64% | 34,982 | 1.11% | 634 | 1.61% |
| Vermont | 624,340 | 1,235 | 0.20% | 10,042 | 1.61% | 32 | 2.62% | 7,496 | 1.20% | 22 | 1.78% |
| Virginia | 8,509,358 | 106,907 | 1.26% | 564,706 | 6.64% | 5,679 | 5.31% | 1,590,974 | 18.70% | 31,331 | 29.31% |
| Washington | 7,512,465 | 67,329 | 0.90% | 656,578 | 8.74% | 5,496 | 8.16% | 279,720 | 3.72% | 3,140 | 4.66% |
| West Virginia | 1,807,426 | 10,588 | 0.59% | 13,948 | 0.77% | 74 | 0.70% | 63,133 | 3.49% | 857 | 8.10% |
| Wisconsin | 5,806,975 | 57,880 | 1.00% | 162,010 | 2.79% | 2,055 | 3.55% | 360,526 | 6.21% | 3,903 | 6.74% |
| Wyoming | 581,348 | 7,219 | 1.24% | 4,746 | 0.82% | 85 | 1.18% | 4,712 | 0.81% | 76 | 1.06% |
| **US nationwide [b]** | **325,149,234** | **5,213,774** | **1.60%** | **17,662,057** | **5.43%** | **337,312** | **6.47%** | **39,969,523** | **12.29%** | **949,881** | **18.22%** |
| **Road** | | | | | | | | | | | |
| Alabama | 4,893,186 | 76,142 | 1.56% | 67,317 | 1.38% | 1,380 | 1.81% | 1,292,950 | 26.42% | 26,636 | 34.98% |
| Alaska | 736,990 | 9,304 | 1.26% | 45,855 | 6.22% | 908 | 9.75% | 22,720 | 3.08% | 498 | 5.36% |
| Arizona | 7,174,064 | 211,571 | 2.95% | 233,048 | 3.25% | 8,160 | 3.86% | 305,973 | 4.26% | 11,287 | 5.33% |
| Arkansas | 3,011,873 | 36,522 | 1.21% | 45,389 | 1.51% | 748 | 2.05% | 455,485 | 15.12% | 7,791 | 21.33% |
| California | 39,346,023 | 1,701,228 | 4.32% | 5,743,983 | 14.60% | 275,208 | 16.18% | 2,142,371 | 5.44% | 105,223 | 6.19% |
| Colorado | 5,684,926 | 121,291 | 2.13% | 178,265 | 3.14% | 4,354 | 3.59% | 224,190 | 3.94% | 6,037 | 4.98% |
| Connecticut | 3,570,549 | 76,334 | 2.14% | 161,787 | 4.53% | 4,265 | 5.59% | 352,036 | 9.86% | 10,791 | 14.14% |
| Delaware | 967,679 | 18,051 | 1.87% | 38,249 | 3.95% | 848 | 4.70% | 208,126 | 21.51% | 4,722 | 26.16% |
| District of Columbia | 701,974 | 38,012 | 5.42% | 28,347 | 4.04% | 1,930 | 5.08% | 312,661 | 44.54% | 14,991 | 39.44% |
| Florida | 21,216,924 | 464,037 | 2.19% | 579,476 | 2.73% | 13,268 | 2.86% | 3,231,108 | 15.23% | 84,988 | 18.31% |
| Georgia | 10,516,579 | 149,958 | 1.43% | 430,473 | 4.09% | 8,136 | 5.43% | 3,275,581 | 31.15% | 57,121 | 38.09% |
| Hawaii | 1,420,074 | 40,059 | 2.82% | 522,125 | 36.77% | 19,189 | 47.90% | 25,130 | 1.77% | 675 | 1.68% |
| Idaho | 1,754,367 | 27,644 | 1.58% | 23,603 | 1.35% | 467 | 1.69% | 10,558 | 0.60% | 223 | 0.81% |
| Illinois | 12,716,164 | 330,485 | 2.60% | 702,457 | 5.52% | 26,151 | 7.91% | 1,766,586 | 13.89% | 56,416 | 17.07% |
| Indiana | 6,696,893 | 127,723 | 1.91% | 157,076 | 2.35% | 3,785 | 2.96% | 622,317 | 9.29% | 17,892 | 14.01% |
| Iowa | 3,150,011 | 53,147 | 1.69% | 78,873 | 2.50% | 1,947 | 3.66% | 114,312 | 3.63% | 3,212 | 6.04% |
| Kansas | 2,912,619 | 41,791 | 1.43% | 85,734 | 2.94% | 1,690 | 4.04% | 159,797 | 5.49% | 3,096 | 7.41% |
| Kentucky | 4,461,952 | 86,389 | 1.94% | 67,524 | 1.51% | 1,965 | 2.27% | 356,048 | 7.98% | 11,973 | 13.86% |
| Louisiana | 4,664,616 | 66,645 | 1.43% | 79,976 | 1.71% | 1,386 | 2.08% | 1,489,071 | 31.92% | 27,432 | 41.16% |
| Maine | 1,340,825 | 15,610 | 1.16% | 15,157 | 1.13% | 302 | 1.94% | 17,761 | 1.32% | 490 | 3.14% |
| Maryland | 6,037,624 | 136,859 | 2.27% | 382,027 | 6.33% | 9,800 | 7.16% | 1,773,702 | 29.38% | 47,284 | 34.55% |
| Massachusetts | 6,873,003 | 241,444 | 3.51% | 462,831 | 6.73% | 21,622 | 8.96% | 466,288 | 6.78% | 19,948 | 8.26% |
| Michigan | 9,973,907 | 219,416 | 2.20% | 314,736 | 3.16% | 8,948 | 4.08% | 1,342,592 | 13.46% | 45,667 | 20.81% |
| Minnesota | 5,600,166 | 93,753 | 1.67% | 273,100 | 4.88% | 6,696 | 7.14% | 354,540 | 6.33% | 10,810 | 11.53% |
| Mississippi | 2,981,835 | 30,670 | 1.03% | 29,561 | 0.99% | 393 | 1.28% | 1,118,689 | 37.52% | 13,413 | 43.73% |
| Missouri | 6,124,160 | 105,419 | 1.72% | 122,506 | 2.00% | 3,004 | 2.85% | 692,510 | 11.31% | 18,206 | 17.27% |
| Montana | 1,061,705 | 13,420 | 1.26% | 8,527 | 0.80% | 138 | 1.03% | 4,931 | 0.46% | 81 | 0.60% |
| Nebraska | 1,923,826 | 32,409 | 1.68% | 47,539 | 2.47% | 1,056 | 3.26% | 89,744 | 4.66% | 2,246 | 6.93% |
| Nevada | 3,030,281 | 62,837 | 2.07% | 246,904 | 8.15% | 5,013 | 7.98% | 271,744 | 8.97% | 6,067 | 9.65% |
| New Hampshire | 1,355,244 | 19,159 | 1.41% | 36,398 | 2.69% | 770 | 4.02% | 18,343 | 1.35% | 488 | 2.55% |
| New Jersey | 8,885,418 | 237,475 | 2.67% | 851,568 | 9.58% | 24,807 | 10.45% | 1,121,134 | 12.62% | 39,246 | 16.53% |
| New Mexico | 2,097,021 | 38,048 | 1.81% | 31,328 | 1.49% | 764 | 2.01% | 37,975 | 1.81% | 888 | 2.33% |
| New York | 19,514,849 | 731,900 | 3.75% | 1,657,284 | 8.49% | 85,064 | 11.62% | 2,737,471 | 14.03% | 122,823 | 16.78% |
| North Carolina | 10,386,227 | 160,131 | 1.54% | 306,140 | 2.95% | 6,545 | 4.09% | 2,182,623 | 21.01% | 42,103 | 26.29% |
| North Dakota | 760,394 | 10,775 | 1.42% | 11,831 | 1.56% | 283 | 2.63% | 23,459 | 3.09% | 600 | 5.56% |
| Ohio | 11,675,275 | 202,249 | 1.73% | 266,601 | 2.28% | 5,707 | 2.82% | 1,422,412 | 12.18% | 36,534 | 18.06% |
| Oklahoma | 3,949,342 | 53,170 | 1.35% | 85,858 | 2.17% | 1,413 | 2.66% | 282,134 | 7.14% | 5,006 | 9.41% |
| Oregon | 4,176,346 | 115,891 | 2.77% | 185,221 | 4.44% | 6,460 | 5.57% | 75,418 | 1.81% | 3,096 | 2.67% |
| Pennsylvania | 12,794,885 | 257,159 | 2.01% | 445,725 | 3.48% | 12,981 | 5.05% | 1,352,329 | 10.57% | 38,352 | 14.91% |
| Rhode Island | 1,057,798 | 28,687 | 2.71% | 35,856 | 3.39% | 1,140 | 3.97% | 58,680 | 5.55% | 2,386 | 8.32% |
| South Carolina | 5,091,517 | 65,606 | 1.29% | 82,800 | 1.63% | 1,367 | 2.08% | 1,337,126 | 26.26% | 19,369 | 29.52% |
| South Dakota | 879,336 | 12,852 | 1.46% | 12,275 | 1.40% | 273 | 2.12% | 18,436 | 2.10% | 525 | 4.09% |
| Tennessee | 6,772,268 | 114,146 | 1.69% | 121,808 | 1.80% | 2,901 | 2.54% | 1,118,600 | 16.52% | 30,056 | 26.33% |
| Texas | 28,635,442 | 667,926 | 2.33% | 1,396,953 | 4.88% | 37,062 | 5.55% | 3,367,449 | 11.76% | 91,425 | 13.69% |
| Utah | 3,151,239 | 82,329 | 2.61% | 72,061 | 2.29% | 2,430 | 2.95% | 34,982 | 1.11% | 1,128 | 1.37% |
| Vermont | 624,340 | 5,943 | 0.95% | 10,042 | 1.61% | 143 | 2.40% | 7,496 | 1.20% | 114 | 1.92% |
| Virginia | 8,509,358 | 173,307 | 2.04% | 564,706 | 6.64% | 16,663 | 9.61% | 1,590,974 | 18.70% | 36,794 | 21.23% |
| Washington | 7,512,465 | 176,478 | 2.35% | 656,578 | 8.74% | 21,342 | 12.09% | 279,720 | 3.72% | 9,194 | 5.21% |
| West Virginia | 1,807,426 | 20,911 | 1.16% | 13,948 | 0.77% | 258 | 1.24% | 63,133 | 3.49% | 1,159 | 5.54% |
| Wisconsin | 5,806,975 | 118,153 | 2.03% | 162,010 | 2.79% | 4,989 | 4.22% | 360,526 | 6.21% | 13,186 | 11.16% |
| Wyoming | 581,348 | 7,641 | 1.31% | 4,746 | 0.82% | 80 | 1.05% | 4,712 | 0.81% | 73 | 0.96% |
| **US nationwide [b]** | **326,569,308** | **7,928,106** | **2.43%** | **18,184,182** | **5.57%** | **666,199** | **8.40%** | **39,994,653** | **12.25%** | **1,109,761** | **14.00%** |

Definition of abbreviation: Pop: population.

[a] Greater than or equal to 54 dB Lden.

[b] Includes the continental US and the states of Alaska and Hawaii.

[c] No rail data available.

**Table S5. Table S4 continued.**

| State | Hispanic | | | | White | | | |
|---|---|---|---|---|---|---|---|---|
| | Pop | % | #HA | %HA | Pop | % | #HA | %HA |
| Alabama | 212,951 | 4.35% | 1,986 | 4.05% | 3,192,147 | 65.24% | 26,058 | 53.13% |
| Alaska | 53,059 | 7.20% | 2,221 | 8.07% | 439,979 | 59.70% | 15,669 | 56.96% |
| Arizona | 2,260,690 | 31.51% | 91,608 | 37.51% | 3,883,722 | 54.14% | 117,639 | 48.17% |
| Arkansas | 229,629 | 7.62% | 3,577 | 16.11% | 2,155,099 | 71.55% | 12,784 | 57.57% |
| California | 15,380,929 | 39.09% | 627,365 | 44.66% | 14,365,145 | 36.51% | 416,756 | 29.67% |
| Colorado | 1,231,126 | 21.66% | 31,759 | 23.39% | 3,837,450 | 67.50% | 84,704 | 62.39% |
| Connecticut | 587,212 | 16.45% | 3,243 | 18.09% | 2,357,942 | 66.04% | 11,399 | 63.58% |
| Delaware | 91,350 | 9.44% | 2,139 | 11.34% | 595,236 | 61.51% | 9,786 | 51.87% |
| District of Columbia | 77,981 | 11.11% | 1,783 | 11.01% | 257,792 | 36.72% | 10,706 | 66.10% |
| Florida | 5,468,826 | 25.78% | 320,367 | 37.61% | 11,331,222 | 53.41% | 319,223 | 37.48% |
| Georgia | 1,013,057 | 9.63% | 24,016 | 15.00% | 5,478,289 | 52.09% | 42,667 | 26.66% |
| Hawaii | 152,566 | 10.74% | 2,425 | 10.68% | 306,360 | 21.57% | 4,300 | 18.93% |
| Idaho | 222,967 | 12.71% | 2,255 | 8.66% | 1,427,529 | 81.37% | 22,105 | 84.90% |
| Illinois | 2,190,696 | 17.23% | 205,631 | 31.35% | 7,737,459 | 60.85% | 315,657 | 48.12% |
| Indiana | 475,475 | 7.10% | 6,511 | 11.52% | 5,253,453 | 78.45% | 36,231 | 64.09% |
| Iowa | 194,407 | 6.17% | 2,235 | 10.31% | 2,677,946 | 85.01% | 17,271 | 79.65% |
| Kansas | 351,602 | 12.07% | 3,967 | 13.57% | 2,193,988 | 75.33% | 21,598 | 73.88% |
| Kentucky | 167,949 | 3.76% | 4,071 | 5.66% | 3,751,738 | 84.08% | 47,715 | 66.33% |
| Louisiana | 243,372 | 5.22% | 4,733 | 11.31% | 2,720,638 | 58.33% | 17,070 | 40.78% |
| Maine | 23,143 | 1.73% | 202 | 2.30% | 1,242,113 | 92.64% | 7,520 | 85.64% |
| Maryland | 619,418 | 10.26% | 11,928 | 11.78% | 3,028,494 | 50.16% | 50,715 | 50.10% |
| Massachusetts | 828,140 | 12.05% | 52,397 | 26.44% | 4,865,022 | 70.78% | 111,568 | 56.30% |
| Michigan | 521,203 | 5.23% | 4,289 | 5.30% | 7,428,622 | 74.48% | 52,594 | 65.03% |
| Minnesota | 307,675 | 5.49% | 9,452 | 8.46% | 4,422,490 | 78.97% | 80,525 | 72.07% |
| Mississippi | 94,342 | 3.16% | 2,158 | 3.70% | 1,680,938 | 56.37% | 32,446 | 55.63% |
| Missouri | 262,677 | 4.29% | 2,519 | 3.98% | 4,826,943 | 78.82% | 27,607 | 43.61% |
| Montana | 41,501 | 3.91% | 1,366 | 5.05% | 908,782 | 85.60% | 23,188 | 85.74% |
| Nebraska | 214,952 | 11.17% | 1,259 | 14.25% | 1,507,127 | 78.34% | 6,362 | 72.02% |
| Nevada | 875,798 | 28.90% | 65,915 | 38.60% | 1,460,159 | 48.19% | 63,398 | 37.13% |
| New Hampshire | 52,792 | 3.90% | 1,315 | 9.91% | 1,214,029 | 89.58% | 10,603 | 79.92% |
| New Jersey | 1,815,078 | 20.43% | 85,429 | 38.13% | 4,858,807 | 54.68% | 82,789 | 36.96% |
| New Mexico | 1,031,788 | 49.20% | 10,698 | 55.35% | 769,139 | 36.68% | 6,231 | 32.24% |
| New York | 3,720,707 | 19.07% | 306,509 | 37.99% | 10,766,297 | 55.17% | 212,400 | 26.33% |
| North Carolina | 991,051 | 9.54% | 15,661 | 9.99% | 6,503,292 | 62.61% | 79,828 | 50.90% |
| North Dakota | 30,325 | 3.99% | 574 | 4.78% | 636,284 | 83.68% | 9,728 | 80.97% |
| Ohio | 459,939 | 3.94% | 7,234 | 6.33% | 9,141,370 | 78.30% | 77,844 | 68.08% |
| Oklahoma | 431,467 | 10.93% | 20,002 | 30.59% | 2,563,119 | 64.90% | 30,381 | 46.46% |
| Oregon | 552,279 | 13.22% | 14,978 | 18.15% | 3,128,494 | 74.91% | 54,360 | 65.85% |
| Pennsylvania | 971,813 | 7.60% | 7,256 | 8.80% | 9,685,118 | 75.70% | 54,550 | 66.13% |
| Rhode Island | 168,007 | 15.88% | 755 | 7.41% | 755,708 | 71.44% | 8,621 | 84.56% |
| South Carolina | 296,897 | 5.83% | 9,422 | 10.26% | 3,230,111 | 63.44% | 50,808 | 55.33% |
| South Dakota | 36,088 | 4.10% | 1,647 | 8.23% | 715,328 | 81.35% | 14,969 | 74.79% |
| Tennessee | 377,162 | 5.57% | 14,499 | 11.97% | 4,969,371 | 73.38% | 47,658 | 39.33% |
| Texas | 11,294,257 | 39.44% | 333,309 | 44.86% | 11,850,477 | 41.38% | 243,597 | 32.79% |
| Utah | 446,067 | 14.16% | 18,665 | 26.87% | 2,455,192 | 77.91% | 42,800 | 61.61% |
| Vermont | 12,518 | 2.00% | 125 | 2.92% | 576,601 | 92.35% | 3,552 | 82.70% |
| Virginia | 810,770 | 9.53% | 15,410 | 10.36% | 5,209,336 | 61.22% | 79,794 | 53.62% |
| Washington | 971,522 | 12.93% | 44,311 | 16.41% | 5,067,909 | 67.46% | 132,273 | 48.97% |
| West Virginia | 28,679 | 1.59% | 193 | 2.35% | 1,654,681 | 91.55% | 6,985 | 84.96% |
| Wisconsin | 408,267 | 7.03% | 6,029 | 8.58% | 4,681,072 | 80.61% | 57,137 | 81.30% |
| Wyoming | 58,854 | 10.12% | 1,651 | 11.34% | 485,816 | 83.57% | 12,054 | 82.81% |
| **US nationwide** | **59,361,020** | **18.18%** | **2,409,049** | **30.68%** | **196,251,375** | **60.09%** | **3,324,223** | **42.34%** |
| **Rail** | | | | | | | | |
| Alabama | 212,951 | 4.35% | 2,683 | 5.66% | 3,192,147 | 65.24% | 22,494 | 47.46% |
| Alaska | 53,059 | 7.20% | 17 | 8.00% | 439,979 | 59.70% | 139 | 67.02% |
| Arizona | 2,260,690 | 31.51% | 10,236 | 38.72% | 3,883,722 | 54.14% | 11,962 | 45.25% |
| Arkansas | 229,629 | 7.62% | 3,884 | 8.87% | 2,155,099 | 71.55% | 23,629 | 53.97% |
| California | 15,380,929 | 39.09% | 310,128 | 48.44% | 14,365,145 | 36.51% | 170,358 | 26.61% |
| Colorado | 1,231,126 | 21.66% | 10,656 | 29.08% | 3,837,450 | 67.50% | 22,210 | 60.61% |
| Connecticut | 587,212 | 16.45% | 12,718 | 28.37% | 2,357,942 | 66.04% | 22,424 | 50.01% |
| Delaware | 91,350 | 9.44% | 1,074 | 10.90% | 595,236 | 61.51% | 4,851 | 49.23% |
| District of Columbia | 77,981 | 11.11% | 1,868 | 9.97% | 257,792 | 36.72% | 5,654 | 30.17% |
| Florida | 5,468,826 | 25.78% | 40,090 | 32.28% | 11,331,222 | 53.41% | 44,147 | 35.54% |
| Georgia | 1,013,057 | 9.63% | 10,062 | 10.53% | 5,478,289 | 52.09% | 40,982 | 42.89% |
| Hawaii | NA | NA | NA | NA | NA | NA | NA | NA |
| Idaho | 222,967 | 12.71% | 1,915 | 18.64% | 1,427,529 | 81.37% | 7,735 | 75.31% |
| Illinois | 2,190,696 | 17.23% | 145,580 | 23.49% | 7,737,459 | 60.85% | 268,814 | 43.37% |
| Indiana | 475,475 | 7.10% | 20,808 | 13.33% | 5,253,453 | 78.45% | 104,740 | 67.11% |
| Iowa | 194,407 | 6.17% | 5,703 | 10.30% | 2,677,946 | 85.01% | 43,353 | 78.28% |
| Kansas | 351,602 | 12.07% | 13,819 | 18.21% | 2,193,988 | 75.33% | 51,975 | 68.49% |
| Kentucky | 167,949 | 3.76% | 3,045 | 5.04% | 3,751,738 | 84.08% | 42,970 | 71.10% |
| Louisiana | 243,372 | 5.22% | 2,746 | 4.83% | 2,720,638 | 58.33% | 24,869 | 43.75% |
| Maine | 23,143 | 1.73% | 99 | 2.49% | 1,242,113 | 92.64% | 3,582 | 89.71% |
| Maryland | 619,418 | 10.26% | 8,029 | 13.16% | 3,028,494 | 50.16% | 23,945 | 39.24% |
| Massachusetts | 828,140 | 12.05% | 27,195 | 18.31% | 4,865,022 | 70.78% | 81,772 | 55.05% |
| Michigan | 521,203 | 5.23% | 2,908 | 7.18% | 7,428,622 | 74.48% | 28,011 | 69.18% |
| Minnesota | 307,675 | 5.49% | 6,223 | 6.74% | 4,422,490 | 78.97% | 64,871 | 70.25% |
| Mississippi | 94,342 | 3.16% | 1,175 | 4.16% | 1,680,938 | 56.37% | 10,850 | 38.42% |
| Missouri | 262,677 | 4.29% | 7,590 | 8.40% | 4,826,943 | 78.82% | 61,156 | 67.73% |
| Montana | 41,501 | 3.91% | 579 | 4.95% | 908,782 | 85.60% | 9,873 | 84.37% |
| Nebraska | 214,952 | 11.17% | 12,111 | 19.28% | 1,507,127 | 78.34% | 45,751 | 72.82% |
| Nevada | 875,798 | 28.90% | 1,403 | 25.68% | 1,460,159 | 48.19% | 2,604 | 47.66% |
| New Hampshire | 52,792 | 3.90% | 134 | 4.50% | 1,214,029 | 89.58% | 2,627 | 88.31% |
| New Jersey | 1,815,078 | 20.43% | 81,775 | 30.16% | 4,858,807 | 54.68% | 108,815 | 40.13% |
| New Mexico | 1,031,788 | 49.20% | 6,593 | 58.78% | 769,139 | 36.68% | 2,914 | 25.98% |
| New York | 3,720,707 | 19.07% | 218,199 | 33.41% | 10,766,297 | 55.17% | 206,366 | 31.60% |
| North Carolina | 991,051 | 9.54% | 3,740 | 12.62% | 6,503,292 | 62.61% | 13,966 | 47.12% |
| North Dakota | 30,325 | 3.99% | 593 | 4.58% | 636,284 | 83.68% | 10,671 | 82.37% |
| Ohio | 459,939 | 3.94% | 15,588 | 6.51% | 9,141,370 | 78.30% | 162,539 | 67.89% |
| Oklahoma | 431,467 | 10.93% | 8,757 | 16.04% | 2,563,119 | 64.90% | 32,039 | 58.68% |
| Oregon | 552,279 | 13.22% | 9,323 | 16.40% | 3,128,494 | 74.91% | 38,846 | 68.35% |
| Pennsylvania | 971,813 | 7.60% | 40,667 | 10.36% | 9,685,118 | 75.70% | 226,105 | 57.61% |
| Rhode Island | 168,007 | 15.88% | 3,022 | 33.03% | 755,708 | 71.44% | 4,170 | 45.59% |
| South Carolina | 296,897 | 5.83% | 2,446 | 8.27% | 3,230,111 | 63.44% | 14,299 | 48.35% |
| South Dakota | 36,088 | 4.10% | 285 | 6.24% | 715,328 | 81.35% | 3,445 | 75.43% |
| Tennessee | 377,162 | 5.57% | 2,195 | 6.70% | 4,969,371 | 73.38% | 19,176 | 58.52% |
| Texas | 11,294,257 | 39.44% | 218,789 | 48.03% | 11,850,477 | 41.38% | 149,177 | 32.75% |
| Utah | 446,067 | 14.16% | 8,144 | 20.74% | 2,455,192 | 77.91% | 27,324 | 69.58% |
| Vermont | 12,518 | 2.00% | 28 | 2.27% | 576,601 | 92.35% | 1,119 | 90.56% |
| Virginia | 810,770 | 9.53% | 10,554 | 9.87% | 5,209,336 | 61.22% | 54,814 | 51.27% |
| Washington | 971,522 | 12.93% | 9,919 | 14.73% | 5,067,909 | 67.46% | 44,035 | 65.40% |
| West Virginia | 28,679 | 1.59% | 267 | 2.52% | 1,654,681 | 91.55% | 8,992 | 84.93% |
| Wisconsin | 408,267 | 7.03% | 6,017 | 10.39% | 4,681,072 | 80.61% | 44,072 | 76.14% |
| Wyoming | 58,854 | 10.12% | 1,166 | 16.15% | 485,816 | 83.57% | 5,606 | 77.66% |
| **US nationwide** | **59,208,454** | **18.21%** | **1,312,545** | **25.17%** | **195,945,015** | **60.26%** | **2,426,838** | **46.55%** |
| **Road** | | | | | | | | |
| Alabama | 212,951 | 4.35% | 3,915 | 5.14% | 3,192,147 | 65.24% | 42,313 | 55.57% |
| Alaska | 53,059 | 7.20% | 821 | 8.82% | 439,979 | 59.70% | 5,050 | 54.27% |
| Arizona | 2,260,690 | 31.51% | 73,095 | 34.55% | 3,883,722 | 54.14% | 107,998 | 51.05% |
| Arkansas | 229,629 | 7.62% | 3,632 | 9.95% | 2,155,099 | 71.55% | 22,661 | 62.05% |
| California | 15,380,929 | 39.09% | 710,592 | 41.77% | 14,365,145 | 36.51% | 538,244 | 31.64% |
| Colorado | 1,231,126 | 21.66% | 29,649 | 24.44% | 3,837,450 | 67.50% | 76,572 | 63.13% |
| Connecticut | 587,212 | 16.45% | 17,545 | 22.98% | 2,357,942 | 66.04% | 41,220 | 54.00% |
| Delaware | 91,350 | 9.44% | 2,099 | 11.63% | 595,236 | 61.51% | 9,765 | 54.10% |
| District of Columbia | 77,981 | 11.11% | 4,498 | 11.83% | 257,792 | 36.72% | 15,228 | 40.06% |
| Florida | 5,468,826 | 25.78% | 150,028 | 32.33% | 11,331,222 | 53.41% | 202,987 | 43.74% |
| Georgia | 1,013,057 | 9.63% | 15,996 | 10.67% | 5,478,289 | 52.09% | 63,866 | 42.59% |
| Hawaii | 152,566 | 10.74% | 3,573 | 8.92% | 306,360 | 21.57% | 6,198 | 15.47% |
| Idaho | 222,967 | 12.71% | 3,625 | 13.11% | 1,427,529 | 81.37% | 22,204 | 80.32% |
| Illinois | 2,190,696 | 17.23% | 64,703 | 19.58% | 7,737,459 | 60.85% | 174,321 | 52.75% |
| Indiana | 475,475 | 7.10% | 11,646 | 9.12% | 5,253,453 | 78.45% | 90,257 | 70.67% |
| Iowa | 194,407 | 6.17% | 4,072 | 7.66% | 2,677,946 | 85.01% | 42,096 | 79.21% |
| Kansas | 351,602 | 12.07% | 6,236 | 14.92% | 2,193,988 | 75.33% | 28,887 | 69.12% |
| Kentucky | 167,949 | 3.76% | 4,304 | 4.98% | 3,751,738 | 84.08% | 65,235 | 75.51% |
| Louisiana | 243,372 | 5.22% | 4,429 | 6.64% | 2,720,638 | 58.33% | 31,574 | 47.38% |
| Maine | 23,143 | 1.73% | 334 | 2.14% | 1,242,113 | 92.64% | 13,914 | 89.14% |
| Maryland | 619,418 | 10.26% | 15,870 | 11.60% | 3,028,494 | 50.16% | 58,563 | 42.79% |
| Massachusetts | 828,140 | 12.05% | 39,387 | 16.31% | 4,865,022 | 70.78% | 151,138 | 62.60% |
| Michigan | 521,203 | 5.23% | 14,206 | 6.47% | 7,428,622 | 74.48% | 142,149 | 64.79% |
| Minnesota | 307,675 | 5.49% | 6,956 | 7.42% | 4,422,490 | 78.97% | 64,744 | 69.06% |
| Mississippi | 94,342 | 3.16% | 1,112 | 3.63% | 1,680,938 | 56.37% | 15,162 | 49.43% |
| Missouri | 262,677 | 4.29% | 5,335 | 5.06% | 4,826,943 | 78.82% | 74,836 | 70.99% |
| Montana | 41,501 | 3.91% | 631 | 4.70% | 908,782 | 85.60% | 11,611 | 86.52% |
| Nebraska | 214,952 | 11.17% | 4,582 | 14.14% | 1,507,127 | 78.34% | 23,354 | 72.06% |
| Nevada | 875,798 | 28.90% | 21,132 | 33.63% | 1,460,159 | 48.19% | 27,135 | 43.18% |
| New Hampshire | 52,792 | 3.90% | 1,389 | 7.25% | 1,214,029 | 89.58% | 15,938 | 83.19% |
| New Jersey | 1,815,078 | 20.43% | 62,360 | 26.26% | 4,858,807 | 54.68% | 104,681 | 44.08% |
| New Mexico | 1,031,788 | 49.20% | 19,716 | 51.82% | 769,139 | 36.68% | 13,813 | 36.30% |
| New York | 3,720,707 | 19.07% | 186,655 | 25.50% | 10,766,297 | 55.17% | 312,234 | 42.66% |
| North Carolina | 991,051 | 9.54% | 17,251 | 10.77% | 6,503,292 | 62.61% | 88,291 | 55.14% |
| North Dakota | 30,325 | 3.99% | 433 | 4.02% | 636,284 | 83.68% | 8,806 | 81.72% |
| Ohio | 459,939 | 3.94% | 10,251 | 5.07% | 9,141,370 | 78.30% | 141,898 | 70.16% |
| Oklahoma | 431,467 | 10.93% | 7,711 | 14.50% | 2,563,119 | 64.90% | 32,215 | 60.59% |
| Oregon | 552,279 | 13.22% | 17,565 | 15.16% | 3,128,494 | 74.91% | 81,860 | 70.64% |
| Pennsylvania | 971,813 | 7.60% | 27,150 | 10.56% | 9,685,118 | 75.70% | 170,873 | 66.45% |
| Rhode Island | 168,007 | 15.88% | 6,601 | 23.01% | 755,708 | 71.44% | 17,281 | 60.24% |
| South Carolina | 296,897 | 5.83% | 4,625 | 7.05% | 3,230,111 | 63.44% | 38,315 | 58.40% |
| South Dakota | 36,088 | 4.10% | 671 | 5.22% | 715,328 | 81.35% | 10,391 | 80.85% |
| Tennessee | 377,162 | 5.57% | 8,448 | 7.40% | 4,969,371 | 73.38% | 69,384 | 60.79% |
| Texas | 11,294,257 | 39.44% | 288,428 | 43.18% | 11,850,477 | 41.38% | 234,047 | 35.04% |
| Utah | 446,067 | 14.16% | 14,206 | 17.25% | 2,455,192 | 77.91% | 60,492 | 73.48% |
| Vermont | 12,518 | 2.00% | 146 | 2.45% | 576,601 | 92.35% | 5,358 | 90.15% |
| Virginia | 810,770 | 9.53% | 21,039 | 12.14% | 5,209,336 | 61.22% | 91,279 | 52.67% |
| Washington | 971,522 | 12.93% | 23,981 | 13.59% | 5,067,909 | 67.46% | 108,672 | 61.58% |
| West Virginia | 28,679 | 1.59% | 399 | 1.91% | 1,654,681 | 91.55% | 18,421 | 88.09% |
| Wisconsin | 408,267 | 7.03% | 12,009 | 10.16% | 4,681,072 | 80.61% | 83,602 | 70.76% |
| Wyoming | 58,854 | 10.12% | 960 | 12.57% | 485,816 | 83.57% | 6,204 | 81.19% |
| **US nationwide** | **59,361,020** | **18.18%** | **1,955,997** | **24.67%** | **196,251,375** | **60.09%** | **3,883,337** | **48.98%** |

Definition of abbreviation: Pop: population.

**Table S6. Table S5 continued.**

| State | AIAN | | | | NHPI | | | | Other | | | |
|---|---|---|---|---|---|---|---|---|---|---|---|---|
| | Pop | % | #HA | %HA | Pop | % | #HA | %HA | Pop | % | #HA | %HA |
| **Aviation** | | | | | | | | | | | | |
| Alabama | 21,297 | 0.44% | 231 | 0.47% | 1,887 | 0.04% | 8 | 0.02% | 104,637 | 2.14% | 822 | 1.68% |
| Alaska | 103,843 | 14.09% | 3,776 | 13.73% | 10,306 | 1.40% | 270 | 0.98% | 61,228 | 8.31% | 2,575 | 9.36% |
| Arizona | 272,294 | 3.80% | 4,500 | 1.84% | 13,323 | 0.19% | 506 | 0.21% | 205,014 | 2.86% | 6,486 | 2.66% |
| Arkansas | 15,912 | 0.53% | 129 | 0.58% | 10,258 | 0.34% | 445 | 2.01% | 100,101 | 3.32% | 854 | 3.85% |
| California | 131,724 | 0.33% | 3,393 | 0.24% | 135,524 | 0.34% | 6,336 | 0.45% | 1,446,347 | 3.68% | 48,637 | 3.46% |
| Colorado | 29,337 | 0.52% | 641 | 0.47% | 7,794 | 0.14% | 288 | 0.21% | 176,764 | 3.11% | 4,974 | 3.66% |
| Connecticut | 4,882 | 0.14% | 20 | 0.11% | 742 | 0.02% | 0 | 0.00% | 105,948 | 2.97% | 739 | 4.12% |
| Delaware | 2,836 | 0.29% | 85 | 0.45% | 290 | 0.03% | 13 | 0.07% | 31,592 | 3.26% | 839 | 4.45% |
| District of Columbia | 1,298 | 0.18% | 29 | 0.18% | 283 | 0.04% | 10 | 0.06% | 23,612 | 3.36% | 567 | 3.50% |
| Florida | 39,070 | 0.18% | 1,087 | 0.13% | 10,889 | 0.05% | 320 | 0.04% | 556,333 | 2.62% | 23,102 | 2.71% |
| Georgia | 17,433 | 0.17% | 353 | 0.22% | 5,441 | 0.05% | 87 | 0.05% | 296,305 | 2.82% | 4,221 | 2.64% |
| Hawaii | 2,100 | 0.15% | 36 | 0.16% | 137,046 | 9.65% | 1,873 | 8.25% | 274,747 | 19.35% | 3,576 | 15.74% |
| Idaho | 18,866 | 1.08% | 164 | 0.63% | 2,611 | 0.15% | 28 | 0.11% | 48,233 | 2.75% | 775 | 2.98% |
| Illinois | 13,301 | 0.10% | 696 | 0.11% | 3,305 | 0.03% | 102 | 0.02% | 302,360 | 2.38% | 13,747 | 2.10% |
| Indiana | 10,363 | 0.15% | 96 | 0.17% | 1,790 | 0.03% | 21 | 0.04% | 176,419 | 2.63% | 2,094 | 3.70% |
| Iowa | 8,055 | 0.26% | 89 | 0.41% | 3,644 | 0.12% | 4 | 0.02% | 72,774 | 2.31% | 571 | 2.63% |
| Kansas | 17,788 | 0.61% | 133 | 0.45% | 1,889 | 0.06% | 10 | 0.03% | 101,821 | 3.50% | 1,100 | 3.76% |
| Kentucky | 7,075 | 0.16% | 133 | 0.19% | 2,691 | 0.06% | 60 | 0.08% | 108,927 | 2.44% | 2,387 | 3.32% |
| Louisiana | 23,328 | 0.50% | 200 | 0.48% | 1,527 | 0.03% | 8 | 0.02% | 106,704 | 2.29% | 1,115 | 2.66% |
| Maine | 8,011 | 0.60% | 57 | 0.65% | 159 | 0.01% | 1 | 0.01% | 34,481 | 2.57% | 247 | 2.81% |
| Maryland | 11,223 | 0.19% | 127 | 0.13% | 1,954 | 0.03% | 25 | 0.03% | 220,806 | 3.66% | 4,948 | 4.89% |
| Massachusetts | 8,943 | 0.13% | 299 | 0.15% | 2,327 | 0.03% | 89 | 0.04% | 239,452 | 3.48% | 5,766 | 2.91% |
| Michigan | 42,931 | 0.43% | 329 | 0.41% | 2,675 | 0.03% | 51 | 0.06% | 321,148 | 3.22% | 2,607 | 3.22% |
| Minnesota | 48,409 | 0.86% | 714 | 0.64% | 1,905 | 0.03% | 15 | 0.01% | 192,047 | 3.43% | 5,052 | 4.52% |
| Mississippi | 12,897 | 0.43% | 173 | 0.30% | 1,090 | 0.04% | 17 | 0.03% | 44,318 | 1.49% | 1,034 | 1.77% |
| Missouri | 20,646 | 0.34% | 162 | 0.26% | 7,876 | 0.13% | 67 | 0.11% | 191,002 | 3.12% | 2,252 | 3.56% |
| Montana | 62,720 | 5.91% | 938 | 3.47% | 552 | 0.05% | 18 | 0.07% | 34,692 | 3.27% | 1,100 | 4.07% |
| Nebraska | 13,378 | 0.70% | 64 | 0.72% | 1,051 | 0.05% | 9 | 0.10% | 50,035 | 2.60% | 308 | 3.49% |
| Nevada | 25,928 | 0.86% | 1,164 | 0.68% | 19,401 | 0.64% | 1,275 | 0.75% | 130,347 | 4.30% | 6,791 | 3.98% |
| New Hampshire | 1,895 | 0.14% | 31 | 0.23% | 320 | 0.02% | 0 | 0.00% | 31,467 | 2.32% | 306 | 2.30% |
| New Jersey | 10,290 | 0.12% | 173 | 0.08% | 2,122 | 0.02% | 59 | 0.03% | 226,419 | 2.55% | 6,690 | 2.99% |
| New Mexico | 179,881 | 8.58% | 758 | 3.92% | 1,408 | 0.07% | 17 | 0.09% | 45,502 | 2.17% | 572 | 2.96% |
| New York | 42,429 | 0.22% | 1,398 | 0.17% | 5,537 | 0.03% | 250 | 0.03% | 585,124 | 3.00% | 20,790 | 2.58% |
| North Carolina | 108,223 | 1.04% | 303 | 0.19% | 6,249 | 0.06% | 95 | 0.06% | 288,649 | 2.78% | 4,243 | 2.71% |
| North Dakota | 37,170 | 4.89% | 315 | 2.62% | 983 | 0.13% | 8 | 0.07% | 20,342 | 2.68% | 387 | 3.22% |
| Ohio | 14,901 | 0.13% | 114 | 0.10% | 3,367 | 0.03% | 13 | 0.01% | 366,685 | 3.14% | 4,858 | 4.25% |
| Oklahoma | 288,801 | 7.31% | 2,505 | 3.83% | 5,820 | 0.15% | 50 | 0.08% | 292,143 | 7.40% | 4,209 | 6.44% |
| Oregon | 35,842 | 0.86% | 851 | 1.03% | 15,614 | 0.37% | 359 | 0.44% | 183,478 | 4.39% | 4,159 | 5.04% |
| Pennsylvania | 12,224 | 0.10% | 69 | 0.08% | 2,968 | 0.02% | 11 | 0.01% | 324,708 | 2.54% | 2,734 | 3.32% |
| Rhode Island | 2,832 | 0.27% | 6 | 0.06% | 648 | 0.06% | 0 | 0.00% | 36,067 | 3.41% | 250 | 2.45% |
| South Carolina | 14,085 | 0.28% | 251 | 0.27% | 3,384 | 0.07% | 73 | 0.08% | 127,114 | 2.50% | 2,802 | 3.05% |
| South Dakota | 72,142 | 8.20% | 719 | 3.59% | 494 | 0.06% | 3 | 0.02% | 24,573 | 2.79% | 677 | 3.38% |
| Tennessee | 13,814 | 0.20% | 135 | 0.11% | 3,439 | 0.05% | 56 | 0.05% | 168,074 | 2.48% | 2,869 | 2.37% |
| Texas | 65,132 | 0.23% | 1,588 | 0.21% | 21,477 | 0.08% | 907 | 0.12% | 639,697 | 2.23% | 17,577 | 2.37% |
| Utah | 27,734 | 0.88% | 563 | 0.81% | 28,820 | 0.91% | 1,482 | 2.13% | 86,383 | 2.74% | 1,864 | 2.68% |
| Vermont | 1,538 | 0.25% | 12 | 0.28% | 192 | 0.03% | 3 | 0.07% | 15,953 | 2.56% | 128 | 2.99% |
| Virginia | 16,856 | 0.20% | 344 | 0.23% | 4,680 | 0.05% | 98 | 0.07% | 312,036 | 3.67% | 6,576 | 4.42% |
| Washington | 75,677 | 1.01% | 1,614 | 0.60% | 49,219 | 0.66% | 3,291 | 1.22% | 411,840 | 5.48% | 16,822 | 6.23% |
| West Virginia | 3,072 | 0.17% | 8 | 0.10% | 453 | 0.03% | 0 | 0.00% | 43,460 | 2.40% | 246 | 2.99% |
| Wisconsin | 43,830 | 0.75% | 250 | 0.36% | 2,174 | 0.04% | 5 | 0.01% | 149,096 | 2.57% | 2,223 | 3.16% |
| Wyoming | 11,596 | 1.99% | 62 | 0.43% | 482 | 0.08% | 14 | 0.09% | 15,142 | 2.60% | 400 | 2.75% |
| **US nationwide** | **2,075,852** | **0.64%** | **31,887** | **0.41%** | **550,080** | **0.17%** | **18,750** | **0.24%** | **10,152,146** | **3.11%** | **250,668** | **3.19%** |
| **Rail** | | | | | | | | | | | | |
| Alabama | 21,297 | 0.44% | 143 | 0.30% | 1,887 | 0.04% | 13 | 0.03% | 104,637 | 2.14% | 1,007 | 2.12% |
| Alaska | 103,843 | 14.09% | 11 | 5.32% | 10,306 | 1.40% | 3 | 1.35% | 61,228 | 8.31% | 19 | 9.02% |
| Arizona | 272,294 | 3.80% | 1,066 | 4.03% | 13,323 | 0.19% | 43 | 0.16% | 205,014 | 2.86% | 915 | 3.46% |
| Arkansas | 15,912 | 0.53% | 182 | 0.41% | 10,258 | 0.34% | 89 | 0.20% | 100,101 | 3.32% | 1,517 | 3.46% |
| California | 131,724 | 0.33% | 1,883 | 0.29% | 135,524 | 0.34% | 2,778 | 0.43% | 1,446,347 | 3.68% | 21,145 | 3.30% |
| Colorado | 29,337 | 0.52% | 189 | 0.52% | 7,794 | 0.14% | 89 | 0.24% | 176,764 | 3.11% | 1,044 | 2.85% |
| Connecticut | 4,882 | 0.14% | 52 | 0.12% | 742 | 0.02% | 11 | 0.02% | 105,948 | 2.97% | 1,363 | 3.04% |
| Delaware | 2,836 | 0.29% | 28 | 0.29% | 290 | 0.03% | 4 | 0.04% | 31,592 | 3.26% | 301 | 3.05% |
| District of Columbia | 1,298 | 0.18% | 47 | 0.25% | 283 | 0.04% | 12 | 0.06% | 23,612 | 3.36% | 551 | 2.94% |
| Florida | 39,070 | 0.18% | 166 | 0.13% | 10,889 | 0.05% | 28 | 0.02% | 556,333 | 2.62% | 2,612 | 2.10% |
| Georgia | 17,433 | 0.17% | 156 | 0.16% | 5,441 | 0.05% | 47 | 0.05% | 296,305 | 2.82% | 2,929 | 3.07% |
| Hawaii | NA | NA | NA | NA | NA | NA | NA | NA | NA | NA | NA | NA |
| Idaho | 18,866 | 1.08% | 97 | 0.94% | 2,611 | 0.15% | 31 | 0.30% | 48,233 | 2.75% | 314 | 3.05% |
| Illinois | 13,301 | 0.10% | 560 | 0.09% | 3,305 | 0.03% | 142 | 0.02% | 302,360 | 2.38% | 14,295 | 2.31% |
| Indiana | 10,363 | 0.15% | 284 | 0.18% | 1,790 | 0.03% | 48 | 0.03% | 176,419 | 2.63% | 4,614 | 2.96% |
| Iowa | 8,055 | 0.26% | 247 | 0.45% | 3,644 | 0.12% | 122 | 0.22% | 72,774 | 2.31% | 1,501 | 2.71% |
| Kansas | 17,788 | 0.61% | 487 | 0.64% | 1,889 | 0.06% | 29 | 0.04% | 101,821 | 3.50% | 2,991 | 3.94% |
| Kentucky | 7,075 | 0.16% | 124 | 0.21% | 2,691 | 0.06% | 49 | 0.08% | 108,927 | 2.44% | 2,139 | 3.54% |
| Louisiana | 23,328 | 0.50% | 171 | 0.30% | 1,527 | 0.03% | 32 | 0.06% | 106,704 | 2.29% | 1,221 | 2.15% |
| Maine | 8,011 | 0.60% | 18 | 0.46% | 159 | 0.01% | 0 | 0.01% | 34,481 | 2.57% | 128 | 3.19% |
| Maryland | 11,223 | 0.19% | 108 | 0.18% | 1,954 | 0.03% | 19 | 0.03% | 220,806 | 3.66% | 2,364 | 3.87% |
| Massachusetts | 8,943 | 0.13% | 212 | 0.14% | 2,327 | 0.03% | 60 | 0.04% | 239,452 | 3.48% | 5,551 | 3.74% |
| Michigan | 42,931 | 0.43% | 138 | 0.34% | 2,675 | 0.03% | 31 | 0.08% | 321,148 | 3.22% | 1,635 | 4.04% |
| Minnesota | 48,409 | 0.86% | 703 | 0.76% | 1,905 | 0.03% | 39 | 0.04% | 192,047 | 3.43% | 3,675 | 3.98% |
| Mississippi | 12,897 | 0.43% | 44 | 0.16% | 1,090 | 0.04% | 5 | 0.02% | 44,318 | 1.49% | 395 | 1.40% |
| Missouri | 20,646 | 0.34% | 349 | 0.39% | 7,876 | 0.13% | 117 | 0.13% | 191,002 | 3.12% | 3,035 | 3.36% |
| Montana | 62,720 | 5.91% | 693 | 5.92% | 552 | 0.05% | 10 | 0.08% | 34,692 | 3.27% | 418 | 3.57% |
| Nebraska | 13,378 | 0.70% | 356 | 0.57% | 1,051 | 0.05% | 40 | 0.06% | 50,035 | 2.60% | 1,388 | 2.21% |
| Nevada | 25,928 | 0.86% | 73 | 1.34% | 19,401 | 0.64% | 23 | 0.43% | 130,347 | 4.30% | 215 | 3.94% |
| New Hampshire | 1,895 | 0.14% | 1 | 0.05% | 320 | 0.02% | 0 | 0.01% | 31,467 | 2.32% | 82 | 2.76% |
| New Jersey | 10,290 | 0.12% | 260 | 0.10% | 2,122 | 0.02% | 71 | 0.03% | 226,419 | 2.55% | 7,531 | 2.78% |
| New Mexico | 179,881 | 8.58% | 1,034 | 9.22% | 1,408 | 0.07% | 7 | 0.06% | 45,502 | 2.17% | 291 | 2.60% |
| New York | 42,429 | 0.22% | 1,254 | 0.19% | 5,537 | 0.03% | 165 | 0.03% | 585,124 | 3.00% | 21,214 | 3.25% |
| North Carolina | 108,223 | 1.04% | 287 | 0.97% | 6,249 | 0.06% | 22 | 0.07% | 288,649 | 2.78% | 872 | 2.94% |
| North Dakota | 37,170 | 4.89% | 334 | 2.58% | 983 | 0.13% | 17 | 0.13% | 20,342 | 2.68% | 446 | 3.44% |
| Ohio | 14,901 | 0.13% | 411 | 0.17% | 3,367 | 0.03% | 61 | 0.03% | 366,685 | 3.14% | 9,144 | 3.82% |
| Oklahoma | 288,801 | 7.31% | 4,032 | 7.38% | 5,820 | 0.15% | 418 | 0.76% | 292,143 | 7.40% | 4,217 | 7.72% |
| Oregon | 35,842 | 0.86% | 413 | 0.73% | 15,614 | 0.37% | 317 | 0.56% | 183,478 | 4.39% | 2,790 | 4.91% |
| Pennsylvania | 12,224 | 0.10% | 558 | 0.14% | 2,968 | 0.02% | 70 | 0.02% | 324,708 | 2.54% | 12,466 | 3.18% |
| Rhode Island | 2,832 | 0.27% | 29 | 0.32% | 648 | 0.06% | 4 | 0.05% | 36,067 | 3.41% | 547 | 5.98% |
| South Carolina | 14,085 | 0.28% | 71 | 0.24% | 3,384 | 0.07% | 21 | 0.07% | 127,114 | 2.50% | 815 | 2.76% |
| South Dakota | 72,142 | 8.20% | 360 | 7.89% | 494 | 0.06% | 13 | 0.28% | 24,573 | 2.79% | 175 | 3.83% |
| Tennessee | 13,814 | 0.20% | 66 | 0.20% | 3,439 | 0.05% | 14 | 0.04% | 168,074 | 2.48% | 833 | 2.54% |
| Texas | 65,132 | 0.23% | 949 | 0.21% | 21,477 | 0.08% | 362 | 0.08% | 639,697 | 2.23% | 8,431 | 1.85% |
| Utah | 27,734 | 0.88% | 256 | 0.65% | 28,820 | 0.91% | 524 | 1.33% | 86,383 | 2.74% | 1,348 | 3.43% |
| Vermont | 1,538 | 0.25% | 1 | 0.12% | 192 | 0.03% | 0 | 0.01% | 15,953 | 2.56% | 33 | 2.64% |
| Virginia | 16,856 | 0.20% | 204 | 0.19% | 4,680 | 0.05% | 58 | 0.05% | 312,036 | 3.67% | 4,266 | 3.99% |
| Washington | 75,677 | 1.01% | 663 | 0.98% | 49,219 | 0.66% | 462 | 0.69% | 411,840 | 5.48% | 3,614 | 5.37% |
| West Virginia | 3,072 | 0.17% | 21 | 0.20% | 453 | 0.03% | 6 | 0.06% | 43,460 | 2.40% | 370 | 3.50% |
| Wisconsin | 43,830 | 0.75% | 284 | 0.49% | 2,174 | 0.04% | 18 | 0.03% | 149,096 | 2.57% | 1,532 | 2.65% |
| Wyoming | 11,596 | 1.99% | 63 | 0.87% | 482 | 0.08% | 8 | 0.12% | 15,142 | 2.60% | 214 | 2.97% |
| **US nationwide** | **2,073,752** | **0.64%** | **20,138** | **0.39%** | **413,034** | **0.13%** | **6,552** | **0.13%** | **9,877,399** | **3.04%** | **160,513** | **3.08%** |
| **Road** | | | | | | | | | | | | |
| Alabama | 21,297 | 0.44% | 221 | 0.29% | 1,887 | 0.04% | 24 | 0.03% | 104,637 | 2.14% | 1,652 | 2.17% |
| Alaska | 103,843 | 14.09% | 843 | 9.06% | 10,306 | 1.40% | 263 | 2.83% | 61,228 | 8.31% | 921 | 9.90% |
| Arizona | 272,294 | 3.80% | 4,082 | 1.93% | 13,323 | 0.19% | 436 | 0.21% | 205,014 | 2.86% | 6,513 | 3.08% |
| Arkansas | 15,912 | 0.53% | 186 | 0.51% | 10,258 | 0.34% | 175 | 0.48% | 100,101 | 3.32% | 1,328 | 3.64% |
| California | 131,724 | 0.33% | 4,290 | 0.25% | 135,524 | 0.34% | 6,255 | 0.37% | 1,446,347 | 3.68% | 61,415 | 3.61% |
| Colorado | 29,337 | 0.52% | 550 | 0.45% | 7,794 | 0.14% | 180 | 0.15% | 176,764 | 3.11% | 3,949 | 3.26% |
| Connecticut | 4,882 | 0.14% | 116 | 0.15% | 742 | 0.02% | 23 | 0.03% | 105,948 | 2.97% | 2,374 | 3.11% |
| Delaware | 2,836 | 0.29% | 48 | 0.27% | 290 | 0.03% | 8 | 0.04% | 31,592 | 3.26% | 561 | 3.11% |
| District of Columbia | 1,298 | 0.18% | 77 | 0.20% | 283 | 0.04% | 22 | 0.06% | 23,612 | 3.36% | 1,267 | 3.33% |
| Florida | 39,070 | 0.18% | 727 | 0.16% | 10,889 | 0.05% | 237 | 0.05% | 556,333 | 2.62% | 11,802 | 2.54% |
| Georgia | 17,433 | 0.17% | 244 | 0.16% | 5,441 | 0.05% | 69 | 0.05% | 296,305 | 2.82% | 4,526 | 3.02% |
| Hawaii | 2,100 | 0.15% | 40 | 0.10% | 137,046 | 9.65% | 3,547 | 8.85% | 274,747 | 19.35% | 6,838 | 17.07% |
| Idaho | 18,866 | 1.08% | 229 | 0.83% | 2,611 | 0.15% | 37 | 0.13% | 48,233 | 2.75% | 860 | 3.11% |
| Illinois | 13,301 | 0.10% | 355 | 0.11% | 3,305 | 0.03% | 70 | 0.02% | 302,360 | 2.38% | 8,470 | 2.56% |
| Indiana | 10,363 | 0.15% | 238 | 0.19% | 1,790 | 0.03% | 43 | 0.03% | 176,419 | 2.63% | 3,862 | 3.02% |
| Iowa | 8,055 | 0.26% | 173 | 0.33% | 3,644 | 0.12% | 82 | 0.15% | 72,774 | 2.31% | 1,565 | 2.95% |
| Kansas | 17,788 | 0.61% | 220 | 0.53% | 1,889 | 0.06% | 25 | 0.06% | 101,821 | 3.50% | 1,638 | 3.92% |
| Kentucky | 7,075 | 0.16% | 137 | 0.16% | 2,691 | 0.06% | 62 | 0.07% | 108,927 | 2.44% | 2,713 | 3.14% |
| Louisiana | 23,328 | 0.50% | 237 | 0.36% | 1,527 | 0.03% | 18 | 0.03% | 106,704 | 2.29% | 1,570 | 2.36% |
| Maine | 8,011 | 0.60% | 67 | 0.43% | 159 | 0.01% | 2 | 0.01% | 34,481 | 2.57% | 501 | 3.21% |
| Maryland | 11,223 | 0.19% | 268 | 0.20% | 1,954 | 0.03% | 48 | 0.04% | 220,806 | 3.66% | 5,026 | 3.67% |
| Massachusetts | 8,943 | 0.13% | 350 | 0.15% | 2,327 | 0.03% | 87 | 0.04% | 239,452 | 3.48% | 8,911 | 3.69% |
| Michigan | 42,931 | 0.43% | 706 | 0.32% | 2,675 | 0.03% | 67 | 0.03% | 321,148 | 3.22% | 7,672 | 3.50% |
| Minnesota | 48,409 | 0.86% | 643 | 0.69% | 1,905 | 0.03% | 33 | 0.04% | 192,047 | 3.43% | 3,870 | 4.13% |
| Mississippi | 12,897 | 0.43% | 74 | 0.24% | 1,090 | 0.04% | 12 | 0.04% | 44,318 | 1.49% | 504 | 1.64% |
| Missouri | 20,646 | 0.34% | 278 | 0.26% | 7,876 | 0.13% | 135 | 0.13% | 191,002 | 3.12% | 3,627 | 3.44% |
| Montana | 62,720 | 5.91% | 458 | 3.41% | 552 | 0.05% | 9 | 0.07% | 34,692 | 3.27% | 493 | 3.67% |
| Nebraska | 13,378 | 0.70% | 162 | 0.50% | 1,051 | 0.05% | 16 | 0.05% | 50,035 | 2.60% | 993 | 3.06% |
| Nevada | 25,928 | 0.86% | 432 | 0.69% | 19,401 | 0.64% | 422 | 0.67% | 130,347 | 4.30% | 2,636 | 4.20% |
| New Hampshire | 1,895 | 0.14% | 24 | 0.13% | 320 | 0.02% | 5 | 0.03% | 31,467 | 2.32% | 544 | 2.84% |
| New Jersey | 10,290 | 0.12% | 266 | 0.11% | 2,122 | 0.02% | 61 | 0.03% | 226,419 | 2.55% | 6,054 | 2.55% |
| New Mexico | 179,881 | 8.58% | 1,886 | 4.96% | 1,408 | 0.07% | 28 | 0.07% | 45,502 | 2.17% | 953 | 2.50% |
| New York | 42,429 | 0.22% | 1,356 | 0.19% | 5,537 | 0.03% | 194 | 0.03% | 585,124 | 3.00% | 23,574 | 3.22% |
| North Carolina | 108,223 | 1.04% | 1,013 | 0.63% | 6,249 | 0.06% | 89 | 0.06% | 288,649 | 2.78% | 4,839 | 3.02% |
| North Dakota | 37,170 | 4.89% | 313 | 2.91% | 983 | 0.13% | 18 | 0.16% | 20,342 | 2.68% | 323 | 3.00% |
| Ohio | 14,901 | 0.13% | 275 | 0.14% | 3,367 | 0.03% | 66 | 0.03% | 366,685 | 3.14% | 7,518 | 3.72% |
| Oklahoma | 288,801 | 7.31% | 2,922 | 5.50% | 5,820 | 0.15% | 109 | 0.20% | 292,143 | 7.40% | 3,795 | 7.14% |
| Oregon | 35,842 | 0.86% | 792 | 0.68% | 15,614 | 0.37% | 648 | 0.56% | 183,478 | 4.39% | 5,470 | 4.72% |
| Pennsylvania | 12,224 | 0.10% | 323 | 0.13% | 2,968 | 0.02% | 64 | 0.02% | 324,708 | 2.54% | 7,416 | 2.88% |
| Rhode Island | 2,832 | 0.27% | 92 | 0.32% | 648 | 0.06% | 21 | 0.07% | 36,067 | 3.41% | 1,165 | 4.06% |
| South Carolina | 14,085 | 0.28% | 148 | 0.23% | 3,384 | 0.07% | 72 | 0.11% | 127,114 | 2.50% | 1,710 | 2.61% |
| South Dakota | 72,142 | 8.20% | 557 | 4.34% | 494 | 0.06% | 10 | 0.08% | 24,573 | 2.79% | 425 | 3.31% |
| Tennessee | 13,814 | 0.20% | 199 | 0.17% | 3,439 | 0.05% | 63 | 0.06% | 168,074 | 2.48% | 3,094 | 2.71% |
| Texas | 65,132 | 0.23% | 1,353 | 0.20% | 21,477 | 0.08% | 548 | 0.08% | 639,697 | 2.23% | 15,065 | 2.26% |
| Utah | 27,734 | 0.88% | 599 | 0.73% | 28,820 | 0.91% | 984 | 1.19% | 86,383 | 2.74% | 2,491 | 3.03% |
| Vermont | 1,538 | 0.25% | 10 | 0.16% | 192 | 0.03% | 3 | 0.04% | 15,953 | 2.56% | 171 | 2.88% |
| Virginia | 16,856 | 0.20% | 331 | 0.19% | 4,680 | 0.05% | 106 | 0.06% | 312,036 | 3.67% | 7,096 | 4.09% |
| Washington | 75,677 | 1.01% | 1,308 | 0.74% | 49,219 | 0.66% | 1,430 | 0.81% | 411,840 | 5.48% | 10,550 | 5.98% |
| West Virginia | 3,072 | 0.17% | 38 | 0.18% | 453 | 0.03% | 10 | 0.05% | 43,460 | 2.40% | 626 | 2.99% |
| Wisconsin | 43,830 | 0.75% | 624 | 0.53% | 2,174 | 0.04% | 54 | 0.05% | 149,096 | 2.57% | 3,689 | 3.12% |
| Wyoming | 11,596 | 1.99% | 102 | 1.34% | 482 | 0.08% | 5 | 0.07% | 15,142 | 2.60% | 216 | 2.83% |
| **US nationwide** | **2,075,852** | **0.64%** | **30,982** | **0.39%** | **550,080** | **0.17%** | **16,995** | **0.21%** | **10,152,146** | **3.11%** | **264,841** | **3.34%** |

Definition of abbreviation: Pop: population.

**Table S7. Population numbers and proportions highly annoyed by different transportation noise sources [a] for the top urban areas.**

| Urban Area | Total | | | Asian | | | | Black | | | |
|---|---|---|---|---|---|---|---|---|---|---|---|
| | Pop | #HA | % | Pop | % | #HA | %HA | Pop | % | #HA | %HA |
| **Aviation** | | | | | | | | | | | |
| Atlanta, GA | 5,522,443 | 134,170 | 2.43% | 357,399 | 6.47% | 4,231 | 3.15% | 1,928,968 | 34.93% | 78,126 | 58.23% |
| Baltimore, MD | 2,406,743 | 77,763 | 3.23% | 154,219 | 6.41% | 7,353 | 9.46% | 809,704 | 33.64% | 19,828 | 25.50% |
| Boston, MA--NH--RI | 4,781,054 | 139,535 | 2.92% | 393,538 | 8.23% | 10,936 | 7.84% | 366,365 | 7.66% | 13,709 | 9.82% |
| Chicago, IL--IN | 8,797,240 | 643,606 | 7.32% | 620,108 | 7.05% | 53,034 | 8.24% | 1,504,768 | 17.11% | 64,252 | 9.98% |
| Dallas--Fort Worth--Arlington, TX | 6,219,214 | 263,820 | 4.24% | 470,804 | 7.57% | 18,662 | 7.07% | 1,069,502 | 17.20% | 46,121 | 17.48% |
| Denver--Aurora, CO | 2,809,068 | 61,975 | 2.21% | 124,465 | 4.43% | 5,258 | 8.48% | 159,965 | 5.69% | 2,448 | 3.95% |
| Detroit, MI | 3,908,573 | 50,687 | 1.30% | 193,059 | 4.94% | 1,195 | 2.36% | 931,422 | 23.83% | 14,164 | 27.94% |
| Houston, TX | 6,040,433 | 187,732 | 3.11% | 524,920 | 8.69% | 14,777 | 7.87% | 1,078,686 | 17.86% | 37,062 | 19.74% |
| Las Vegas--Henderson, NV | 2,167,403 | 126,699 | 5.85% | 215,148 | 9.93% | 11,034 | 8.71% | 252,780 | 11.66% | 18,442 | 14.56% |
| Los Angeles--Long Beach--Anaheim, CA | 12,692,809 | 627,470 | 4.94% | 2,124,208 | 16.74% | 47,214 | 7.52% | 800,480 | 6.31% | 69,225 | 11.03% |
| Miami, FL | 6,087,385 | 356,670 | 5.86% | 150,433 | 2.47% | 6,709 | 1.88% | 1,204,856 | 19.79% | 61,807 | 17.33% |
| Minneapolis--St. Paul, MN--WI | 3,025,526 | 106,418 | 3.52% | 234,255 | 7.74% | 6,708 | 6.30% | 297,269 | 9.83% | 9,099 | 8.55% |
| New York--Newark, NY--NJ--CT | 19,059,660 | 944,749 | 4.96% | 2,194,397 | 11.51% | 110,473 | 11.69% | 2,994,531 | 15.71% | 189,200 | 20.03% |

| | | | | | | | | | | |
|---|---|---|---|---|---|---|---|---|---|---|
| Philadelphia, PA--NJ--DE--MD | 5,861,646 | 55,272 | 0.94% | 361,801 | 6.17% | 2,363 | 4.27% | 1,195,277 | 20.39% | 14,822 | 26.82% |
| Phoenix--Mesa, AZ | 4,319,365 | 177,909 | 4.12% | 178,821 | 4.14% | 10,702 | 6.02% | 225,687 | 5.23% | 9,568 | 5.38% |
| Portland, OR--WA | 2,176,518 | 60,296 | 2.77% | 164,780 | 7.57% | 5,047 | 8.37% | 65,958 | 3.03% | 2,794 | 4.63% |
| Riverside--San Bernardino, CA | 2,205,194 | 29,113 | 1.32% | 156,217 | 7.08% | 5,323 | 18.28% | 174,626 | 7.92% | 2,131 | 7.32% |
| San Antonio, TX | 2,190,236 | 40,202 | 1.84% | 61,274 | 2.80% | 745 | 1.85% | 151,595 | 6.92% | 2,108 | 5.24% |
| San Diego, CA | 3,242,514 | 228,352 | 7.04% | 387,063 | 11.94% | 23,767 | 10.41% | 152,123 | 4.69% | 13,463 | 5.90% |
| San Francisco--Oakland, CA | 3,633,940 | 123,726 | 3.40% | 1,050,708 | 28.91% | 46,788 | 37.82% | 265,149 | 7.30% | 6,670 | 5.39% |
| Seattle, WA | 3,637,300 | 228,946 | 6.29% | 542,287 | 14.91% | 42,192 | 18.43% | 224,175 | 6.16% | 27,837 | 12.16% |
| St. Louis, MO--IL | 2,287,681 | 53,570 | 2.34% | 71,723 | 3.14% | 983 | 1.84% | 490,046 | 21.42% | 30,619 | 57.16% |
| Tampa--St. Petersburg, FL | 2,875,484 | 49,523 | 1.72% | 105,419 | 3.67% | 2,500 | 5.05% | 346,907 | 12.06% | 5,706 | 11.52% |
| Washington, DC--VA--MD | 5,292,177 | 118,109 | 2.23% | 618,270 | 11.68% | 19,472 | 16.49% | 1,348,602 | 25.48% | 12,382 | 10.48% |
| **Rail** | | | | | | | | | | | |
| Atlanta, GA | 5,522,443 | 54,856 | 0.99% | 357,399 | 6.47% | 3,838 | 7.00% | 1,928,968 | 34.93% | 22,221 | 40.51% |
| Baltimore, MD | 2,406,743 | 28,881 | 1.20% | 154,219 | 6.41% | 1,244 | 4.31% | 809,704 | 33.64% | 13,671 | 47.34% |
| Boston, MA--NH--RI | 4,781,054 | 134,670 | 2.82% | 393,538 | 8.23% | 14,817 | 11.00% | 366,365 | 7.66% | 17,242 | 12.80% |
| Chicago, IL--IN | 8,797,240 | 552,695 | 6.28% | 620,108 | 7.05% | 32,458 | 5.87% | 1,504,768 | 17.11% | 149,253 | 27.00% |
| Dallas--Fort Worth--Arlington, TX | 6,219,214 | 113,786 | 1.83% | 470,804 | 7.57% | 5,449 | 4.79% | 1,069,502 | 17.20% | 18,183 | 15.98% |
| Denver--Aurora, CO | 2,809,068 | 22,409 | 0.80% | 124,465 | 4.43% | 702 | 3.13% | 159,965 | 5.69% | 1,174 | 5.24% |
| Detroit, MI | 3,908,573 | 19,485 | 0.50% | 193,059 | 4.94% | 504 | 2.58% | 931,422 | 23.83% | 4,325 | 22.19% |
| Houston, TX | 6,040,433 | 106,996 | 1.77% | 524,920 | 8.69% | 4,989 | 4.66% | 1,078,686 | 17.86% | 20,404 | 19.07% |
| Las Vegas--Henderson, NV | 2,167,403 | 3,227 | 0.15% | 215,148 | 9.93% | 391 | 12.13% | 252,780 | 11.66% | 588 | 18.22% |
| Los Angeles--Long Beach--Anaheim, CA | 12,692,809 | 163,912 | 1.29% | 2,124,208 | 16.74% | 18,941 | 11.56% | 800,480 | 6.31% | 11,328 | 6.91% |
| Miami, FL | 6,087,385 | 84,496 | 1.39% | 150,433 | 2.47% | 1,166 | 1.38% | 1,204,856 | 19.79% | 26,562 | 31.44% |
| Minneapolis--St. Paul, MN--WI | 3,025,526 | 57,937 | 1.91% | 234,255 | 7.74% | 6,080 | 10.49% | 297,269 | 9.83% | 8,506 | 14.68% |
| New York--Newark, NY--NJ--CT | 19,059,660 | 822,455 | 4.32% | 2,194,397 | 11.51% | 100,729 | 12.25% | 2,994,531 | 15.71% | 159,814 | 19.43% |
| Philadelphia, PA--NJ--DE--MD | 5,861,646 | 224,696 | 3.83% | 361,801 | 6.17% | 14,318 | 6.37% | 1,195,277 | 20.39% | 80,117 | 35.66% |
| Phoenix--Mesa, AZ | 4,319,365 | 10,085 | 0.23% | 178,821 | 4.14% | 492 | 4.88% | 225,687 | 5.23% | 731 | 7.25% |
| Portland, OR--WA | 2,176,518 | 41,249 | 1.90% | 164,780 | 7.57% | 2,714 | 6.58% | 65,958 | 3.03% | 2,140 | 5.19% |
| Riverside--San Bernardino, CA | 2,205,194 | 42,925 | 1.95% | 156,217 | 7.08% | 2,122 | 4.94% | 174,626 | 7.92% | 2,447 | 5.70% |
| San Antonio, TX | 2,190,236 | 42,907 | 1.96% | 61,274 | 2.80% | 435 | 1.01% | 151,595 | 6.92% | 2,968 | 6.92% |
| San Diego, CA | 3,242,514 | 22,447 | 0.69% | 387,063 | 11.94% | 1,552 | 6.91% | 152,123 | 4.69% | 1,104 | 4.92% |
| San Francisco--Oakland, CA | 3,633,940 | 111,667 | 3.07% | 1,050,708 | 28.91% | 30,274 | 27.11% | 265,149 | 7.30% | 12,561 | 11.25% |
| Seattle, WA | 3,637,300 | 34,366 | 0.94% | 542,287 | 14.91% | 4,653 | 13.54% | 224,175 | 6.16% | 2,650 | 7.71% |
| St. Louis, MO--IL | 2,287,681 | 39,012 | 1.71% | 71,723 | 3.14% | 834 | 2.14% | 490,046 | 21.42% | 14,732 | 37.76% |
| Tampa--St. Petersburg, FL | 2,875,484 | 2,916 | 0.10% | 105,419 | 3.67% | 91 | 3.13% | 346,907 | 12.06% | 559 | 19.17% |
| Washington, DC--VA--MD | 5,292,177 | 74,155 | 1.40% | 618,270 | 11.68% | 7,039 | 9.49% | 1,348,602 | 25.48% | 22,194 | 29.93% |
| **Road** | | | | | | | | | | | |
| Atlanta, GA | 5,522,443 | 97,511 | 1.77% | 357,399 | 6.47% | 7,167 | 7.35% | 1,928,968 | 34.93% | 38,990 | 39.99% |
| Baltimore, MD | 2,406,743 | 68,373 | 2.84% | 154,219 | 6.41% | 4,541 | 6.64% | 809,704 | 33.64% | 26,235 | 38.37% |
| Boston, MA--NH--RI | 4,781,054 | 186,819 | 3.91% | 393,538 | 8.23% | 19,758 | 10.58% | 366,365 | 7.66% | 16,387 | 8.77% |
| Chicago, IL--IN | 8,797,240 | 276,238 | 3.14% | 620,108 | 7.05% | 24,425 | 8.84% | 1,504,768 | 17.11% | 50,832 | 18.40% |
| Dallas--Fort Worth--Arlington, TX | 6,219,214 | 179,846 | 2.89% | 470,804 | 7.57% | 13,907 | 7.73% | 1,069,502 | 17.20% | 34,730 | 19.31% |
| Denver--Aurora, CO | 2,809,068 | 75,599 | 2.69% | 124,465 | 4.43% | 3,276 | 4.33% | 159,965 | 5.69% | 4,585 | 6.06% |
| Detroit, MI | 3,908,573 | 117,370 | 3.00% | 193,059 | 4.94% | 5,824 | 4.96% | 931,422 | 23.83% | 34,305 | 29.23% |
| Houston, TX | 6,040,433 | 161,813 | 2.68% | 524,920 | 8.69% | 13,520 | 8.36% | 1,078,686 | 17.86% | 30,033 | 18.56% |
| Las Vegas--Henderson, NV | 2,167,403 | 46,805 | 2.16% | 215,148 | 9.93% | 4,190 | 8.95% | 252,780 | 11.66% | 5,635 | 12.04% |
| Los Angeles--Long Beach--Anaheim, CA | 12,692,809 | 751,370 | 5.92% | 2,124,208 | 16.74% | 118,454 | 15.77% | 800,480 | 6.31% | 50,912 | 6.78% |
| Miami, FL | 6,087,385 | 198,775 | 3.27% | 150,433 | 2.47% | 4,535 | 2.28% | 1,204,856 | 19.79% | 42,678 | 21.47% |

| | | | | | | | | | | |
|---|---|---|---|---|---|---|---|---|---|---|
| Minneapolis--St. Paul, MN--WI | 3,025,526 | 71,015 | 2.35% | 234,255 | 7.74% | 6,136 | 8.64% | 297,269 | 9.83% | 9,810 | 13.81% |
| New York--Newark, NY--NJ--CT | 19,059,660 | 811,952 | 4.26% | 2,194,397 | 11.51% | 102,940 | 12.68% | 2,994,531 | 15.71% | 139,349 | 17.16% |
| Philadelphia, PA--NJ--DE--MD | 5,861,646 | 153,134 | 2.61% | 361,801 | 6.17% | 11,016 | 7.19% | 1,195,277 | 20.39% | 35,401 | 23.12% |
| Phoenix--Mesa, AZ | 4,319,365 | 160,008 | 3.70% | 178,821 | 4.14% | 6,873 | 4.30% | 225,687 | 5.23% | 9,431 | 5.89% |
| Portland, OR--WA | 2,176,518 | 76,315 | 3.51% | 164,780 | 7.57% | 5,756 | 7.54% | 65,958 | 3.03% | 2,854 | 3.74% |
| Riverside--San Bernardino, CA | 2,205,194 | 70,347 | 3.19% | 156,217 | 7.08% | 4,448 | 6.32% | 174,626 | 7.92% | 5,745 | 8.17% |
| San Antonio, TX | 2,190,236 | 57,378 | 2.62% | 61,274 | 2.80% | 1,568 | 2.73% | 151,595 | 6.92% | 4,073 | 7.10% |
| San Diego, CA | 3,242,514 | 128,496 | 3.96% | 387,063 | 11.94% | 14,866 | 11.57% | 152,123 | 4.69% | 6,830 | 5.32% |
| San Francisco--Oakland, CA | 3,633,940 | 198,279 | 5.46% | 1,050,708 | 28.91% | 58,674 | 29.59% | 265,149 | 7.30% | 15,993 | 8.07% |
| Seattle, WA | 3,637,300 | 110,701 | 3.04% | 542,287 | 14.91% | 19,144 | 17.29% | 224,175 | 6.16% | 7,958 | 7.19% |
| St. Louis, MO--IL | 2,287,681 | 55,978 | 2.45% | 71,723 | 3.14% | 1,955 | 3.49% | 490,046 | 21.42% | 13,465 | 24.05% |
| Tampa--St. Petersburg, FL | 2,875,484 | 60,109 | 2.09% | 105,419 | 3.67% | 2,225 | 3.70% | 346,907 | 12.06% | 8,647 | 14.39% |
| Washington, DC--VA--MD | 5,292,177 | 164,281 | 3.10% | 618,270 | 11.68% | 19,779 | 12.04% | 1,348,602 | 25.48% | 42,722 | 26.01% |

Definition of abbreviation: Pop: population.

[a] Greater than or equal to 54 dB Lden.

**Table S8. Table S7 continued.**

| Urban Area | Hispanic | | | | White | | | | AIAN | | | |
|---|---|---|---|---|---|---|---|---|---|---|---|---|
| | Pop | % | #HA | %HA | Pop | % | #HA | %HA | Pop | % | #HA | %HA |
| **Aviation** | | | | | | | | | | | | |
| Atlanta, GA | 614,403 | 11.13% | 19,148 | 14.27% | 2,439,901 | 44.18% | 28,753 | 21.43% | 8,310 | 0.15% | 314 | 0.23% |
| Baltimore, MD | 158,559 | 6.59% | 6,664 | 8.57% | 1,189,788 | 49.44% | 39,982 | 51.41% | 4,620 | 0.19% | 121 | 0.16% |
| Boston, MA--NH--RI | 555,601 | 11.62% | 39,858 | 28.57% | 3,291,310 | 68.84% | 70,671 | 50.65% | 5,309 | 0.11% | 216 | 0.15% |
| Chicago, IL--IN | 2,007,048 | 22.81% | 205,246 | 31.89% | 4,447,686 | 50.56% | 307,017 | 47.70% | 8,087 | 0.09% | 664 | 0.10% |
| Dallas--Fort Worth--Arlington, TX | 1,914,761 | 30.79% | 94,670 | 35.88% | 2,577,740 | 41.45% | 96,683 | 36.65% | 14,624 | 0.24% | 647 | 0.25% |
| Denver--Aurora, CO | 669,986 | 23.85% | 10,859 | 17.52% | 1,751,784 | 62.36% | 40,968 | 66.10% | 10,658 | 0.38% | 207 | 0.33% |
| Detroit, MI | 183,734 | 4.70% | 2,339 | 4.61% | 2,468,701 | 63.16% | 31,430 | 62.01% | 9,357 | 0.24% | 146 | 0.29% |
| Houston, TX | 2,359,738 | 39.07% | 86,195 | 45.91% | 1,925,596 | 31.88% | 45,290 | 24.12% | 10,182 | 0.17% | 237 | 0.13% |
| Las Vegas--Henderson, NV | 686,583 | 31.68% | 49,996 | 39.46% | 886,192 | 40.89% | 40,103 | 31.65% | 9,626 | 0.44% | 589 | 0.47% |
| Los Angeles--Long Beach--Anaheim, CA | 5,888,756 | 46.39% | 339,096 | 54.04% | 3,443,866 | 27.13% | 151,835 | 24.20% | 23,341 | 0.18% | 1,155 | 0.18% |
| Miami, FL | 2,752,279 | 45.21% | 213,715 | 59.92% | 1,847,912 | 30.36% | 69,096 | 19.37% | 6,454 | 0.11% | 280 | 0.08% |
| Minneapolis--St. Paul, MN--WI | 197,076 | 6.51% | 9,353 | 8.79% | 2,157,705 | 71.32% | 75,727 | 71.16% | 13,086 | 0.43% | 632 | 0.59% |
| New York--Newark, NY--NJ--CT | 4,717,835 | 24.75% | 382,023 | 40.44% | 8,578,756 | 45.01% | 236,639 | 25.05% | 26,652 | 0.14% | 1,430 | 0.15% |
| Philadelphia, PA--NJ--DE--MD | 572,794 | 9.77% | 5,195 | 9.40% | 3,548,830 | 60.54% | 31,063 | 56.20% | 7,044 | 0.12% | 37 | 0.07% |
| Phoenix--Mesa, AZ | 1,305,362 | 30.22% | 49,759 | 27.97% | 2,396,042 | 55.47% | 99,074 | 55.69% | 74,673 | 1.73% | 3,055 | 1.72% |
| Portland, OR--WA | 271,231 | 12.46% | 11,495 | 19.06% | 1,542,869 | 70.89% | 37,168 | 61.64% | 10,601 | 0.49% | 333 | 0.55% |
| Riverside--San Bernardino, CA | 1,271,256 | 57.65% | 14,527 | 49.90% | 533,254 | 24.18% | 5,861 | 20.13% | 6,058 | 0.27% | 27 | 0.09% |
| San Antonio, TX | 1,256,159 | 57.35% | 21,290 | 52.96% | 666,022 | 30.41% | 15,098 | 37.55% | 3,241 | 0.15% | 37 | 0.09% |
| San Diego, CA | 1,101,719 | 33.98% | 77,910 | 34.12% | 1,444,090 | 44.54% | 101,563 | 44.48% | 8,630 | 0.27% | 622 | 0.27% |
| San Francisco--Oakland, CA | 795,635 | 21.89% | 36,339 | 29.37% | 1,298,892 | 35.74% | 26,009 | 21.02% | 8,170 | 0.22% | 256 | 0.21% |
| Seattle, WA | 379,345 | 10.43% | 33,398 | 14.59% | 2,202,497 | 60.55% | 106,054 | 46.32% | 20,992 | 0.58% | 1,202 | 0.52% |
| St. Louis, MO--IL | 77,261 | 3.38% | 1,698 | 3.17% | 1,570,831 | 68.66% | 18,312 | 34.18% | 3,370 | 0.15% | 112 | 0.21% |
| Tampa--St. Petersburg, FL | 590,013 | 20.52% | 17,339 | 35.01% | 1,740,235 | 60.52% | 22,462 | 45.36% | 5,667 | 0.20% | 38 | 0.08% |
| Washington, DC--VA--MD | 911,064 | 17.22% | 17,494 | 14.81% | 2,188,896 | 41.36% | 63,314 | 53.61% | 9,376 | 0.18% | 217 | 0.18% |
| **Rail** | | | | | | | | | | | | |
| Atlanta, GA | 614,403 | 11.13% | 5,376 | 9.80% | 2,439,901 | 44.18% | 21,374 | 38.96% | 8,310 | 0.15% | 91 | 0.17% |
| Baltimore, MD | 158,559 | 6.59% | 2,021 | 7.00% | 1,189,788 | 49.44% | 10,800 | 37.40% | 4,620 | 0.19% | 71 | 0.25% |
| Boston, MA--NH--RI | 555,601 | 11.62% | 23,423 | 17.39% | 3,291,310 | 68.84% | 73,795 | 54.80% | 5,309 | 0.11% | 188 | 0.14% |
| Chicago, IL--IN | 2,007,048 | 22.81% | 146,895 | 26.58% | 4,447,686 | 50.56% | 211,370 | 38.24% | 8,087 | 0.09% | 478 | 0.09% |
| Dallas--Fort Worth--Arlington, TX | 1,914,761 | 30.79% | 49,512 | 43.51% | 2,577,740 | 41.45% | 37,833 | 33.25% | 14,624 | 0.24% | 252 | 0.22% |
| Denver--Aurora, CO | 669,986 | 23.85% | 6,765 | 30.19% | 1,751,784 | 62.36% | 13,045 | 58.21% | 10,658 | 0.38% | 99 | 0.44% |
| Detroit, MI | 183,734 | 4.70% | 1,201 | 6.16% | 2,468,701 | 63.16% | 12,566 | 64.49% | 9,357 | 0.24% | 71 | 0.37% |
| Houston, TX | 2,359,738 | 39.07% | 56,493 | 52.80% | 1,925,596 | 31.88% | 23,447 | 21.91% | 10,182 | 0.17% | 155 | 0.14% |
| Las Vegas--Henderson, NV | 686,583 | 31.68% | 879 | 27.24% | 886,192 | 40.89% | 1,174 | 36.40% | 9,626 | 0.44% | 20 | 0.63% |
| Los Angeles--Long Beach--Anaheim, CA | 5,888,756 | 46.39% | 104,176 | 63.56% | 3,443,866 | 27.13% | 25,600 | 15.62% | 23,341 | 0.18% | 307 | 0.19% |
| Miami, FL | 2,752,279 | 45.21% | 32,852 | 38.88% | 1,847,912 | 30.36% | 22,302 | 26.39% | 6,454 | 0.11% | 80 | 0.09% |
| Minneapolis--St. Paul, MN--WI | 197,076 | 6.51% | 4,209 | 7.26% | 2,157,705 | 71.32% | 35,973 | 62.09% | 13,086 | 0.43% | 424 | 0.73% |
| New York--Newark, NY--NJ--CT | 4,717,835 | 24.75% | 285,322 | 34.69% | 8,578,756 | 45.01% | 249,680 | 30.36% | 26,652 | 0.14% | 1,268 | 0.15% |
| Philadelphia, PA--NJ--DE--MD | 572,794 | 9.77% | 27,027 | 12.03% | 3,548,830 | 60.54% | 95,825 | 42.65% | 7,044 | 0.12% | 391 | 0.17% |
| Phoenix--Mesa, AZ | 1,305,362 | 30.22% | 4,012 | 39.79% | 2,396,042 | 55.47% | 4,302 | 42.66% | 74,673 | 1.73% | 179 | 1.77% |
| Portland, OR--WA | 271,231 | 12.46% | 5,826 | 14.12% | 1,542,869 | 70.89% | 27,979 | 67.83% | 10,601 | 0.49% | 238 | 0.58% |
| Riverside--San Bernardino, CA | 1,271,256 | 57.65% | 29,583 | 68.92% | 533,254 | 24.18% | 7,819 | 18.22% | 6,058 | 0.27% | 145 | 0.34% |
| San Antonio, TX | 1,256,159 | 57.35% | 29,631 | 69.06% | 666,022 | 30.41% | 9,115 | 21.24% | 3,241 | 0.15% | 50 | 0.12% |
| San Diego, CA | 1,101,719 | 33.98% | 7,786 | 34.69% | 1,444,090 | 44.54% | 11,063 | 49.28% | 8,630 | 0.27% | 57 | 0.25% |
| San Francisco--Oakland, CA | 795,635 | 21.89% | 36,430 | 32.62% | 1,298,892 | 35.74% | 25,939 | 23.23% | 8,170 | 0.22% | 315 | 0.28% |
| Seattle, WA | 379,345 | 10.43% | 3,925 | 11.42% | 2,202,497 | 60.55% | 20,234 | 58.88% | 20,992 | 0.58% | 324 | 0.94% |
| St. Louis, MO--IL | 77,261 | 3.38% | 1,259 | 3.23% | 1,570,831 | 68.66% | 20,943 | 53.68% | 3,370 | 0.15% | 67 | 0.17% |
| Tampa--St. Petersburg, FL | 590,013 | 20.52% | 838 | 28.75% | 1,740,235 | 60.52% | 1,335 | 45.79% | 5,667 | 0.20% | 7 | 0.24% |
| Washington, DC--VA--MD | 911,064 | 17.22% | 12,672 | 17.09% | 2,188,896 | 41.36% | 29,055 | 39.18% | 9,376 | 0.18% | 142 | 0.19% |
| **Road** | | | | | | | | | | | | |
| Atlanta, GA | 614,403 | 11.13% | 11,300 | 11.59% | 2,439,901 | 44.18% | 36,744 | 37.68% | 8,310 | 0.15% | 154 | 0.16% |
| Baltimore, MD | 158,559 | 6.59% | 4,538 | 6.64% | 1,189,788 | 49.44% | 30,414 | 44.48% | 4,620 | 0.19% | 148 | 0.22% |
| Boston, MA--NH--RI | 555,601 | 11.62% | 28,021 | 15.00% | 3,291,310 | 68.84% | 115,458 | 61.80% | 5,309 | 0.11% | 259 | 0.14% |
| Chicago, IL--IN | 2,007,048 | 22.81% | 62,220 | 22.52% | 4,447,686 | 50.56% | 131,635 | 47.65% | 8,087 | 0.09% | 287 | 0.10% |
| Dallas--Fort Worth--Arlington, TX | 1,914,761 | 30.79% | 58,423 | 32.48% | 2,577,740 | 41.45% | 67,412 | 37.48% | 14,624 | 0.24% | 402 | 0.22% |
| Denver--Aurora, CO | 669,986 | 23.85% | 19,518 | 25.82% | 1,751,784 | 62.36% | 45,395 | 60.05% | 10,658 | 0.38% | 319 | 0.42% |
| Detroit, MI | 183,734 | 4.70% | 6,153 | 5.24% | 2,468,701 | 63.16% | 67,064 | 57.14% | 9,357 | 0.24% | 288 | 0.25% |
| Houston, TX | 2,359,738 | 39.07% | 68,687 | 42.45% | 1,925,596 | 31.88% | 45,666 | 28.22% | 10,182 | 0.17% | 253 | 0.16% |
| Las Vegas--Henderson, NV | 686,583 | 31.68% | 16,391 | 35.02% | 886,192 | 40.89% | 17,970 | 38.39% | 9,626 | 0.44% | 198 | 0.42% |
| Los Angeles--Long Beach--Anaheim, CA | 5,888,756 | 46.39% | 367,360 | 48.89% | 3,443,866 | 27.13% | 189,201 | 25.18% | 23,341 | 0.18% | 1,406 | 0.19% |
| Miami, FL | 2,752,279 | 45.21% | 95,563 | 48.08% | 1,847,912 | 30.36% | 51,921 | 26.12% | 6,454 | 0.11% | 203 | 0.10% |
| Minneapolis--St. Paul, MN--WI | 197,076 | 6.51% | 5,702 | 8.03% | 2,157,705 | 71.32% | 45,682 | 64.33% | 13,086 | 0.43% | 436 | 0.61% |
| New York--Newark, NY--NJ--CT | 4,717,835 | 24.75% | 230,320 | 28.37% | 8,578,756 | 45.01% | 313,399 | 38.60% | 26,652 | 0.14% | 1,217 | 0.15% |
| Philadelphia, PA--NJ--DE--MD | 572,794 | 9.77% | 17,939 | 11.71% | 3,548,830 | 60.54% | 83,888 | 54.78% | 7,044 | 0.12% | 229 | 0.15% |
| Phoenix--Mesa, AZ | 1,305,362 | 30.22% | 52,892 | 33.06% | 2,396,042 | 55.47% | 82,781 | 51.74% | 74,673 | 1.73% | 2,683 | 1.68% |
| Portland, OR--WA | 271,231 | 12.46% | 10,220 | 13.39% | 1,542,869 | 70.89% | 52,695 | 69.05% | 10,601 | 0.49% | 411 | 0.54% |
| Riverside--San Bernardino, CA | 1,271,256 | 57.65% | 42,729 | 60.74% | 533,254 | 24.18% | 15,307 | 21.76% | 6,058 | 0.27% | 209 | 0.30% |
| San Antonio, TX | 1,256,159 | 57.35% | 35,287 | 61.50% | 666,022 | 30.41% | 15,112 | 26.34% | 3,241 | 0.15% | 87 | 0.15% |
| San Diego, CA | 1,101,719 | 33.98% | 46,754 | 36.39% | 1,444,090 | 44.54% | 54,075 | 42.08% | 8,630 | 0.27% | 301 | 0.23% |
| San Francisco--Oakland, CA | 795,635 | 21.89% | 44,766 | 22.58% | 1,298,892 | 35.74% | 66,333 | 33.45% | 8,170 | 0.22% | 459 | 0.23% |
| Seattle, WA | 379,345 | 10.43% | 12,428 | 11.23% | 2,202,497 | 60.55% | 62,101 | 56.10% | 20,992 | 0.58% | 636 | 0.57% |
| St. Louis, MO--IL | 77,261 | 3.38% | 2,017 | 3.60% | 1,570,831 | 68.66% | 36,589 | 65.36% | 3,370 | 0.15% | 86 | 0.15% |
| Tampa--St. Petersburg, FL | 590,013 | 20.52% | 13,327 | 22.17% | 1,740,235 | 60.52% | 33,949 | 56.48% | 5,667 | 0.20% | 113 | 0.19% |
| Washington, DC--VA--MD | 911,064 | 17.22% | 28,401 | 17.29% | 2,188,896 | 41.36% | 66,587 | 40.53% | 9,376 | 0.18% | 297 | 0.18% |

Definition of abbreviation: Pop: population.

[a] Greater than or equal to 54 dB Lden.

**Table S9. Table S7 Continued.**

| Urban Area | NHPI | | | | Other | | | |
|---|---|---|---|---|---|---|---|---|
| | Pop | % | #HA | %HA | Pop | % | #HA | %HA |
| **Aviation** | | | | | | | | |
| Atlanta, GA | 1,907 | 0.03% | 61 | 0.05% | 171,555 | 3.11% | 3,537 | 2.64% |
| Baltimore, MD | 855 | 0.04% | 13 | 0.02% | 88,998 | 3.70% | 3,803 | 4.89% |
| Boston, MA--NH--RI | 1,530 | 0.03% | 58 | 0.04% | 167,401 | 3.50% | 4,086 | 2.93% |
| Chicago, IL--IN | 2,249 | 0.03% | 99 | 0.02% | 207,294 | 2.36% | 13,295 | 2.07% |
| Dallas--Fort Worth--Arlington, TX | 5,813 | 0.09% | 252 | 0.10% | 165,970 | 2.67% | 6,784 | 2.57% |
| Denver--Aurora, CO | 3,664 | 0.13% | 115 | 0.19% | 88,546 | 3.15% | 2,120 | 3.42% |
| Detroit, MI | 934 | 0.02% | 17 | 0.03% | 121,366 | 3.11% | 1,397 | 2.76% |
| Houston, TX | 2,449 | 0.04% | 84 | 0.04% | 138,862 | 2.30% | 4,087 | 2.18% |
| Las Vegas--Henderson, NV | 15,721 | 0.73% | 1,050 | 0.83% | 101,353 | 4.68% | 5,485 | 4.33% |
| Los Angeles--Long Beach--Anaheim, CA | 30,295 | 0.24% | 1,427 | 0.23% | 381,863 | 3.01% | 17,519 | 2.79% |
| Miami, FL | 1,984 | 0.03% | 45 | 0.01% | 123,467 | 2.03% | 5,018 | 1.41% |
| Minneapolis--St. Paul, MN--WI | 852 | 0.03% | 11 | 0.01% | 125,283 | 4.14% | 4,889 | 4.59% |
| New York--Newark, NY--NJ--CT | 5,059 | 0.03% | 278 | 0.03% | 542,430 | 2.85% | 24,706 | 2.62% |
| Philadelphia, PA--NJ--DE--MD | 1,468 | 0.03% | 19 | 0.03% | 174,432 | 2.98% | 1,773 | 3.21% |
| Phoenix--Mesa, AZ | 8,074 | 0.19% | 428 | 0.24% | 130,706 | 3.03% | 5,323 | 2.99% |
| Portland, OR--WA | 11,522 | 0.53% | 327 | 0.54% | 109,557 | 5.03% | 3,131 | 5.19% |
| Riverside--San Bernardino, CA | 5,454 | 0.25% | 151 | 0.52% | 58,329 | 2.65% | 1,092 | 3.75% |
| San Antonio, TX | 1,849 | 0.08% | 61 | 0.15% | 50,096 | 2.29% | 865 | 2.15% |
| San Diego, CA | 12,479 | 0.38% | 1,257 | 0.55% | 136,410 | 4.21% | 9,770 | 4.28% |
| San Francisco--Oakland, CA | 26,275 | 0.72% | 2,087 | 1.69% | 189,111 | 5.20% | 5,577 | 4.51% |
| Seattle, WA | 31,800 | 0.87% | 3,064 | 1.34% | 236,204 | 6.49% | 15,199 | 6.64% |
| St. Louis, MO--IL | 511 | 0.02% | 17 | 0.03% | 73,939 | 3.23% | 1,828 | 3.41% |
| Tampa--St. Petersburg, FL | 1,952 | 0.07% | 8 | 0.02% | 85,291 | 2.97% | 1,469 | 2.97% |
| Washington, DC--VA--MD | 2,668 | 0.05% | 96 | 0.08% | 213,301 | 4.03% | 5,134 | 4.35% |
| **Rail** | | | | | | | | |
| Atlanta, GA | 1,907 | 0.03% | 26 | 0.05% | 171,555 | 3.11% | 1,931 | 3.52% |
| Baltimore, MD | 855 | 0.04% | 9 | 0.03% | 88,998 | 3.70% | 1,064 | 3.69% |
| Boston, MA--NH--RI | 1,530 | 0.03% | 53 | 0.04% | 167,401 | 3.50% | 5,152 | 3.83% |
| Chicago, IL--IN | 2,249 | 0.03% | 116 | 0.02% | 207,294 | 2.36% | 12,125 | 2.19% |
| Dallas--Fort Worth--Arlington, TX | 5,813 | 0.09% | 135 | 0.12% | 165,970 | 2.67% | 2,422 | 2.13% |
| Denver--Aurora, CO | 3,664 | 0.13% | 20 | 0.09% | 88,546 | 3.15% | 605 | 2.70% |
| Detroit, MI | 934 | 0.02% | 21 | 0.11% | 121,366 | 3.11% | 797 | 4.09% |
| Houston, TX | 2,449 | 0.04% | 24 | 0.02% | 138,862 | 2.30% | 1,485 | 1.39% |
| Las Vegas--Henderson, NV | 15,721 | 0.73% | 16 | 0.49% | 101,353 | 4.68% | 158 | 4.90% |
| Los Angeles--Long Beach--Anaheim, CA | 30,295 | 0.24% | 360 | 0.22% | 381,863 | 3.01% | 3,200 | 1.95% |
| Miami, FL | 1,984 | 0.03% | 12 | 0.01% | 123,467 | 2.03% | 1,522 | 1.80% |
| Minneapolis--St. Paul, MN--WI | 852 | 0.03% | 23 | 0.04% | 125,283 | 4.14% | 2,722 | 4.70% |
| New York--Newark, NY--NJ--CT | 5,059 | 0.03% | 202 | 0.02% | 542,430 | 2.85% | 25,439 | 3.09% |
| Philadelphia, PA--NJ--DE--MD | 1,468 | 0.03% | 48 | 0.02% | 174,432 | 2.98% | 6,969 | 3.10% |
| Phoenix--Mesa, AZ | 8,074 | 0.19% | 15 | 0.15% | 130,706 | 3.03% | 352 | 3.49% |
| Portland, OR--WA | 11,522 | 0.53% | 273 | 0.66% | 109,557 | 5.03% | 2,081 | 5.04% |
| Riverside--San Bernardino, CA | 5,454 | 0.25% | 72 | 0.17% | 58,329 | 2.65% | 737 | 1.72% |
| San Antonio, TX | 1,849 | 0.08% | 13 | 0.03% | 50,096 | 2.29% | 694 | 1.62% |
| San Diego, CA | 12,479 | 0.38% | 112 | 0.50% | 136,410 | 4.21% | 774 | 3.45% |
| San Francisco--Oakland, CA | 26,275 | 0.72% | 879 | 0.79% | 189,111 | 5.20% | 5,269 | 4.72% |
| Seattle, WA | 31,800 | 0.87% | 315 | 0.92% | 236,204 | 6.49% | 2,265 | 6.59% |
| St. Louis, MO--IL | 511 | 0.02% | 8 | 0.02% | 73,939 | 3.23% | 1,170 | 3.00% |
| Tampa--St. Petersburg, FL | 1,952 | 0.07% | 1 | 0.04% | 85,291 | 2.97% | 84 | 2.88% |
| Washington, DC--VA--MD | 2,668 | 0.05% | 44 | 0.06% | 213,301 | 4.03% | 3,010 | 4.06% |
| **Road** | | | | | | | | |
| Atlanta, GA | 1,907 | 0.03% | 33 | 0.03% | 171,555 | 3.11% | 3,122 | 3.20% |
| Baltimore, MD | 855 | 0.04% | 29 | 0.04% | 88,998 | 3.70% | 2,467 | 3.61% |
| Boston, MA--NH--RI | 1,530 | 0.03% | 72 | 0.04% | 167,401 | 3.50% | 6,865 | 3.67% |
| Chicago, IL--IN | 2,249 | 0.03% | 54 | 0.02% | 207,294 | 2.36% | 6,785 | 2.46% |
| Dallas--Fort Worth--Arlington, TX | 5,813 | 0.09% | 195 | 0.11% | 165,970 | 2.67% | 4,777 | 2.66% |
| Denver--Aurora, CO | 3,664 | 0.13% | 106 | 0.14% | 88,546 | 3.15% | 2,400 | 3.17% |
| Detroit, MI | 934 | 0.02% | 31 | 0.03% | 121,366 | 3.11% | 3,703 | 3.16% |
| Houston, TX | 2,449 | 0.04% | 63 | 0.04% | 138,862 | 2.30% | 3,592 | 2.22% |
| Las Vegas--Henderson, NV | 15,721 | 0.73% | 323 | 0.69% | 101,353 | 4.68% | 2,098 | 4.48% |
| Los Angeles--Long Beach--Anaheim, CA | 30,295 | 0.24% | 1,801 | 0.24% | 381,863 | 3.01% | 22,237 | 2.96% |
| Miami, FL | 1,984 | 0.03% | 75 | 0.04% | 123,467 | 2.03% | 3,800 | 1.91% |
| Minneapolis--St. Paul, MN--WI | 852 | 0.03% | 23 | 0.03% | 125,283 | 4.14% | 3,226 | 4.54% |
| New York--Newark, NY--NJ--CT | 5,059 | 0.03% | 217 | 0.03% | 542,430 | 2.85% | 24,510 | 3.02% |
| Philadelphia, PA--NJ--DE--MD | 1,468 | 0.03% | 47 | 0.03% | 174,432 | 2.98% | 4,615 | 3.01% |
| Phoenix--Mesa, AZ | 8,074 | 0.19% | 337 | 0.21% | 130,706 | 3.03% | 5,011 | 3.13% |
| Portland, OR--WA | 11,522 | 0.53% | 460 | 0.60% | 109,557 | 5.03% | 3,918 | 5.13% |
| Riverside--San Bernardino, CA | 5,454 | 0.25% | 142 | 0.20% | 58,329 | 2.65% | 1,768 | 2.51% |
| San Antonio, TX | 1,849 | 0.08% | 47 | 0.08% | 50,096 | 2.29% | 1,204 | 2.10% |
| San Diego, CA | 12,479 | 0.38% | 459 | 0.36% | 136,410 | 4.21% | 5,210 | 4.05% |
| San Francisco--Oakland, CA | 26,275 | 0.72% | 1,632 | 0.82% | 189,111 | 5.20% | 10,422 | 5.26% |
| Seattle, WA | 31,800 | 0.87% | 1,054 | 0.95% | 236,204 | 6.49% | 7,380 | 6.67% |
| St. Louis, MO--IL | 511 | 0.02% | 14 | 0.02% | 73,939 | 3.23% | 1,853 | 3.31% |
| Tampa--St. Petersburg, FL | 1,952 | 0.07% | 43 | 0.07% | 85,291 | 2.97% | 1,806 | 3.00% |
| Washington, DC--VA--MD | 2,668 | 0.05% | 96 | 0.06% | 213,301 | 4.03% | 6,399 | 3.89% |

Definition of abbreviation: Pop: population.